\documentclass[oldversion]{aa}  
\usepackage{graphicx}
\usepackage{savesymbol}
\usepackage{amsmath}
\savesymbol{iint}
\usepackage{txfonts}
\restoresymbol{TXF}{iint}
\usepackage{txfonts}
\usepackage{float}
\usepackage{placeins}
\usepackage{lscape}
\usepackage{natbib}

\bibpunct{(}{)}{;}{a}{}{,}
\DeclareGraphicsExtensions{.eps}
\graphicspath{{./pub_graphics/}}

\renewcommand{\deg}{\hbox{$^\circ$}}
\newcommand{\hmsm}[4]{{#1}^{\rm h}{#2}^{\rm m}{#3}^{\rm s}\!\!.\:\!{#4}}
\newcommand{\dmsm}[4]{\, {#1}^{\circ}{#2}^{\prime}{#3}^{\prime\prime}\!\!\!.\;\!{#4}}

\newcommand{\tth}{\cdot 10^{19}\,{\rm atoms}\,{\rm cm}^{-2}}
\newcommand{\mJyb}{\,{\rm mJy}/{\rm beam}}

\def\fm{\hbox{$.\!\!^{\rm m}$}}

\begin{document}
   \title{Kinematic modeling of disk galaxies}

   \subtitle{III. The warped ``Spindle'' NGC~2685}

      \author{G. I. G. J\'ozsa\thanks{\email{gjozsa@astro.uni-bonn.de}}
          \inst{1}
                    \and
T. A. Oosterloo\thanks{\email{oosterloo@astron.nl}}
\inst{1,3}
                    \and
R. Morganti\thanks{\email{morganti@astron.nl}}
\inst{1,3}
    \and
          U. Klein\thanks{\email{uklein@astro.uni-bonn.de}}
          \inst{2}
    \and
          T. Erben\thanks{\email{terben@astro.uni-bonn.de}}
          \inst{2}
      }
   \offprints{G. J\'ozsa}

   \institute{Netherlands Institute for Radio Astronomy, Postbus 2,
 7990 AA Dwingeloo, The Netherlands\\
\and
Argelander Institute for Astronomy (AIfA), Univ. Bonn,
              Auf dem H\"ugel 71, D-53121 Bonn\\
         \and
              Kapteyn Astronomical Institute, Univ. Groningen, Postbus 800, 9700 AV Groningen, The Netherlands}

   \date{Received 9 January 2008 / Accepted 10 October 2008}

\abstract { 
This is the third of a series of papers in which the structure and
kinematics of disk galaxies is studied. Employing direct tilted-ring
fits to the data cube as introduced in Paper I, we perform an analysis
of the ``Spindle'' galaxy NGC~2685, previously
regarded as two-ringed polar ring galaxy.

Deep \ion{H}{i} and optical (i$^{\prime}$-band) observations are
presented. The \ion{H}{i} observations strongly suggest that the
gaseous structure of NGC~2685 does not consist of two separate
mutually inclined regions, but forms a coherent, extremely warped
disk, the appearance of two rings being due to projection effects. By
comparing the \ion{H}{i} total-intensity maps with the optical image
we demonstrate that at large radii a faint stellar disk is well
aligned with the outer \ion{H}{i} disk. The shape of the dust-lanes
obscuring the NE part of the inner stellar body indicates that also at
smaller radii NGC~2685 possesses a disk containing gas, dust, and
stars in which the various constituents are aligned.

At smaller radii, this disk is kinematically decoupled from the
central stellar body. Hence, in the region of the bright, central
stellar body, NGC~2685 appears to consist of two disks that share a
common centre, but have different orientation: a bright stellar
lenticular body apparently devoid of dust and gas, and a heavily
warped low-surface brightness disk containing stars, gas and dust. The
low-surface-brightness disk changes its orientation gradually and at
large radii assumes the orientation of the central stellar S0
disk. Since, according to our analysis, the intrinsic orientation of
the low-surface-brightness disk changes through $70\deg$, the gaseous
disk is coherent, and is at no radius orientated perpendicularly with
respect to the central stellar body, NGC~2685 is likely not a
classical polar-ring galaxy.
}

\keywords{Galaxies: kinematics and dynamics -- Galaxies: structure -- Galaxies: ISM -- Galaxies: peculiar Galaxies: individual: NGC~2685}
   \maketitle
%
\section{Introduction}
\label{Sect_3.1}
NGC~2685 obtained the name ``Spindle Galaxy'' from
\citet[][]{Sandage61} who referred to this object as ``perhaps the
most unusual galaxy in the Shapely-Ames catalogue''. The image
consists of a bright central object elongated from NE to SW. The whole
of the NE side of the central object is obscured by dust-lanes with
associated star-forming regions. Dust-lanes and {H\,{\small II}}
regions are missing on the SW side. In projection the dust-lanes
resemble a helical structure, hence NGC~2685 is also often referred to
as the ``Helix''. At larger radii an outer stellar ring is visible on
optical images that appears to have the same orientation as the
central object. Basic properties of NGC~2685 are
listed in Table~\ref{Tab_1} where distance dependent quantities
taken from the literature were corrected for differences in the
assumed distances (see Sect.~\ref{Sect_5}).

An inner ionised gas component, rotating rapidly on orbits
perpendicular to the major axis of the central stellar object, was
detected by \citet[][]{Ulrich75}. She showed that inside radii of
$4\arcsec$ to $10\arcsec$ this component is rotating differentially
with a projected amplitude of about $125\,{\rm km}\,{\rm
s}^{-1}$. This was confirmed by \citet[][]{Schechter78}, who analysed
stellar absorption- and emission spectra employing long-slit
observations along the projected major- and minor axes of the central
body. For the stellar component they found a velocity gradient along
the major axis. They observed that the central galaxy shows
non-solid-body rotation with a projected amplitude of $115\,{\rm
km}\,{\rm s}^{-1}$ and concluded that it is a lenticular galaxy
\citep[see also][]{Whitmore90,Emsellem04}. \citet[][]{Whitmore90}
showed that the inner ring structure has an excess luminosity with
respect to the central body, indicating the existence of a stellar
component in the inner ring and the helical structure.
%
%
\begin{table*}
\caption{
Basic properties of NGC~2685.}
\label{Tab_1}
\begin{center}
\begin{tabular}{llrr}
\hline
\hline
Description                                                                                & Parameter name                                &                             & Reference                          \\
\hline
Classification of galaxy                                                                   & Type                                          &    SB0+                      &  \citealt{Vaucouleurs91}        \\
Right Ascension (J2000) (NED)                                                              & RA                                            & $ \hmsm{08}{55}{34}{75}      $ &  NED                            \\
Declination (J2000) (NED)                                                                  & Dec                                           & $ \dmsm{+58}{44}{03}{9}      $ &  NED                            \\
Optical heliocentric systemic velocity ($\rm km\,{\rm s}^{-1}$)                            & $V_{\rm sys}$                                 & $     875.2   \,\pm\,  2.0   $ &  this work                      \\
Distance of object ($\rm Mpc$)                                                             & $\rm D$                                       & $      15.2   \,\pm\,  3.8   $ & this work \\
Scale between distance on sky and true distance (${\rm pc}/\,^{\prime\prime}$).            & $sc$                                          & $      74     \,\pm\, 19     $ & this work                       \\
Apparent B-band magnitude ($\rm mag$).                                                     & $m_{\rm B}$                                   & $      12.05  \,\pm\,  0.15  $ & LEDA \citep[][]{LEDA}           \\
Apparent I-band magnitude ($\rm mag$).                                                     & $m_{\rm I}$                                   & $       9.90  \,\pm\,  0.10  $ & LEDA \citep[][]{LEDA}           \\
Extinction-corrected apparent B-band magnitude ($\rm mag$).                                & $m_{\rm B}^{\rm c}$                           & $      11.77  \,\pm\,  0.15  $ & NED                             \\
Extinction-corrected apparent I-band magnitude ($\rm mag$).                                & $m_{\rm I}^{\rm c}$                           & $       9.54  \,\pm\,  0.12  $ & this work                       \\
Absolute B-band magnitude ($\rm mag$).                                                     & $M_{\rm B}$                                   & $     -19.1   \,\pm\,  0.7   $ & this work                       \\
Absolute I-band magnitude ($\rm mag$).                                                     & $M_{\rm I}$                                   & $     -21.4   \,\pm\,  0.6   $ & this work                       \\
B-band luminosity ($ 10^9\, {L}_\odot $).                                                  & $L_{\rm B}$                                   & $       7.0   \,\pm\,  3.7   $ & this work                       \\
I-band luminosity ($ 10^9\, {L}_\odot $).                                                  & $L_{\rm I}$                                   & $      15.2   \,\pm\,  7.7   $ & this work                       \\
Total {\ion{H}{i}} flux (${\rm Jy}\,{\rm km}\,{\rm s}^{-1}$).                              & $F_{\rm {\ion{H}{i}}}$                        & $      31.4   \,\pm\,  3.1   $ & this work                       \\
{\ion{H}{i}} mass ($10^9\,{M}_\odot$).                                                     & $M_{\rm {\ion{H}{i}}}$                        & $       1.7   \,\pm\,  0.9   $ & this work                       \\
B-band optical radius ($\,^{\prime\prime}$).                                               & $r_{25}$                                      & $     147     \,\pm\, 30     $ & LEDA \citep[][]{LEDA}           \\
{\ion{H}{i}} radius ($\,^{\prime\prime}$).                                                 & $r_{\rm {\ion{H}{i}}}$                        & $     200     \,\pm\, 40     $ & this work                       \\
Terminal radius ($\,^{\prime\prime}$).                                                     & $r_{\rm t}$                                   & $     420     \,\pm\, 30     $ & this work                       \\
B-band optical radius ($\rm kpc$).                                                         & $R_{25}$                                      & $      10.8   \,\pm\,  2.7   $ & this work                       \\ 
{\ion{H}{i}} radius ($\rm kpc$).                                                           & $R_{\rm {\ion{H}{i}}}$                        & $      14.8   \,\pm\,  3.7   $ & this work                       \\
Terminal radius ($\rm kpc$).                                                               & $R_{\rm t}$                                   & $      31.0   \,\pm\,  7.7   $ & this work                       \\
Rotation velocity at terminal radius (${\rm km}\,{\rm s}^{-1}$).                           & $V_{\rm t}$                                   & $     147     \,\pm\, 15     $ & this work                       \\
Rotation period at terminal radius (${\rm Gyr}$).                                          & $t_{\rm t}$                                   & $       1.297 \,\pm\,  0.030 $ & this work                       \\
Dynamical mass ($10^9\,{M}_\odot$).                                                        & $M_{\rm dyn}$                                 & $ >\, 155     \,\pm\, 50     $ & this work                       \\
Ratio of {\ion{H}{i}} mass and B-band luminosity (${M}_\odot\,{L}_\odot^{-1}$).            & $\frac{M_{\rm {\ion{H}{i}}}}{L_{\rm B}}$      & $       0.2   \,\pm\,  0.03  $ & this work                       \\
Ratio of {\ion{H}{i}} mass and I-band luminosity (${M}_\odot\,{L}_\odot^{-1}$).            & $\frac{M_{\rm {\ion{H}{i}}}}{L_{\rm I}}$      & $       0.1   \,\pm\,  0.01  $ & this work                       \\
Ratio of dynamical mass and {\ion{H}{i}} mass.                                             & $\frac{M_{\rm dyn}}{M_{\rm {\ion{H}{i}}}}$    & $ >\,  91     \,\pm\, 25     $ & this work                       \\
Ratio of dynamical mass and B-band luminosity (${M}_\odot\,{L}_\odot^{-1}$).               & $\frac{M_{\rm dyn}}{L_{\rm B}}$               & $ >\,  22     \,\pm\,  7.3   $ & this work                       \\
Ratio of dynamical mass and I-band luminosity (${M}_\odot\,{L}_\odot^{-1}$).               & $\frac{M_{\rm dyn}}{L_{\rm I}}$               & $ >\,  10.2   \,\pm\,  3.4   $ & this work                       \\
\hline
\end{tabular}
\end{center}
\end{table*}

%
%
Hence, NGC~2685 is claimed to be a polar-ring galaxy with a gaseous-
and stellar component that rotates at rather small radii on a polar
orbit about a central lenticular galaxy.

Today's picture of NGC~2685 as containing two kinematically
independent, clearly separated kinematic systems in the neutral gas
was drawn by Shane \citeyearpar{Shane77,Shane80} and sustained in
other studies by \citet[][]{Mahon92} and
\citet[][]{Schinnerer02}. Examining \ion{H}{i} synthesis observations
of the galaxy, they concluded that the neutral gas resides in two
separate rings. One is supposedly connected with the dust-lane
structure seen in the galaxy. The other one at large radii is assumed
to be orientated perpendicularly with respect to the first ring. It
seems to be aligned with the central stellar body and to rotate at a
speed of about $150\,{\rm km}\,{\rm s}^{-1}$.

The purpose of this paper is to present new observations conducted
with the Westerbork Synthesis Radio Telescope (WSRT, Westerbork, The
Netherlands) and the 2.5m Isaac Newton Telescope (INT, Roque de los
Muchachos, Spain). Using our \ion{H}{i} observations to perform a
simple tilted-ring analysis, we argue that the picture of NGC 2685
containing two seperate \ion{H}{i} rings is probably not tenable: we
reproduce the \ion{H}{i} observations by assuming a coherent, warped
disk structure of the \ion{H}{i} component, in which the orbits are
circular. Comparing our model to the observations, there is no
evidence for significant deviations from circular orbits, except
possibly at the edges of the projected central body. Based on these
findings we suggest a different scenario: in addition to the central
stellar body NGC~2685 contains an extremely warped disk of low surface
brightness, consisting of gas, dust, and stars.  Only in projection
the galaxy appears to have two rings and a central ``helix''. At large
radii the gaseous disk shares its orientation with the central
lenticular galaxy. Since the single low-surface brightness disk is
coherent and changing its orientation gradually, NGC~2685 does not
contain two clearly separated systems with nearly orthogonal spin
vectors. Hence, NGC~2685 is probably not a classic polar ring galaxy
as defined by \citet[][ a polar ring galaxy contains two rotating
sub-systems, a disk and a ring or an additional disk, of comparable
sizes with orthogonal angular momentum vectors of similar amplitude;
both systems share the centre and the systemic velocity and the ring
is nearly planar.]{Whitmore90}: The rotation curve derived from the
\ion{H}{i} data resembles a rotation curve of a spiral galaxy, staying
relatively flat out to the last measured data point. The shape of
rotation curve, in conjunction with the extreme warp geometry,
indicates that the overall potential is not extremely flattened but
close to spherical.
%
%
\begin{table*}[htbp]
\caption{Summary of observations and data reduction.}
\label{Tab_2}
\begin{center}
\begin{tabular}{l r r r}
\hline
\hline
Name of telescope                                                                                 &                                             & \multicolumn{2}{c}{WSRT}            \\
Total on-source integration time ($\rm h$)                                                        & $t_{\rm obs}$                               & \multicolumn{2}{c}{$4\times 12$}    \\
Observation dates                                                                                 &                                             \multicolumn{3}{c}{19-20/12/02, 21-22/12/02, 7-8/01/03, 8-9/01/03}     \\
\hline                                                                           
Resolution of data cube                                                                           &                                             & high            & low                 \\
\hline                                                                           
Weighting scheme                                                                                  &                                             & uniform         & robust 0.4          \\
Pixel size ($\,\arcsec$)                                                                          & $dx$                                        & $ 4     $       & $ 7     $           \\
Channel width (${\rm km}\,{{\rm s}^{-1}}$)                                                        & $dv$                                        & $ 2.06  $       & $ 4.12  $           \\
Velocity resolution (${\rm km}\,{{\rm s}^{-1}}$)                                                  & $FWHM_{\rm V}$                              & $ 4.12  $       & $ 8.24  $           \\
Half-power-beam-width along the beam major axis ($\,\arcsec$)                                     & ${\rm HPBW}_{\rm maj}$                      & $ 13.7  $       & $ 28.4  $           \\
Half-power-beam-width along the beam minor axis ($\,\arcsec$)                                     & ${\rm HPBW}_{\rm min}$                      & $ 12.0  $       & $ 25.6  $           \\
Beam position angle ($\,\deg$)                                                                    & ${\rm PA}_{\rm beam}$                       & $ -0.6  $       & $ 10.6  $           \\
rms noise (${\rm mJy}/{\rm beam}$) in the data cubes                                              & $\sigma_{\rm rms}$                          & $ 0.42  $       & $ 0.24  $           \\
rms noise per velocity resolution element ($10^{19}\,{\rm atoms}\,{\rm cm}^{-2}$)                 & $\sigma_{\rm rms}^{N}$                      & $ 1.16  $       & $ 0.30  $           \\
\hline                               
Name of telescope/instrument                                                                      &                                             & \multicolumn{2}{c}{INT WFC}         \\
Total on-source integration time ($\rm s$)                                                        & $t_{\rm obs}$                               & \multicolumn{2}{c}{$13350$}        \\
Observing dates                                                                                   &                                             & \multicolumn{2}{c}{26/02-04/03/04}  \\
rms noise in the optical image (${\rm mag}\,{\rm arcsec}^{-2}$)                                   & $\sigma_{{\rm i}^\prime} $                  & \multicolumn{2}{c}{$ 25.8 $}        \\
rms noise in the optical image, $5\,\sigma_{\rm rms}$-level for a $2\arcsec$ aperture ($\rm mag$) & $\sigma_{{\rm i}^\prime,2^{\prime\prime}} $ & \multicolumn{2}{c}{$ 21.3 $}        \\
\hline
\end{tabular}
\end{center}
\end{table*}
%
%

The paper is laid out as follows: In Sect.~\ref{Sect_2} our
observations and the data reduction are described. In
Sect.~\ref{Sect_3} we compare the appearance of NGC~2685 in the optical
and in the \ion{H}{i}.  In Sect.~\ref{Sect_4} we present basic
arguments in favour of NGC~2685 possessing one coherent, warped disk,
rather than two separate disks. In Sect.~\ref{Sect_5} the data
analysis via a direct tilted-ring modelling is described, the results
of which are discussed in Sect.~\ref{Sect_6}. In
Sect.~\ref{Sect_7} the possible presence and location of
non-circular motions is discussed. In Sect.~\ref{Sect_8} we briefly
discuss dynamical implications of our kinematical
model. Section~\ref{Sect_9} discusses the detection of possible
companions to NGC~2685. Section~\ref{Sect_10} summarises our results
and a discusses possible formation scenarios for the peculiar
structure of NGC~2685. Supplementary figures and tables are placed in
the appendix.
\section{Observations and data reduction}
\label{Sect_2}
\subsection{Optical observations}
\label{Sect_2.1}
NGC~2685 was observed with the INT in the i$^\prime$-band from
February 26th until March 4th 2004 under non-photometric weather
conditions. Each night, sky flats and bias frames were obtained.
Using a series of exposures with a maximum of 400s single exposure
time while using a wide dithering scheme the total on-source observing
time amounts to $13350\,\rm s$.  To obtain a coadded image from the
data we made use of the GaBoDS Wide Field Imager reduction pipeline
\citep[][]{Schirmer03,Erben05}. After the overscan- and bias
correction, the images of each chip were flat-fielded,
super-flat-fielded and defringed on a per-night basis. Non-Gaussian
noise-features (cosmic rays, hot and cold pixels) were detected and
masked, and the single frames visually inspected to apply a
manual masking or, in some cases, to reject single frames. The images
were photometrically (relatively) calibrated,
background-subtracted and co-added, taking into account the sky
background variation in the single chips and solving for astrometric
distortion and for the relative sensitivity of the chips by making use
of the USNO-A2 standard catalog \citep[][]{Monet98}. Using average
zero-points for the WFC, as published in the Internet by the CASU INT
Wide Field Survey team \citep[][]{McMahon01} the resulting unbinned
coadded map can roughly be estimated to have a noise level of
$25.8\,{\rm mag}\,{\rm arcsec}^{-2}$ (see Table~\ref{Tab_2}. The
obtained image is shown in Fig.~\ref{Fig_01} (upper left panel).
%
%
\subsection{\ion{H}{i} observations}
\label{Sect_2.2}
\begin{figure*}[htbp]
\begin{center}
\includegraphics[angle=270,width=0.49\textwidth]{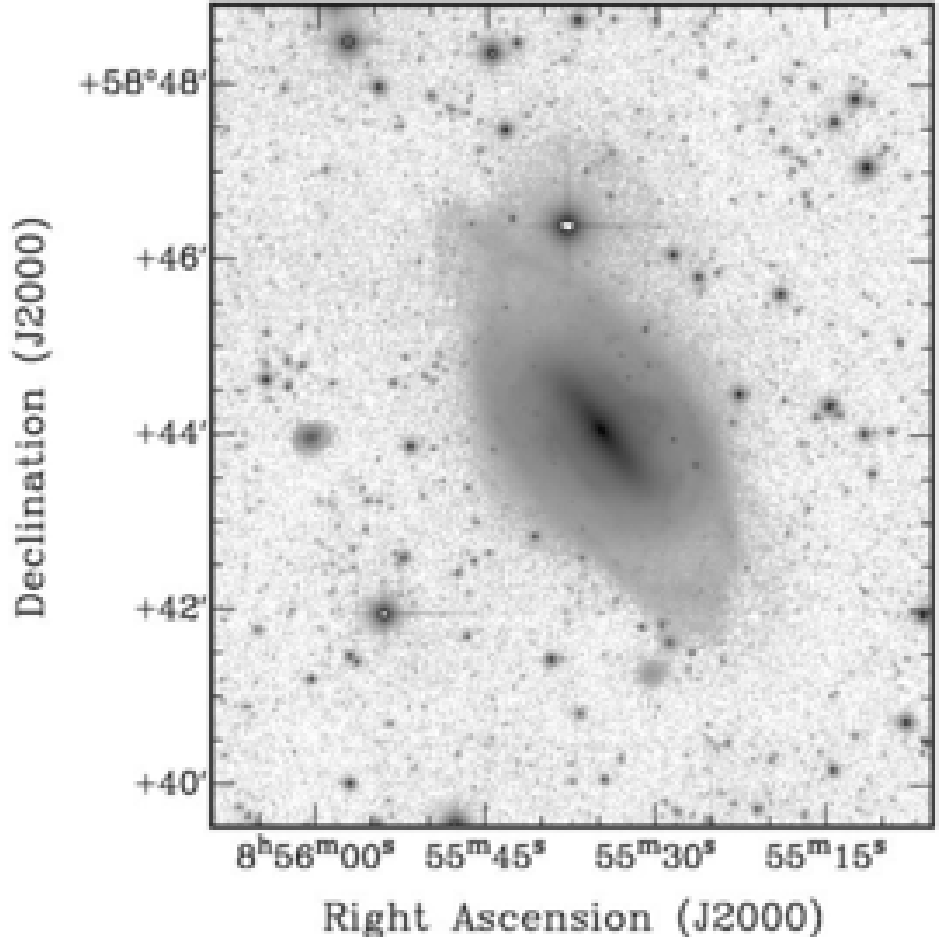}
\includegraphics[angle=270,width=0.49\textwidth]{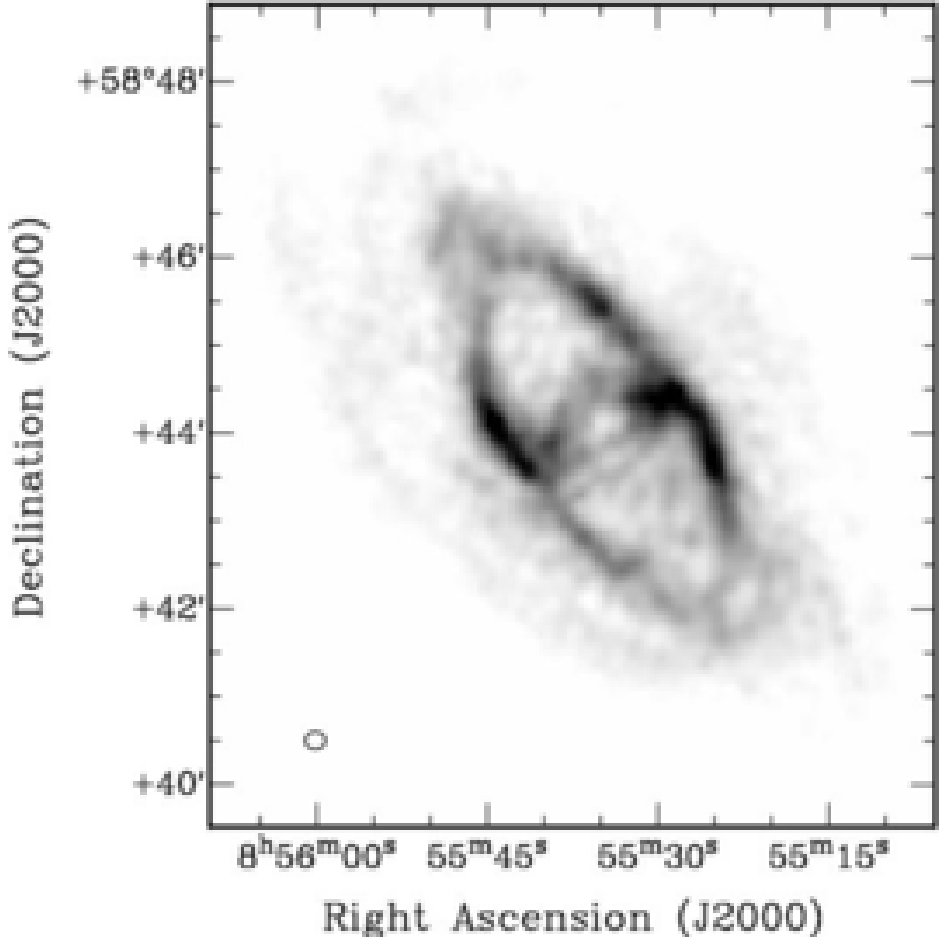}
\includegraphics[angle=270,width=0.49\textwidth]{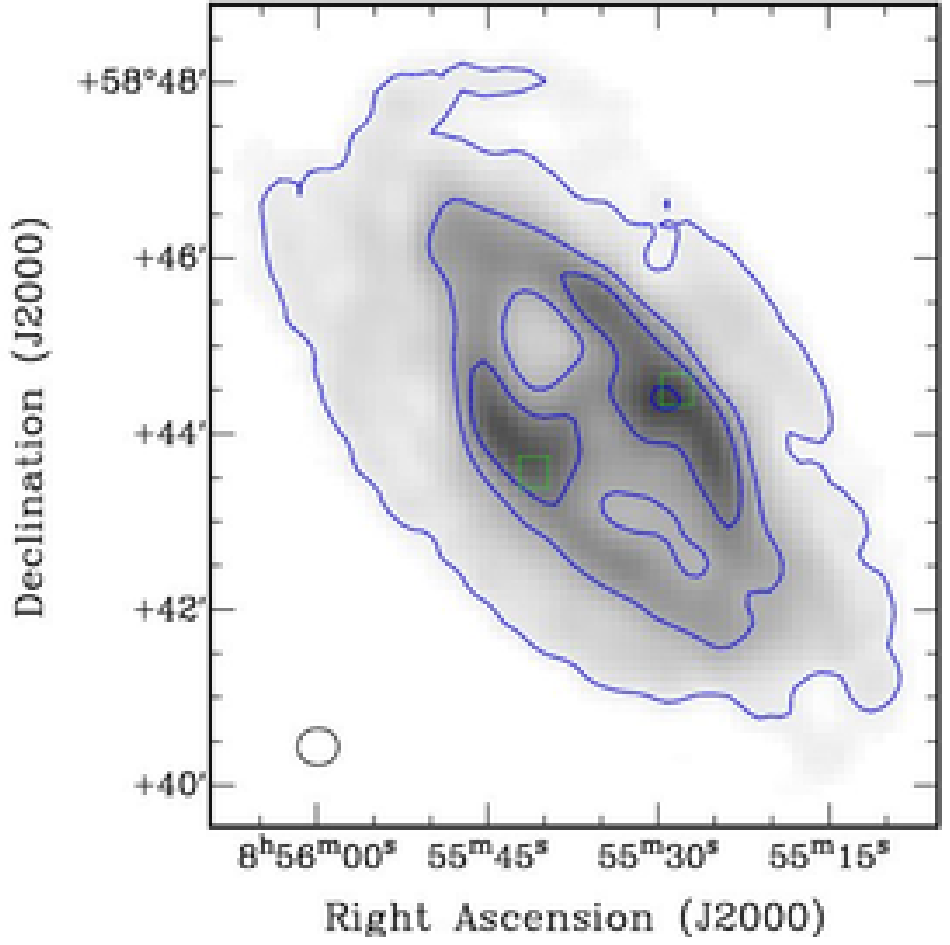}
\includegraphics[angle=270,width=0.49\textwidth]{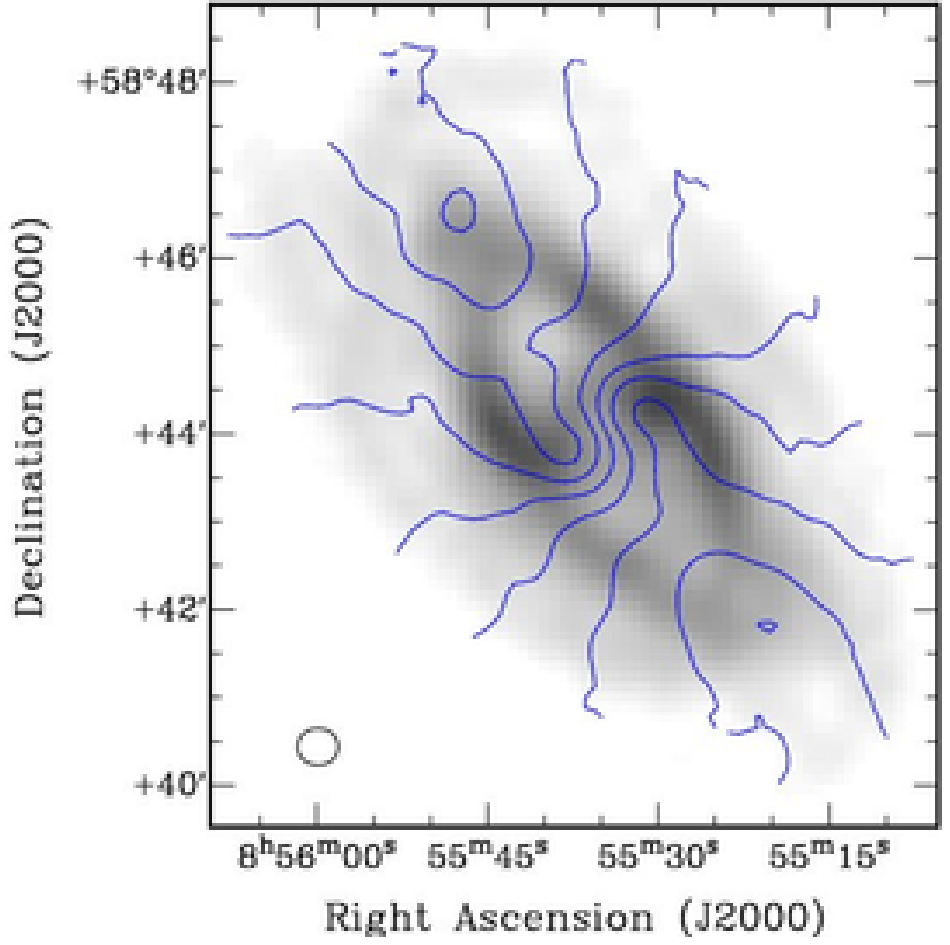}
\caption{Top, left panel: logarithmic grey-scale plot of the i'-band image
obtained with the INT. Top, right panel: Grey-scale plot of the
high-resolution \ion{H}{i} total-intensity map.  Bottom, left panel:
Low-resolution total-intensity \ion{H}{i} map (contours and grey
scale). Contours represent the $5, 40, 80, 160 \tth$ levels (see
Table~\ref{Tab_2}).  The boxes indicate the region where spectra were
taken to support the hypothesis that NGC~2685 contains a warped disk
(see Sect.~\ref{Sect_4}).  Bottom, right panel: first-moment velocity
field (contours $V_{\rm sys} \pm 0, 40, 80, 120, 140\,{\rm km}\,{\rm
s}^{-1}$) derived from the low-resolution data cube, overlaid on
low-resolution total-intensity map (grey scale). The approaching side
is orientated towards NE. Ellipses in the lower left corner of the
\ion{H}{i} maps represent the clean beam HPBW. All images are on the same spatial scale.}
\label{Fig_01}
\end{center}
\end{figure*}

NGC~2685 was observed with the WSRT in December 2002 (2$\times$12h)
and January 2004 (2$\times$12h, see Table~\ref{Tab_2}). To ensure a
good central uv-coverage all observations were done in maxi-short
configuration. We used a total bandwidth of $10\,\rm MHz$, two
parallel polarisations, and 1024 channels in total.  The data
underwent a standard data reduction with the Miriad
\citep[][]{Sault95} software package (for a detailed description see
\citealt{Jozsa06}). The continuum data resulting from a second-order
continuum subtraction with the Miriad-task uvlin were used to correct
the frequency independent gains by means of self-calibration. To
enhance the signal-to-noise ratio, we performed a second continuum
subtraction on the self-calibrated data. We used the clean table of
our continuum image to subtract a continuum model from the
visibilities (Miriad task uvmodel). Then, we performed a first-order
continuum subtraction using 550 line-free channels (Miriad task
uvlin).  Before inverting the visibilities we applied a Hanning
smoothing using the miriad task uvfil.  Two data cubes, a
high-resolution (uniform weighting using all baselines) and a
low-resolution one (Robust weighting of 0.4, using baselines of length
$<6.4\,{\rm k\lambda}$) were produced. The data cubes were deconvolved
using an iterative CLEAN with appropriate masks. The derived total
flux for NGC~2685 is in good agreement with single-dish measurements
\citep[][]{Richter94}, showing that no substantial fraction of the
flux is resolved out in the \ion{H}{i} observations.  Representative
examples of resulting channel maps are shown in Figs.~\ref{Fig_A1} and
\ref{Fig_A2}. Total intensity maps and first-moment maps were
generated by first masking the cubes with the finally applied clean
masks. The resulting high-resolution total-intensity map, the
low-resolution total-intensity map, and a low-resolution moment-1 map
are shown in Fig.~\ref{Fig_01}. A second plot of the high-resolution
total intensity map and the corresponding velocity field are shown in
\ref{Fig_A3}.
%
%
%
%
Table~\ref{Tab_2} gives an overview of the observations and related
parameters.  The quality of our \ion{H}{i} observations is
significantly better than that of former ones by \citet[][]{Shane80}
and \citet[][]{Mahon92}. \citet[][]{Shane80} obtains a sensitivity of
$4.5\cdot10^{19}\,{\rm atoms\,{\rm cm}^{-2}}$ per velocity resolution
element with a velocity resolution of $27\,{\rm km}{\rm s}^{-1}$
(FWHM) and a beam of $49\,\arcsec\times57\,\arcsec$ (HPBW). The high
sensitivity is reached at the expense of spatial resolution. The
observations of \citet[][]{Mahon92} yielded data cubes
\citep[cf.][]{Schinnerer02} with a velocity resolution of $41.4\,{\rm
km}\,{\rm s}^{-1}$ (FWHM) and a sensitivity of $1.2\cdot10^{19}\,{\rm
atoms\,{\rm cm}^{-2}}$ ($1.7\cdot10^{20}\,{\rm atoms\,{\rm cm}^{-2}}$)
per velocity resolution element, depending on the spatial resolution
(HPBW: $35\,\arcsec\times34\,\arcsec$, and
$13\,\arcsec\times11\,\arcsec$ respectively). While the spatial
resolution of our observations is comparable to that of the
observations of Mahon, the noise level of our data cubes is better by
a factor of 4 and the velocity resolution increased by a factor of 5
(10 for the high-resolution data cube, see Table~\ref{Tab_2}).
%
%
\section{The relation between stars and ISM in NGC~2685}
\label{Sect_3}
%
%
\begin{figure*}[htbp]
\includegraphics[width=0.32\textwidth]{xx_02a.eps}
\includegraphics[width=0.32\textwidth]{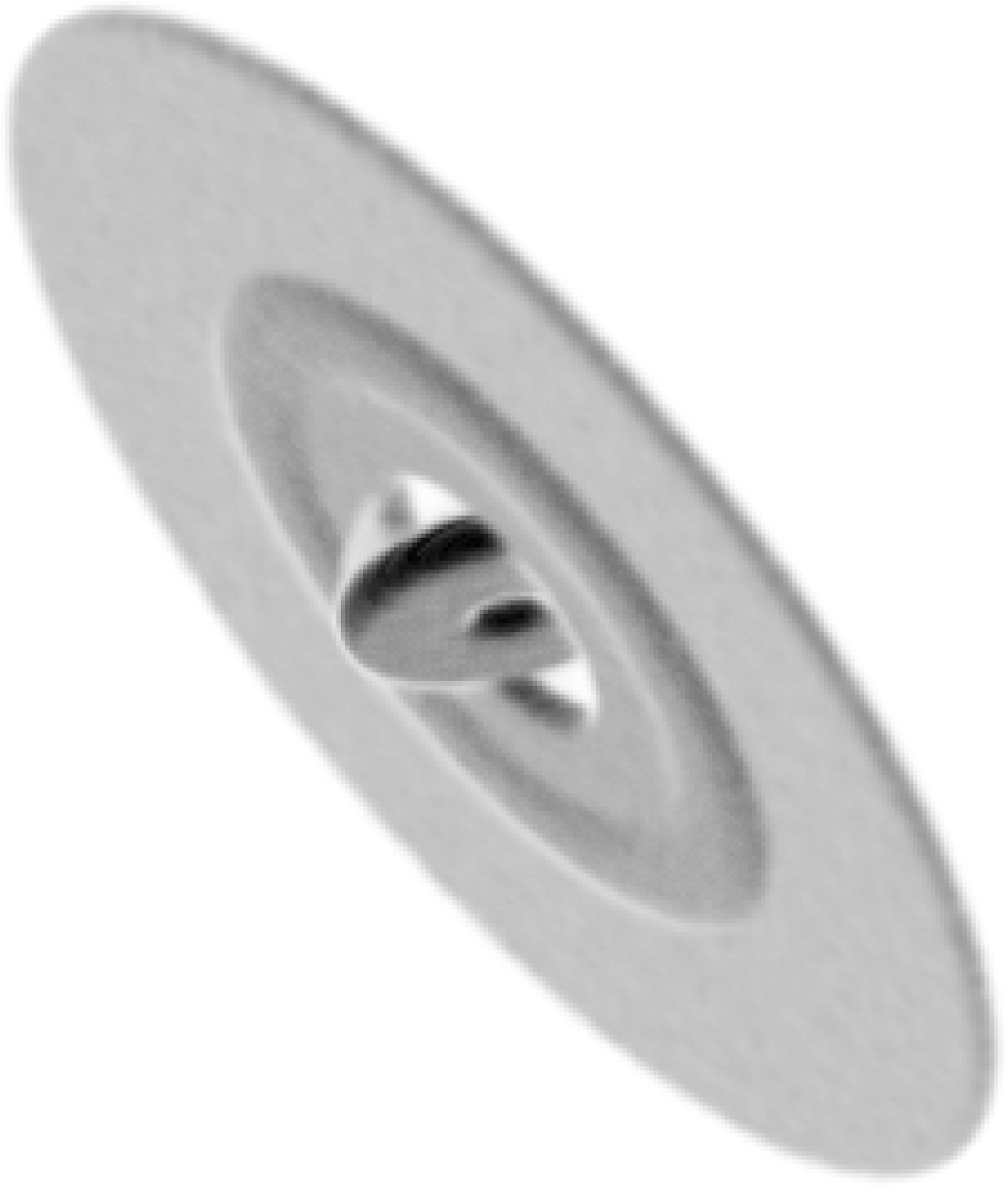}
\includegraphics[width=0.32\textwidth]{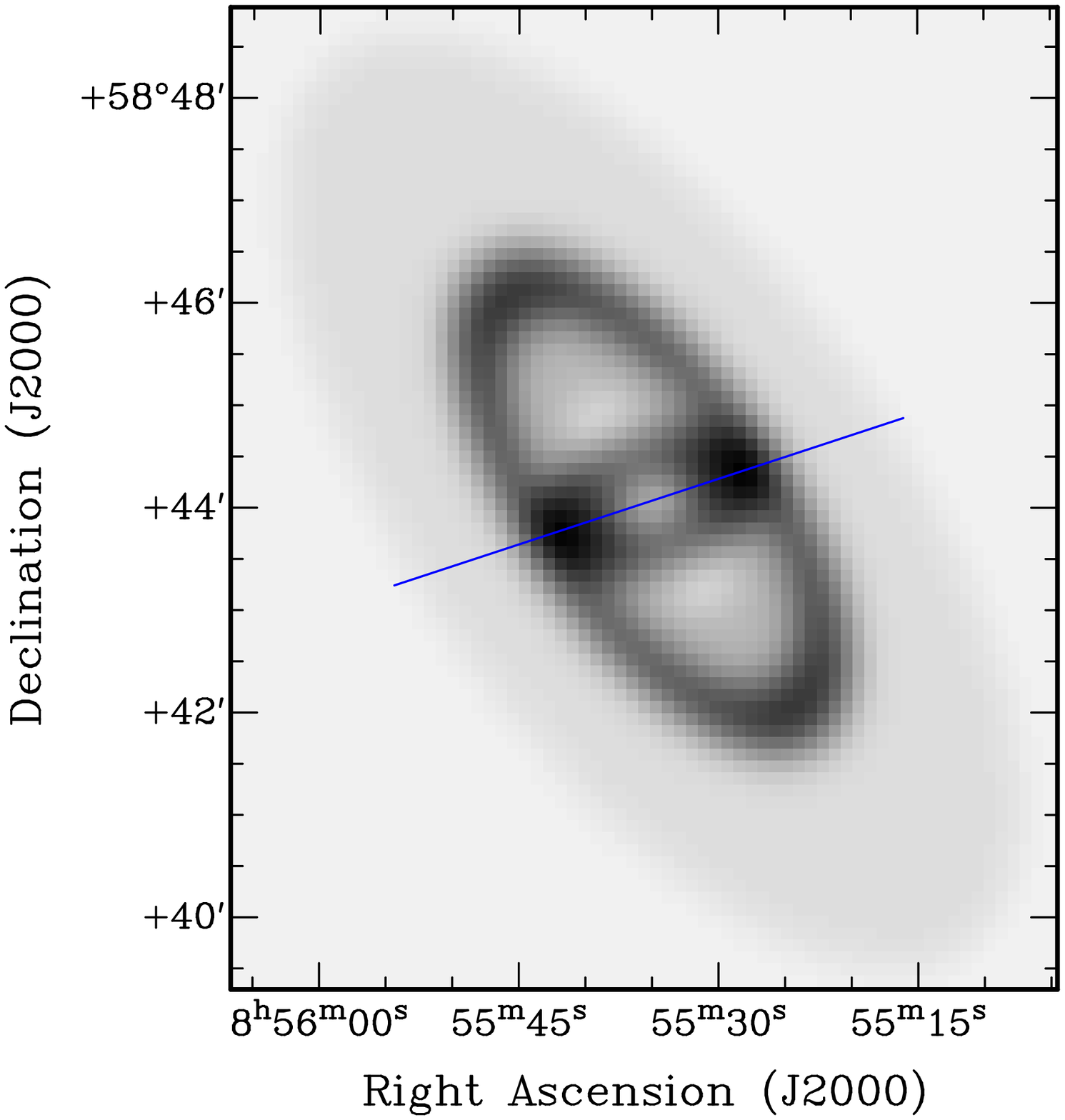}\\
\includegraphics[width=0.32\textwidth]{xx_02b.eps}
\includegraphics[width=0.32\textwidth]{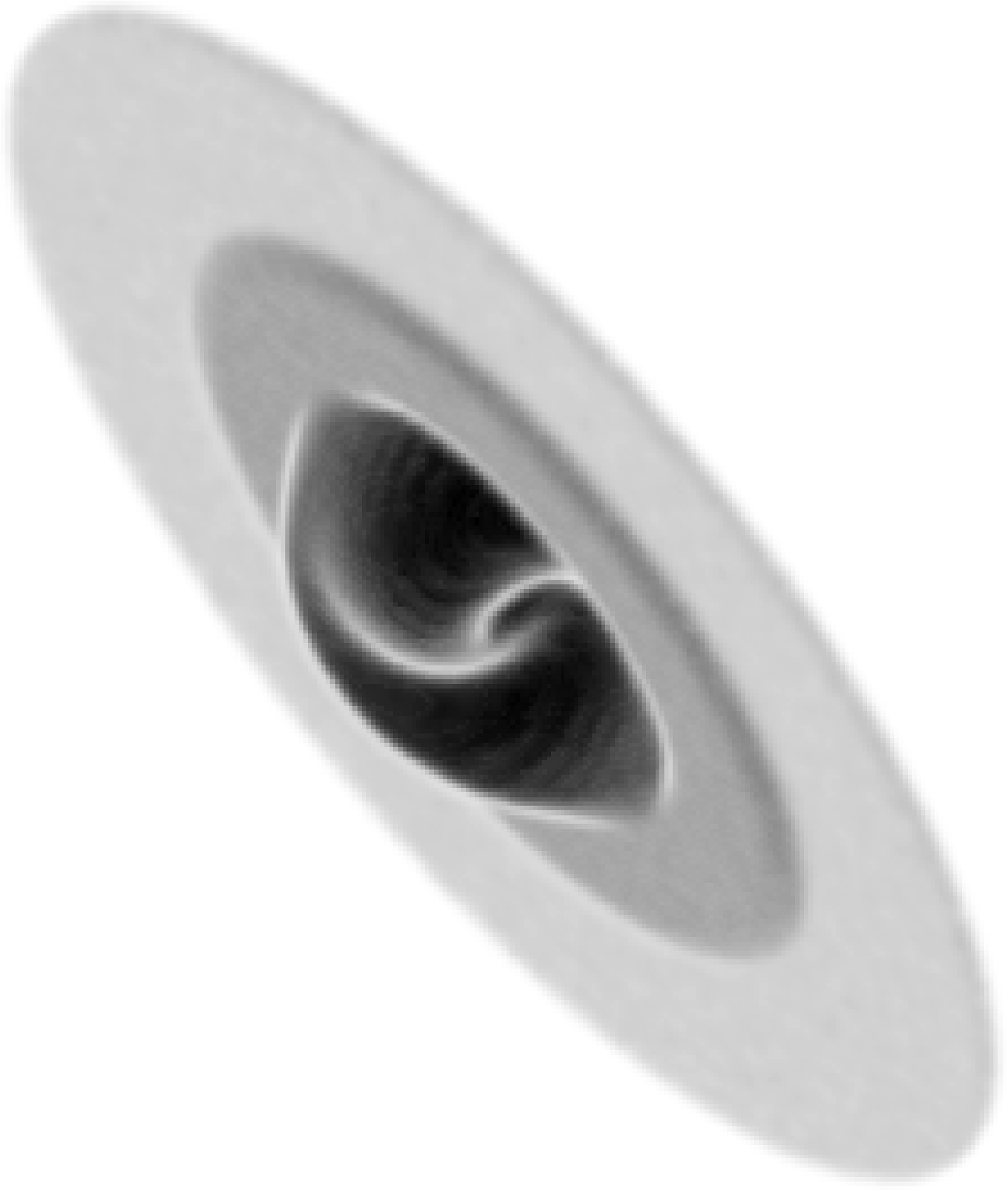}
\includegraphics[width=0.32\textwidth]{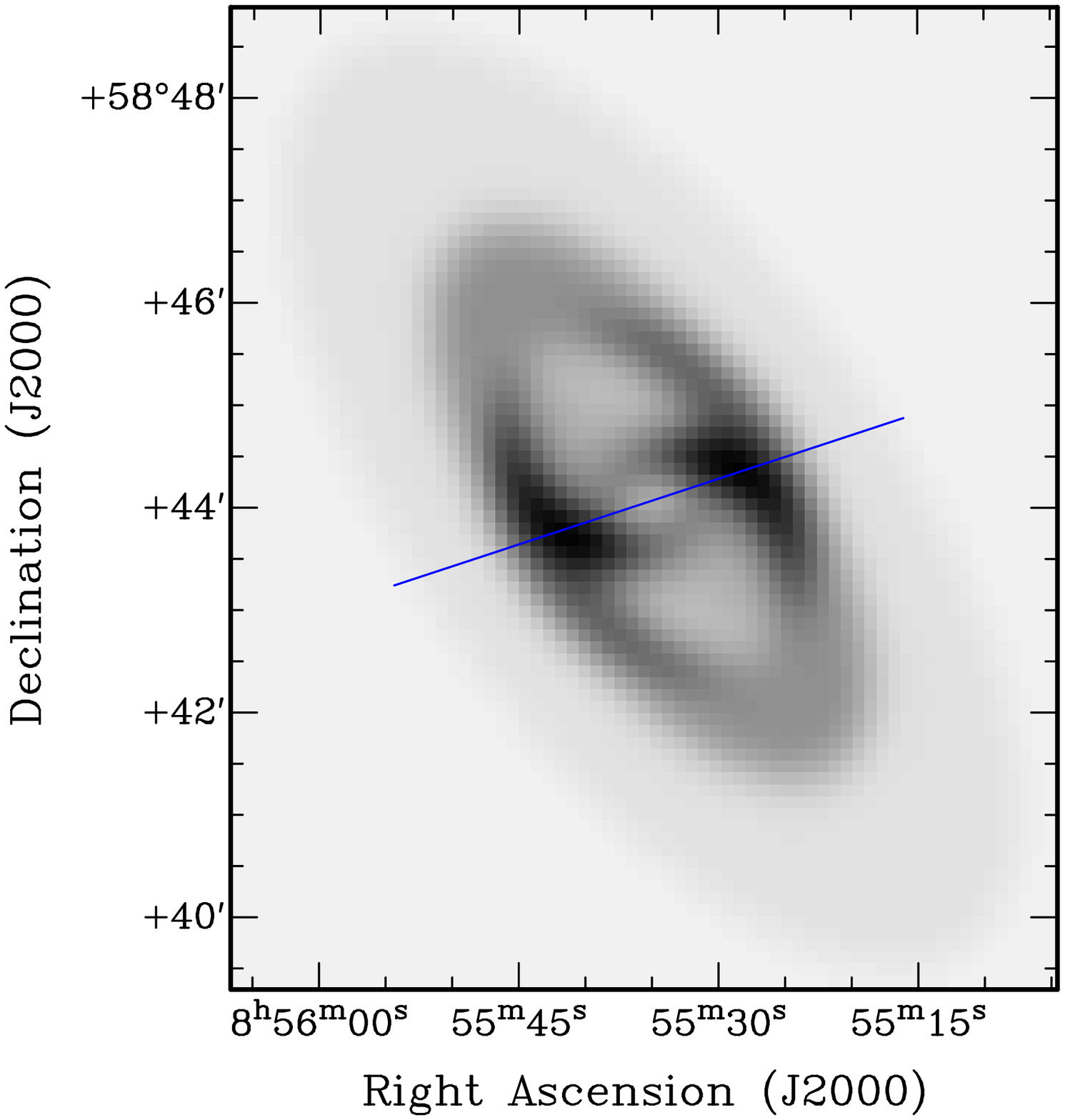}
\caption{
Parametrisations of toy model data cubes and toy model total-intensity
maps. Top panels: ``Flip model''.  The orientation of the disk flips
at a certain radius, the minimum requirement in order to speak of a
two-ring or two-disk system in NGC~2685. Bottom panels: ``Warp
model''. Here a coherent but heavily warped disk is assumed. Left:
surface brightness profile, rotation curve and disk orientation. SD:
\ion{H}{i} face-on column density/surface mass density. VROT: rotation
velocity. INCL: inclination. PA: position angle. R: radius. Middle:
three-dimensional visualisations of the disks. Right: Toy model
total-intensity maps. The lines trace the track along which the
spectra in the position-velocity diagrams (PV-diagrams) shown in
Fig.~\ref{Fig_03} are traced.}
\label{Fig_02}
\end{figure*}
%
%
The upper left panel in Fig.~\ref{Fig_01} shows the optical image of
the Spindle obtained with the INT WFC, while in the top right panel a
grey-scale plot of the \ion{H}{i} high-resolution total-intensity map
is shown, using the same spatial scale.  A comparison by eye is
sufficient to identify several structures to be present in all maps.

Firstly, apart from showing the central lenticular body, the optical
image resembles the integrated \ion{H}{i} map in such great detail
that there is little doubt that both stars and gas belong to the same
structure that exists in addition to the central stellar body.
Especially the small deviations from axisymmetry, such as a hook-like
structure to the NE side of the galaxy and a more diffuse structure
towards the SW side are visible in all images \citep[see
also][]{Shane77,Shane80}.

Secondly, there is no region completely devoid of stars or gas within
the projected outer ring. Thanks to the high sensitivity of our
observations, this is also evident in the high-resolution \ion{H}{i}
map (Fig.~\ref{Fig_01}, top right panel), where beam-smearing effects
can be excluded to cause the appearance of a filled inner region.
This means that the galaxy possesses an extended disk or, being a
rather implausible option, a stellar and gaseous halo that does not
belong to its inner bright lenticular stellar body.

Thirdly, looking at the high-resolution \ion{H}{i} map
(Fig.~\ref{Fig_01}, top right panel), it seems that the helical
structure is also present in the \ion{H}{i} component. Apparently, the
\ion{H}{i} follows the dust-lane structure like \ion{H}{i} follows
spiral arms in spiral galaxies. As the \ion{H}{i} component is
optically thin, the spiral-arm structures can be seen on both sides of
the galaxy. Using integral-field emission line spectra from the SAURON
survey \citep[][]{Emsellem04} and the \ion{H}{i} data presented here,
\citet[][]{Morganti06} show that the neutral gas 
has an ionised counterpart towards the centre \citep[see
also][]{Ulrich75} of the galaxy and that in the overlap region the
ionised gas shares the kinematical structure of the \ion{H}{i}.

It can be concluded that the \ion{H}{i} component in the galaxy
belongs to one or two structures, distinct from the central stellar
body, which contain molecular, neutral, and ionised gas, dust, and
stars. The distribution of all components follows the \ion{H}{i}
distribution. At the largest radii beyond the outer projected ring,
the \ion{H}{i} maps show a disk-like extension that cannot be seen in
the optical image, possibly due to insufficient sensitivity in the
optical observations. The outer disk is not detected in the observations
of \citet[][]{Shane80} but is evident in the work of
\citet[][]{Mahon92} and \citet[][]{Schinnerer02}.
%
%
\section{The warped structure of the gas disk of NGC~2685}
\label{Sect_4}
It has been argued that the \ion{H}{i} component of NGC~2685 consists
of two rings \citep[][]{Shane77,Shane80,Mahon92,Schinnerer02}. For a
number of reasons, we believe that this is probably not the case.
%
%
\paragraph{The spectral and integrated appearance of the {H\,{\small I}}
component\\} 
Our main point is that the total \ion{H}{i} appearance
and the kinematics as observed in the \ion{H}{i} component of NGC~2685
are not well matched with two disjunct rings or disks.

To demonstrate this, we constructed two tilted-ring models using our
software TiRiFiC \citep[][ hereafter Paper I]{Jozsa07a}. These
represent a galaxy as a set of thin, mutually inclined rings projected
on a data cube. The model parametrisations and 3-d realisations of the
toy models are shown in Fig.~\ref{Fig_02}. For the first toy model
(``flip model'') it is assumed that the \ion{H}{i} disk of NGC~2685
flips at a certain radius. The other model (``warp model''), has a
smooth transition from one disk orientation to the
other. Figure~\ref{Fig_02} displays the total-intensity maps of the
two models. Both models were kept as simple as possible, while by
changing the model parameters by hand a rough match with the
observations was achieved. Both models require an
\ion{H}{i} deficiency towards the centre to reproduce the central
\ion{H}{i} ``hole''. Comparing to the original total-intensity map
(Fig.~\ref{Fig_01}), the shape of the outer ring and the intensity
distribution along the ring is clearly better represented by the warp
model. While in the flip model an elliptical ring is produced, the
shape of the outer ring is properly reproduced and deviates from a
perfect ellipse in the warp model. This is due to projection effects
only. The intensity maxima on the minor axis of the ring are highly
concentrated at the position where the two disks adjoin in the flip
model, while in the warp model they are smoothed out along the outer
ring as is the case in the observed total-intensity map. The diffuse
appearance of the projected surface brightness at the major axis
positions is also seen in the warp model, while it is not in the flip
model.

\begin{figure}[htbp]
\resizebox{0.87\hsize}{!}{\includegraphics[angle=270]{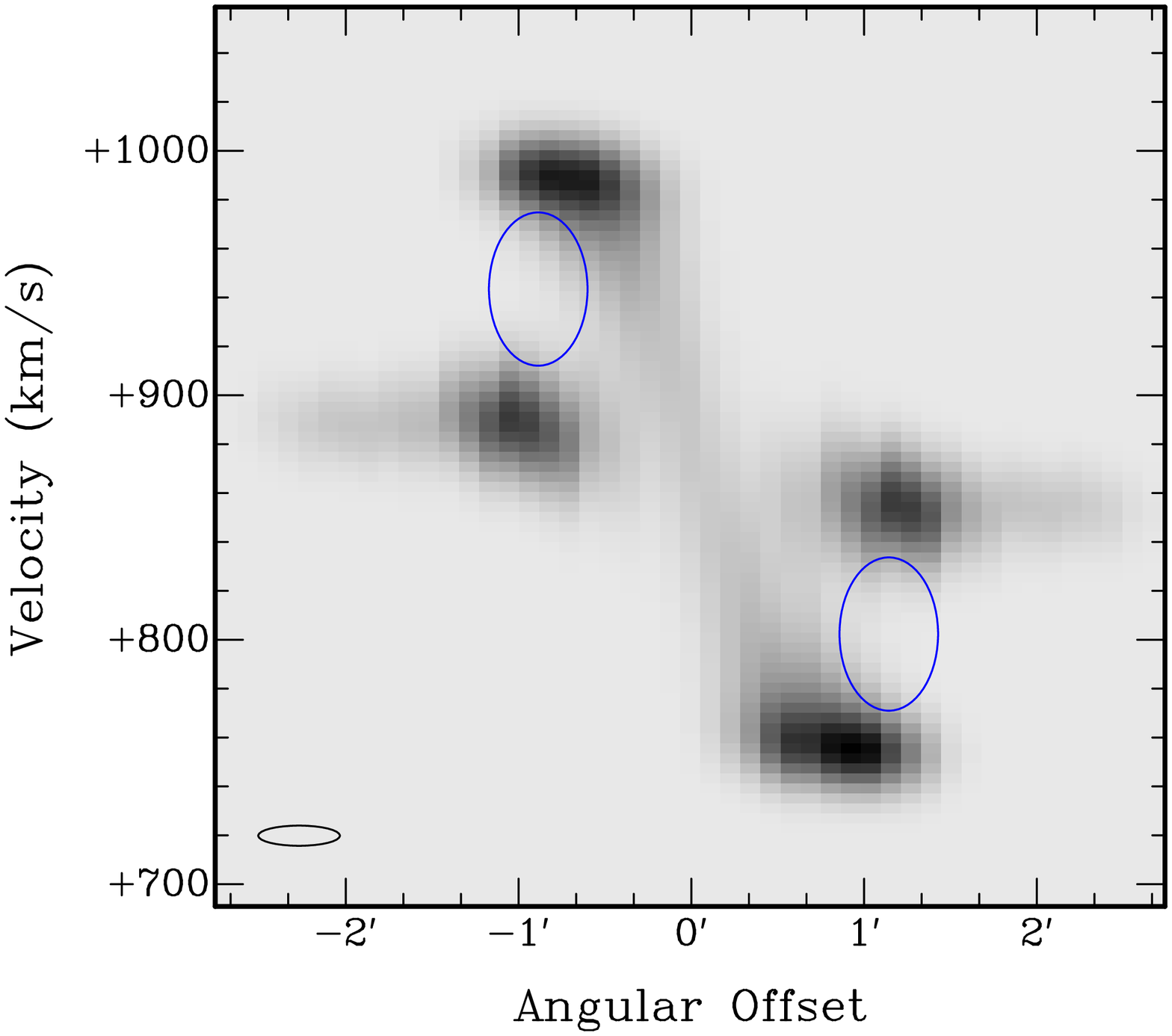}}\\
\resizebox{0.87\hsize}{!}{\includegraphics[angle=270]{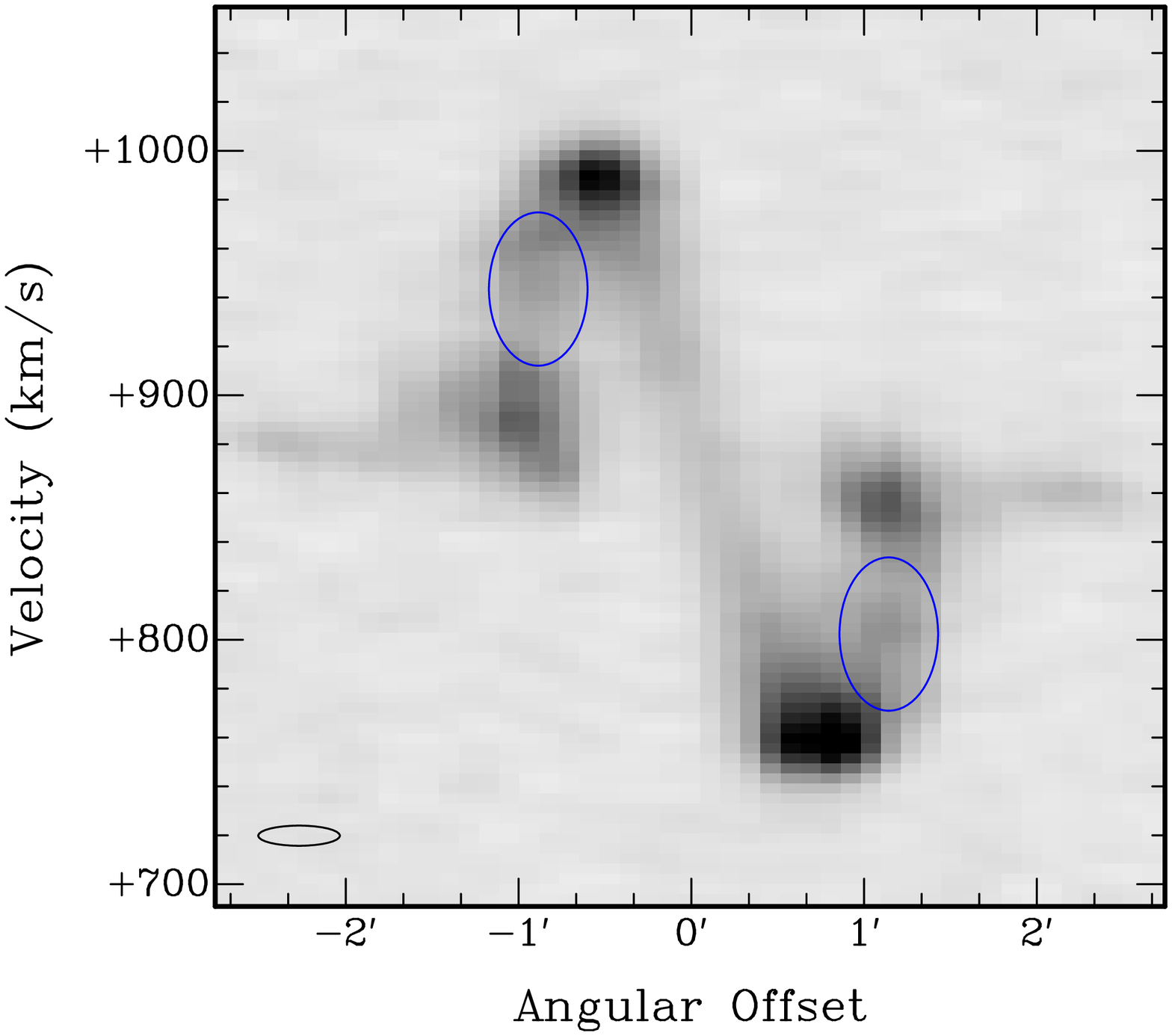}}\\
\resizebox{0.87\hsize}{!}{\includegraphics[angle=270]{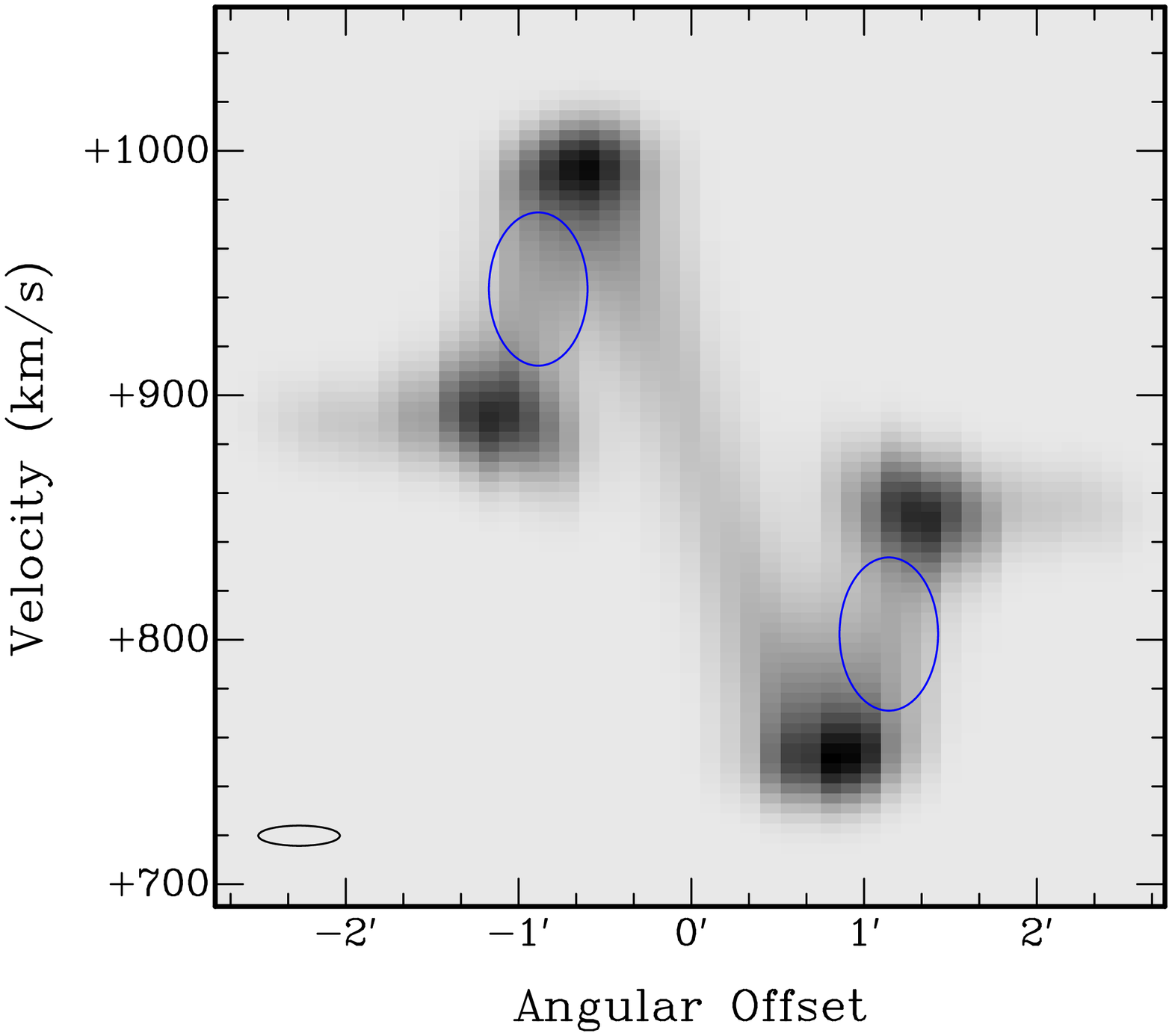}}
\caption{
PV-diagrams along the lines across the minor axis of the inner ring
shown in Fig.~\ref{Fig_02}. Top panel: taken from the flip-model data
cube. Middle panel: taken from the original data cube. Bottom panel:
taken from the warp-model data cube. The ellipses in the lower left
corners denote the clean beam major axis (HPBW) and the velocity
resolution (FWHM). The warp model is able to reproduce the connecting
feature indicated by the ellipses, while the flip model is not.
}
\label{Fig_03}
\end{figure}

The strongest argument against a sudden flip in the \ion{H}{i} disk of
NGC~2685 concerns the kinematics. Figure~\ref{Fig_03} shows
position-velocity diagrams (PV-diagrams) along the lines shown in
Fig.~\ref{Fig_02} calculated from the original low-resolution data
cube and the two toy model data cubes. Every discontinuity in the disk
orientation causes a gap between the spectral features that belong to
the single constituents, as is the case for the flip
%
%
%
%
%
%
\begin{figure}[htbp]
  \begin{center}
\resizebox{0.9\hsize}{!}{\includegraphics{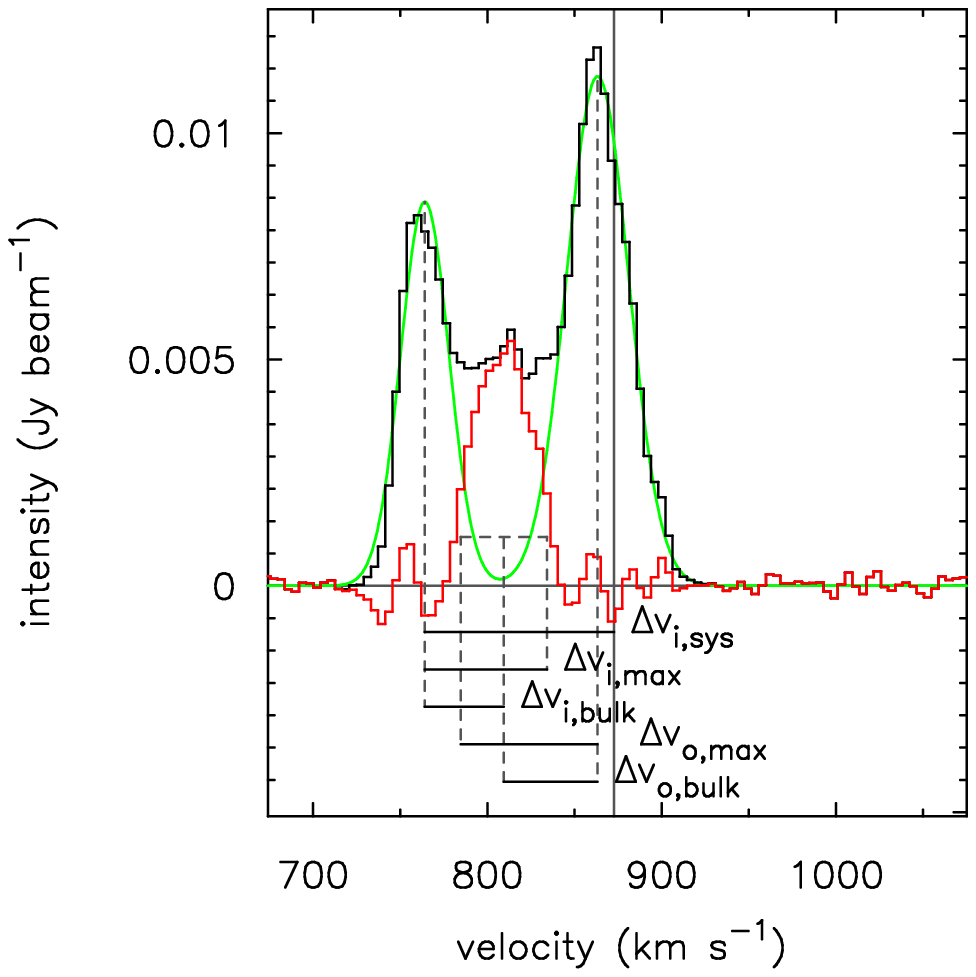}}
\resizebox{0.9\hsize}{!}{\includegraphics{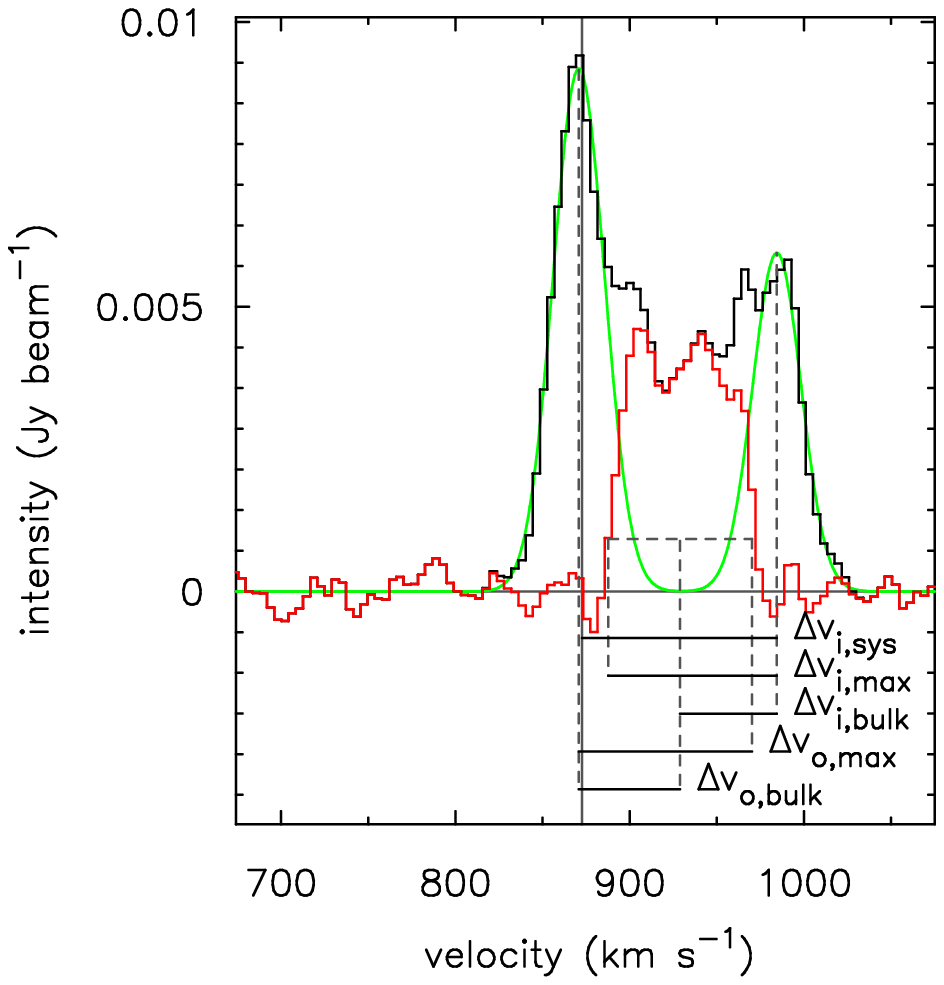}}
\caption{
Spectra taken from the low-resolution data cube at the positions
indicated by the boxes in Fig.~\ref{Fig_01}. Black: the original
spectrum. Light grey (green in online version): fitted Gaussians. Dark
grey (red in online version): residual after subtracting the fitted
Gaussians from the original profile. The velocity differences $\Delta
V_{\rm i,sys}$, $\Delta V_{\rm i,max}$, $\Delta V_{\rm i,bulk}$,
$\Delta V_{\rm o,max}$, $\Delta V_{\rm o, bulk}$ are used to quantify
the amplitude of hypothetical streaming motions (see text). Top:
spectrum derived at RA $ \hmsm{08}{55}{28}{304} $, Dec $
\dmsm{58}{44}{30}{98} $ (J2000). Bottom: spectrum derived at RA $
\hmsm{08}{55}{40}{893} $, Dec $ \dmsm{58}{43}{34}{99}$ (J2000).
}
\label{Fig_04}
\end{center}
\end{figure}
%
%
model. There is no connecting feature to be seen between the intensity
maxima belonging to the ``outer ring'' (intensity peaks closer to the
systemic velocity) and the intensity maxima belonging to the ``inner
ring''. The PV-diagrams taken from the original data cube and from the
warp model show this connection. The superior quality of our
observations, in this case the high spectral resolution, enables us to
ensure that the connection of the intensity peaks in the PV-diagram of
the original data is a real feature and not due to resolution effects.

Under the assumption of the presence of two kinematically disjunct
rings, strong streaming motions have to be invoked to give rise to the
connecting feature (Fig.~\ref{Fig_03}).  We analysed 18 spectra at
equidistant positions on either side of the centre of the galaxy. They
were chosen to show the connecting feature in the PV-diagram of
Fig.~\ref{Fig_03}. The spectra were taken from our low-resolution data
cube, at the positions of $2\times 9$ adjacent pixels indicated by
boxes in Fig.~\ref{Fig_01}. The co-ordinates of the corresponding
positions are listed in Table~\ref{Tab_3a} in the
appendix. Fig.~\ref{Fig_04} shows two example spectra, extracted at
the centre of the boxes. Like the example spectra, all spectra show
the same features, two intensity peaks are visible, connected at a
lower intensity level. Under the assumption that two separate rings
are observed in NGC~2685, the positions of the peak closer to the
systemic velocity mark the line-of-sight velocity of the outer ring
and the other peak positions the line-of-sight velocity of the inner
ring.  If all emission would stem from gas on regular orbits, one
expects the shape of both emission features to resemble a
Gaussian. The rotation velocity of NGC~2685 at a distance of
$10\arcsec$ from the centre is already comparable to the overall
rotational amplitude of the galaxy
\citep[][]{Ulrich75,Shane80}. Inside that radius NGC~2685 is devoid of
\ion{H}{i}. Thus, the assumption is well justified.  In the two-ringed
picture, gas at intermediate velocities belongs to one of the rings,
but does not follow the regular orbits of the inner or the outer
ring. To separate the anomalous gas component from the regular gas
component we fitted two Gaussians to the peaks, and subtracted them
from the spectra (Fig.~\ref{Fig_04}). From the peak positions of the
fitted Gaussians and the velocities at 20 percent of the peak
intensity level of the residuum we calculated the projected bulk- and
maximal excess velocities of the anomalous component with respect to
the assumed regular components (see Fig.~\ref{Fig_04}). We also
calculated the total flux contained in the residuum and in the fitted
components (see Table~\ref{Tab_3a} in the appendix).  On average, if
the excess component would belong to the outer ring, our analysis
would indicate a bulk excess velocity of the anomalous component of
$\overline{\Delta V_{\rm o,bulk}} = 56\pm 6{\,\rm km\,s^{-1}} $,
reaching on average $\overline{\Delta V_{\rm o,max}} = 88\pm 12{\,\rm
km\,s^{-1}}$. If it would belong to the inner ring, the average bulk
excess velocity would amount to $\overline{\Delta V_{\rm i,bulk}} =
48\pm 6{\,\rm km\,s^{-1}}$, reaching on average $\overline{\Delta
V_{\rm i,max}} = 80\pm 13{\,\rm km\,s^{-1}}$. The mass of the
anomalous component compared to the mass contained in the outer ring
averages to $\overline{F_{\rm
\ion{H}{i},e}/F_{\rm \ion{H}{i},o}} = 0.6\pm 0.4$. Compared to the
mass contained in the inner ring it is on average $\overline{F_{\rm
\ion{H}{i},e}/F_{\rm \ion{H}{i},i}} = 1.1\pm 0.6$.  The
\textit{projected} velocities of the assumed excess component reach
more than two thirds of the assumed projected rotation velocity of
either ring and are on average at a level of nearly half the assumed
rotation velocity (for the inner ring we assumed $130\,\rm km\,s^{-1}$
as the projected rotation velocity, for the inner ring we directly
determined $\Delta V_{\rm i,sys}$ from the spectra, see
Fig.~\ref{Fig_04}).  We thus conclude that the hypothesis that the
\ion{H}{i} in NGC~2685 has the structure of two disconnected rings or
disks would require the presence of massive streaming motions to
explain the spectral appearance of the \ion{H}{i}. This would,
however, contradict the lack of enhanced star formation in the galaxy
\citep[][]{Schinnerer02}. Hence, an analysis of our spectral data cube
renders this hypothesis improbable.  We also remark that it would
state a strong coincidence to detect gas with such strong deviations
from the regular component at the junction of the projected
rings. This would not contradict the two-ring hypothesis but weakens
it even further.  The assumption of a strongly warped \ion{H}{i} disk
qualitatively fits the observations. We thus conclude that the
(spectral) appearance of NGC~2685 is best compatible with the
assumption that it is caused by geometrical effects.  In the following
section (Sect.~\ref{Sect_5}) we show that we can reproduce the data
cube in great detail under that assumption.

For the ionised gas disk of NGC~2685, the warped structure of the
helix has already been documented by
\citet[][]{Nicholson87}.

Further support for the warp hypothesis comes from optical
measurements, mainly showing that the \ion{H}{i} structure is also
found to be present in stars and dust (see also Sect.~\ref{Sect_3}).
%
%
\paragraph{The dust lanes\\}
The appearance of the dust-lanes obscuring the whole NE part of the
central body is inconsistent with a single central ring. The simplest
way to explain the shape of the dust-lanes would be to assume a strong
warp in the associated component.  This was already realised by other
authors \citep[e.g.][]{Simonson83,Nicholson87,Taniguchi90} and agrees
well with our suggestion that the inner low-surface brightness
component of NGC~2685 is a heavily warped disk.
%
%
\paragraph{The outer disk and the filled inner ring\\}
Beyond the outer apparent \ion{H}{i}-ring NGC~2685 possesses a diffuse,
extended structure, most likely a disk (since it shows a rotational
signature in the velocity field, Fig.~\ref{Fig_01}).  The outer disk
has already been detected by \citet[][]{Mahon92}. As mentioned above,
another feature directly evident in the {H\,{\small I}} maps is that
inside the outer apparent \ion{H}{i}-ring no empty regions are found
(Sect.~\ref{Sect_3}). This is especially true for the optical image
and the high-resolution {H\,{\small I}} total-intensity map
(Fig.~\ref{Fig_01}, upper right panel), which have a sufficient
resolution and sensitivity to detect such holes. If the central
\ion{H}{i} structure was a polar ring, we have to assume in our flip
model a filled inner region of low-surface brightness, orientated like
the outer ring, leading to a quite implausible surface-brightness
profile in the flip-model (Fig.~\ref{Fig_02}, top left). While it has
been suggested that the outer projected ring could be a resonance
feature reacting to the central stellar body, which would then be a
bar-like structure \citep[][]{Schinnerer02}, the stellar velocity
field published by \citet[][]{Emsellem04} leaves not much doubt that
the central stellar body is a differentially rotating disk as has
already been found by \citet[][]{Schechter78}. Under the assumption of
a polar ring, the same very peculiar surface brightness profile in the
gas and in the stars would be required to fill in projection the area
inside the projected outer ring. Assuming a warp, this is not the
case, the required schematic surface brightness profile (as shown in
Fig.~\ref{Fig_02}, bottom left panel) resembles that of a normal disk
galaxy.

The arguments given above make it plausible to draw the following
picture: NGC~2685 contains a stellar- and gaseous disk that shares
typical characteristics with the disks of spiral galaxies. However,
the disk is extremely warped and NGC~2685 owes its two-ringed
appearance to projection effects. Apparently, the helical structure
consists of dust-features made up by spiral arms that are wrapped
around the central body in a warped structure.

%
%
\begin{figure}[htbp]
  \begin{center}
\resizebox{0.9\hsize}{!}{\includegraphics{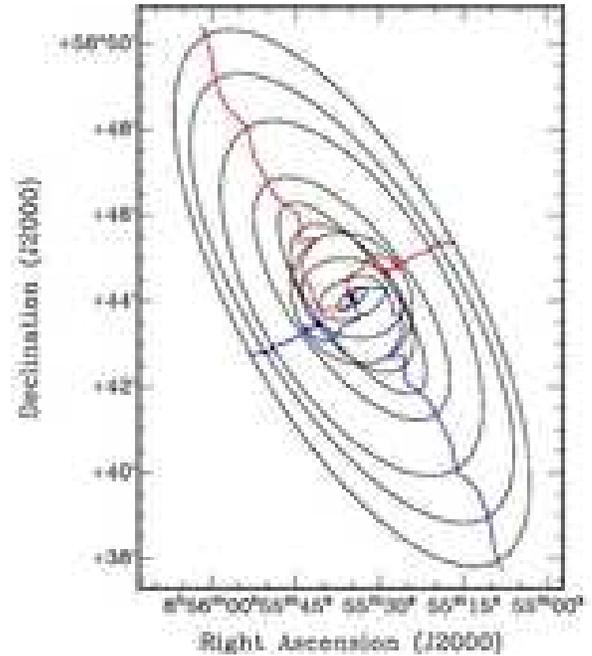}}
\caption{
``Inclinogram'' of the NGC~2685 overlaid on the original total
intensity maps: the ellipses mark the projection of the rings at radii
of $12\arcsec,36\arcsec,60\arcsec,84\arcsec,108\arcsec,140\arcsec,
180\arcsec,240\arcsec,320\arcsec,$ and $400\arcsec$ on the celestial
plane, grey lines mark the kinematical major and minor axes. In light
grey (red in the online version) the kinematical major axis in the
receding part of the galaxy is plotted, in dark grey (blue in online
version) the kinematical major axis of the approaching part.
}
\label{Fig_05}
\end{center}
\end{figure}
%
%
Cases of extreme warps are not unknown. \citet[][]{Schwarz85}
quantified the structure of the warp in the large spiral galaxy
NGC~3718. It shows an extreme warp in its stellar and gaseous disk
that changes its orientation by nearly $90\deg$. In this galaxy, that
has a different orientation and does not appear ring-like in
projection, the warp is already evident through the appearance of
dust-lanes in optical photographs. In a similar approach as presented
above, \citet[][]{Arnaboldi93} showed that the galaxy NGC~660 probably
has a very simple disk structure with a warp of a large amplitude.
Several cases of extremely warped \ion{H}{i} disks in gas-rich early
type galaxies are known
\citep[][]{Morganti97,Oosterloo02,Serra06,Oosterloo07,Sparke08},
suggesting that warps of large amplitude could be a common feature
among this type of galaxy. Also, the kinematical decoupling of the
inner stellar- and gaseous components is not an uncommon feature for
early-type galaxies \citep[][]{Sarzi06}.
%
%
\begin{figure*}[htbp]
  \begin{center}
\includegraphics[angle=270,width=\textwidth]{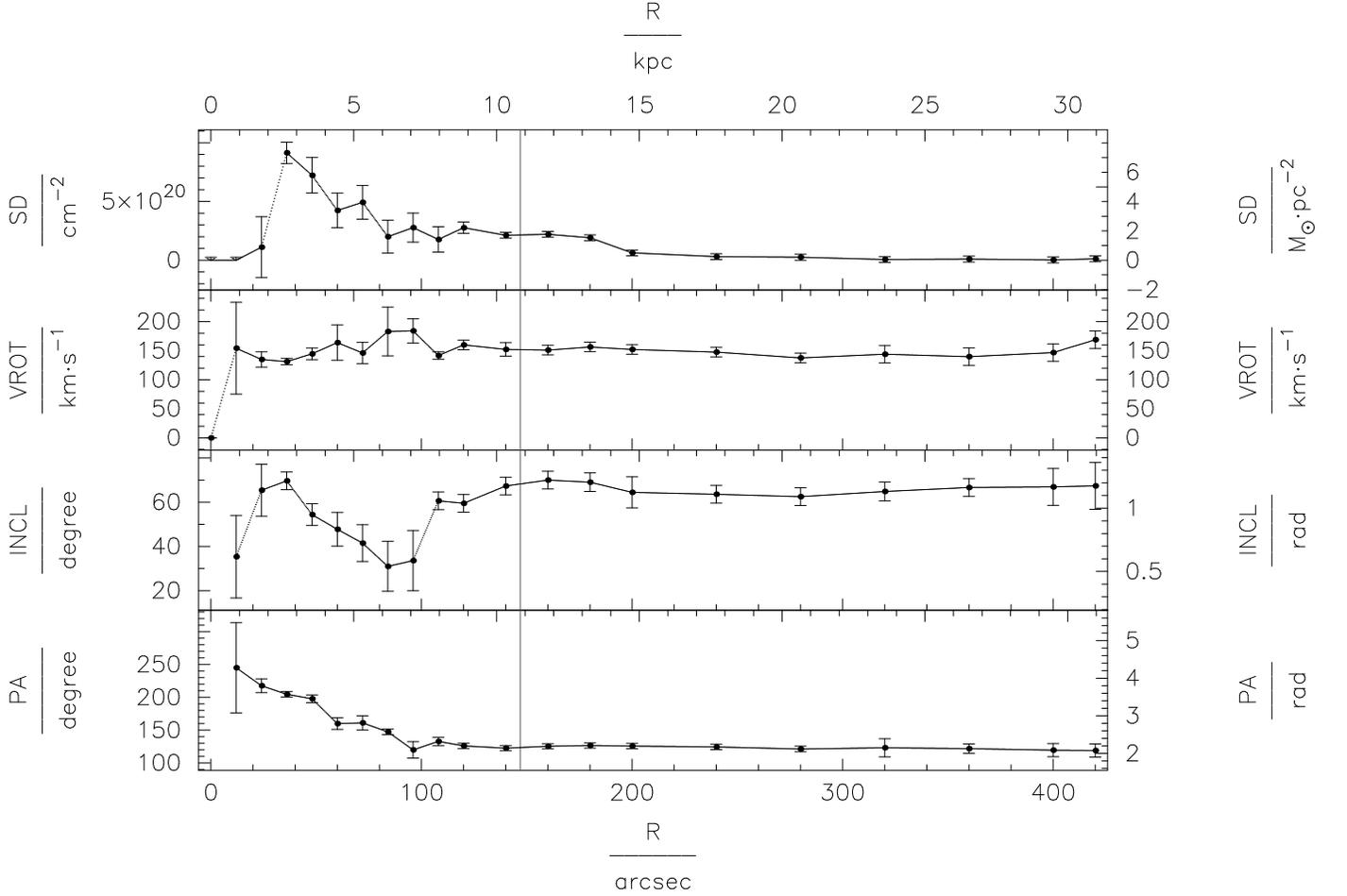}
\caption{
Tilted-ring parametrisation of the Spindle galaxy. SBR: surface
brightness (upper limits only for the first two data points). VROT:
rotation velocity. INCL: inclination (unconstrained at origin). PA:
position angle (unconstrained at origin). R: radius. The vertical line
denotes the optical radius $r_{25}$.}
\label{Fig_06}
\end{center}
\end{figure*}
%
%
%
\section{Tilted-ring modelling}
\label{Sect_5}
%
%

\begin{figure*}[htbp]
  \begin{center}
  \includegraphics[angle=270,width=0.48\textwidth]{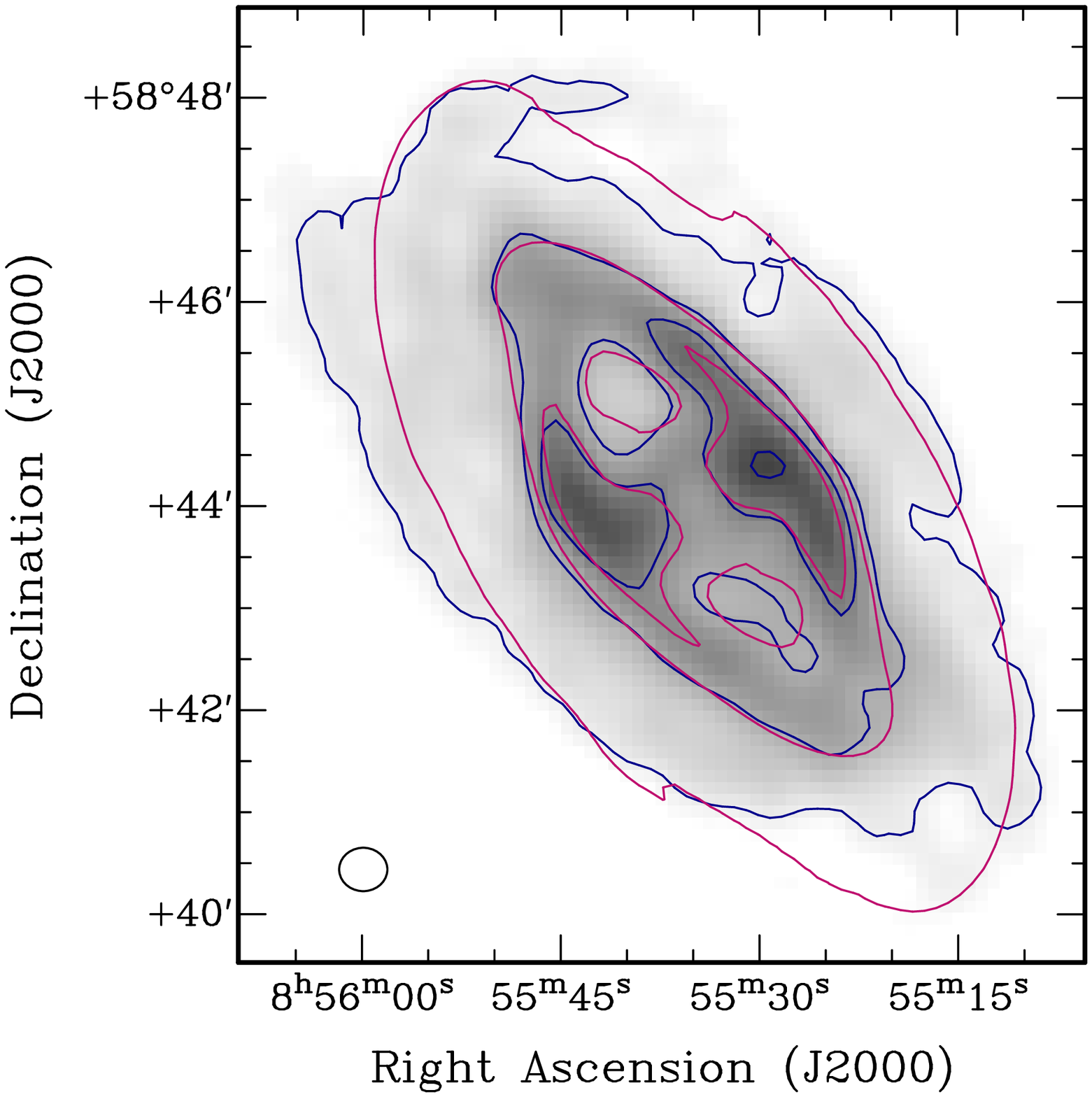}
  \includegraphics[angle=270,width=0.48\textwidth]{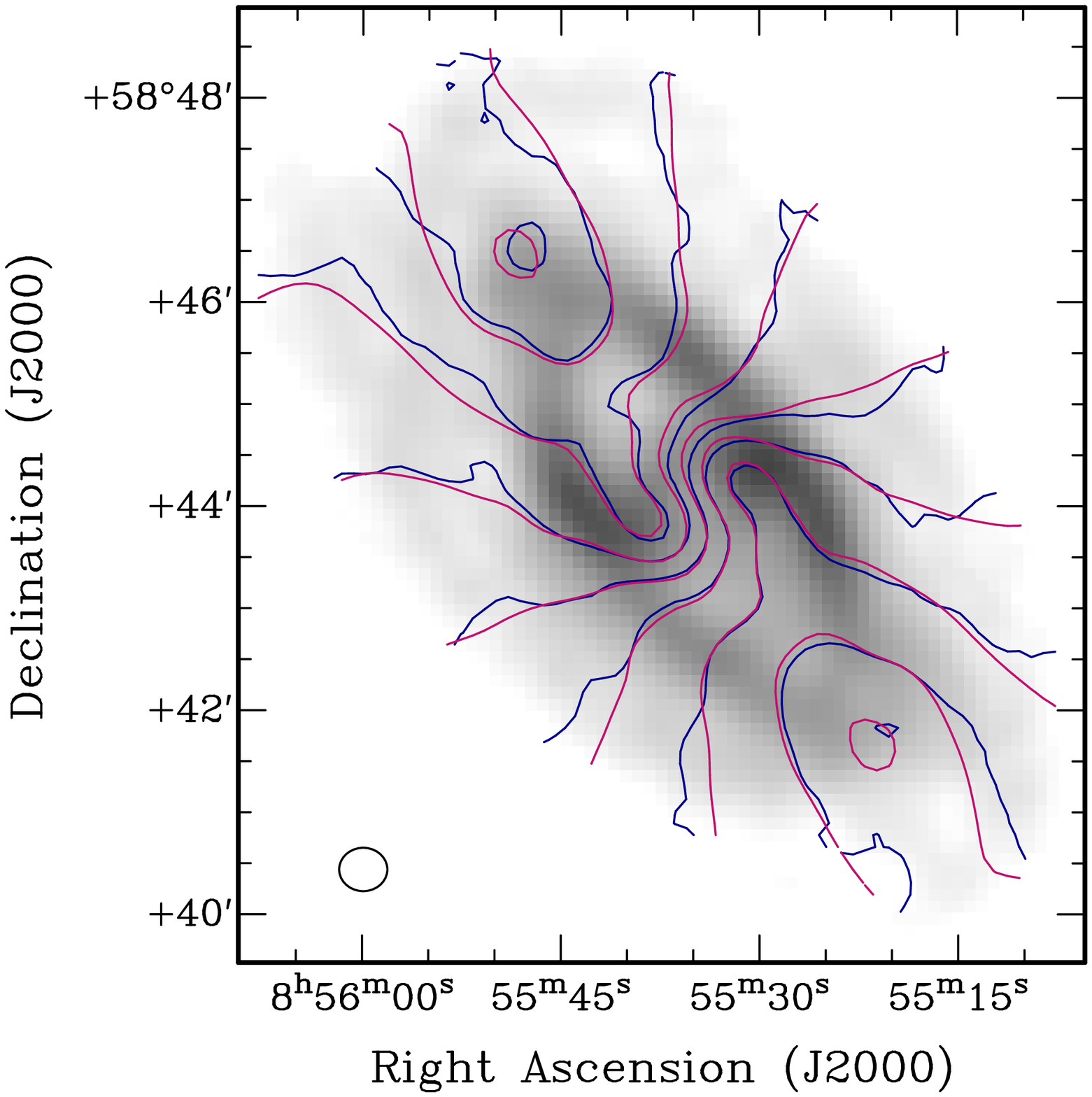}\\
  \includegraphics[angle=270,width=0.48\textwidth]{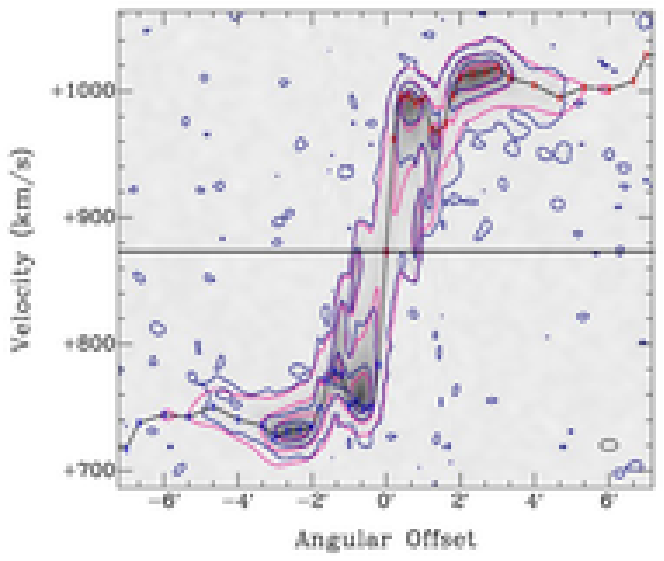}
  \includegraphics[angle=270,width=0.48\textwidth]{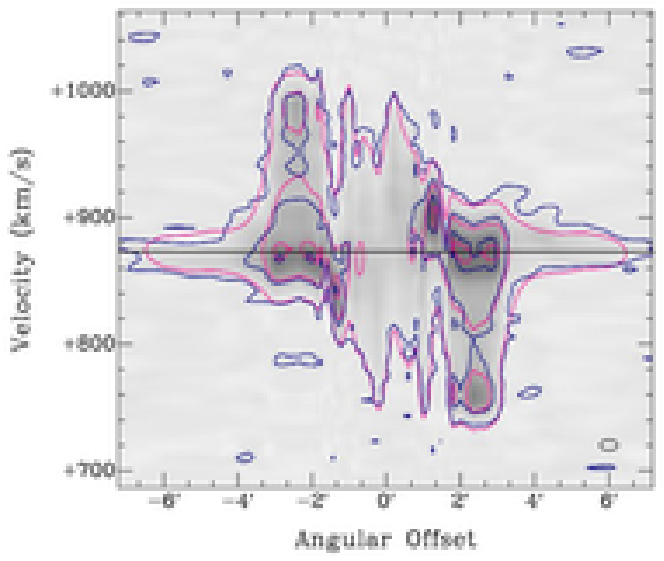}
\caption{
Top panels: total-intensity contour plots (left panel) and
first-moment velocity fields (right panel) derived from the
low-resolution data cubes (dark-grey, blue in online version:
original, light grey, pink in online version: model) overlaid on a
total-intensity grey-scale map derived from the observed data
cube. The ellipse in the lower left corner represents the clean beam.
The contours represent the $5, 20, 80, 160 \tth$ levels and the
$V_{\rm sys} \pm 0,40,80,120,140\,{\rm km}\,{\rm s}^{-1}$ levels
respectively (see Tab.~\ref{Tab_2}). Bottom panels: PV-diagrams
comparing the final best-fit models (light grey contours, pink in
online version) with the original low-resolution data (dark grey
contours, blue in online version, and grey scale image). The vertical
line marks the systemic velocity. The ellipses in the lower right
corners denote the clean beam major axis (HPBW) and the velocity
resolution (FWHM). Bottom left: PV-diagram along the kinematical major
axis (see Fig.~\ref{Fig_05}), the boxes connected with the lines
denote the rotation curve corrected for the inclination. Bottom right:
PV-diagram along the kinematical minor axis (see
Fig.~\ref{Fig_05}). Contour levels are $0.5, 4, 10\,\mJyb$.
}
\label{Fig_07}
\end{center}
\end{figure*}
%
%
\begin{table*}[htbp]
\begin{center}
\caption{
Global best-fit tilted-ring parameters of NGC~2685. Error in systemic
velocity and dispersion: $2.0\,{\rm km}\,{\rm s}^{-1}$. Error in
position and scale height: $6.0^{\prime\prime}\,\hat{=}\,450\,\rm
pc$.}
\label{Tab_3}
\begin{tabular}{l l c c}
\hline
\hline
Descr.                                                                            & Par.\\
\hline
Right Ascension (J2000), fitting results                                          & RA$_{\rm fit}$     & \multicolumn{2}{c}{$ \hmsm{08}{55}{34}{92} $} \\
Declination (J2000), fitting results                                              & Dec$_{\rm fit}$    & \multicolumn{2}{c}{$ \dmsm{+58}{44}{04}{4} $} \\
Right Ascension (J2000) from NED                                                  & RA$_{\rm NED}$     & \multicolumn{2}{c}{$ \hmsm{08}{55}{34}{75} $} \\
Declination (J2000) from NED                                                      & Dec$_{\rm NED}$    & \multicolumn{2}{c}{$ \dmsm{+58}{44}{03}{9} $} \\
Heliocentric optical systemic velocity (${\rm km}\,{\rm s}^{-1}$)                 & $V_{\rm sys}$      & \multicolumn{2}{c}{$ 875.2 $}                 \\
Type of data cube used for analysis, high-resolution or low-resolution            &                    & high                  & low                     \\
Dispersion (${\rm km}\,{\rm s}^{-1}$)                                             & $V_{\rm disp,tot}$ & $ 9.8  $              & $ 10.2 $                \\
Internal dispersion, instrumental dispersion subtracted (${\rm km}\,{\rm s}^{-1}$)& $V_{\rm disp,int}$ & $ 9.6  $              & $ 9.6  $                \\
Global scaleheight ($\arcsec$)                                                    & $z_0$              & $ 5.4  $              & $ 3.0  $                \\
Global scaleheight ($\rm pc$)                                                     & $z_0^{\rm pc}$     & $ 400  $              & $ 220  $                \\
\hline                  
\end{tabular}
\end{center}
\end{table*}
%
%
After having argued that there are reasons to believe that the
\ion{H}{i} (-and optical) disk of NGC~2685 is heavily warped, we
present our attempts to fit a full tilted-ring model to the data
cube. Due to the assumed warp, the projections of the orbits are very
crowded at some positions. It is impossible to disentangle this orbit
crowding employing any tilted-ring analysis of a velocity field (see
discussion in Paper I). Instead, we perform a direct, automated fit to
the data cube with our software TiRiFiC. Since the software is
described in great detail in Papers I and II
\citep[][]{Jozsa07a,Jozsa07b}, we refer to these papers and to
\citet[][]{Jozsa06} for details of the analysis procedure. The
low-resolution data cube was analysed first using a spacing of ring
radii of $20\,\arcsec$.  The centre of the galaxy, the systemic
velocity, the velocity dispersion, and the scale height were kept as
free parameters, but were not allowed to vary independently from ring
to ring. For each ring, the surface-brightness, the orientational
parameters, and the rotation velocity were varied independently.  To
estimate errors, independent fits to the receding and approaching side
were performed.  The errors were calculated as the maximum deviation
of the parameters from the fits to the approaching and the receding
side from the earlier fit to the entire galaxy. The errors of the
central position, scale height, the systemic velocity and the global
dispersion of the galaxy can be very roughly estimated to be less than
$6\,\arcsec$ and $2\,\,{\rm km}\,{\rm s}^{-1}$ respectively, roughly
corresponding to half the (minor-axis) FWHM of the resolution for the
high-resolution data cube (concerning the accuracy of the applied
method see Paper I). In a few cases, where the errors calculated were
obviously too large or too small, the data were visually inspected and
errors were estimated by varying the parameter in question and
comparing resulting model data cubes with the original data cube. This
was only necessary for some data points at the largest and the
smallest radii. To refine the spatial resolution in the inner regions
of the galaxy, the high-resolution data cube was used. Up to a radius
of $108\arcsec$, where all parameters vary rapidly with radius, the
spacing of the ring radii was refined to $12\arcsec$. The finally
resulting fit is shown in Fig.~\ref{Fig_06}. Figure \ref{Fig_05}
illustrates the geometry of the fitted disk by means of an
``inclinogram'', projecting the rings onto the celestial plane.
Analysing the high-resolution data cube, it was found that the
\ion{H}{i} surface brightness inside a radius of
$10\,\arcsec-15\,\arcsec$ was negligibly small.  Therefore, the
orientational parameters are not constrained at the origin. We
consider our model and the formal errors reliable beyond the third
data point (at a radius of $36^{\prime\prime}$).  Figure~\ref{Fig_A1}
in the appendix shows contour-plots overlaying the low-resolution data
cube with the corresponding synthetic data cube.  Figure~\ref{Fig_07}
shows overlays of corresponding moment maps of the original data and
the synthetic ones and PV-diagrams along the kinematical major and
minor axes as traced in Fig.~\ref{Fig_05}. All comparisons show a very
good agreement of measurement and model out to largest radii until the
{H\,{\small I}} distribution in the galaxy becomes asymmetric. The
tilted-ring model is able to reproduce both the kinematics and the
surface-brightness profile of the {H\,{\small I}} component in
NGC~2685 in great detail, despite the peculiar structure of NGC~2685.
Table~\ref{Tab_3} lists the finally adopted global parameters of the
model. Internal velocity dispersion (the velocity dispersion of the
\ion{H}{i} disk itself after subtracting instrumental effects) and
scale height are in good agreement for both fits. Table~\ref{Tab_5}
contains the radially dependent best-fit parameters; in addition, for
a simplified further use, it lists the normal vectors of the rings as
derived quantities.

To estimate some of the global physical parameters shown in
Tab.~\ref{Tab_1}, the Tully-Fisher relation was applied to derive the
distance of NGC~2685 following \citet[][Sak00]{Sakai00} , treating the
galaxy like a spiral galaxy.
%
%
\begin{figure*}[htbp]
\begin{center}
\includegraphics[width=0.34\textwidth]{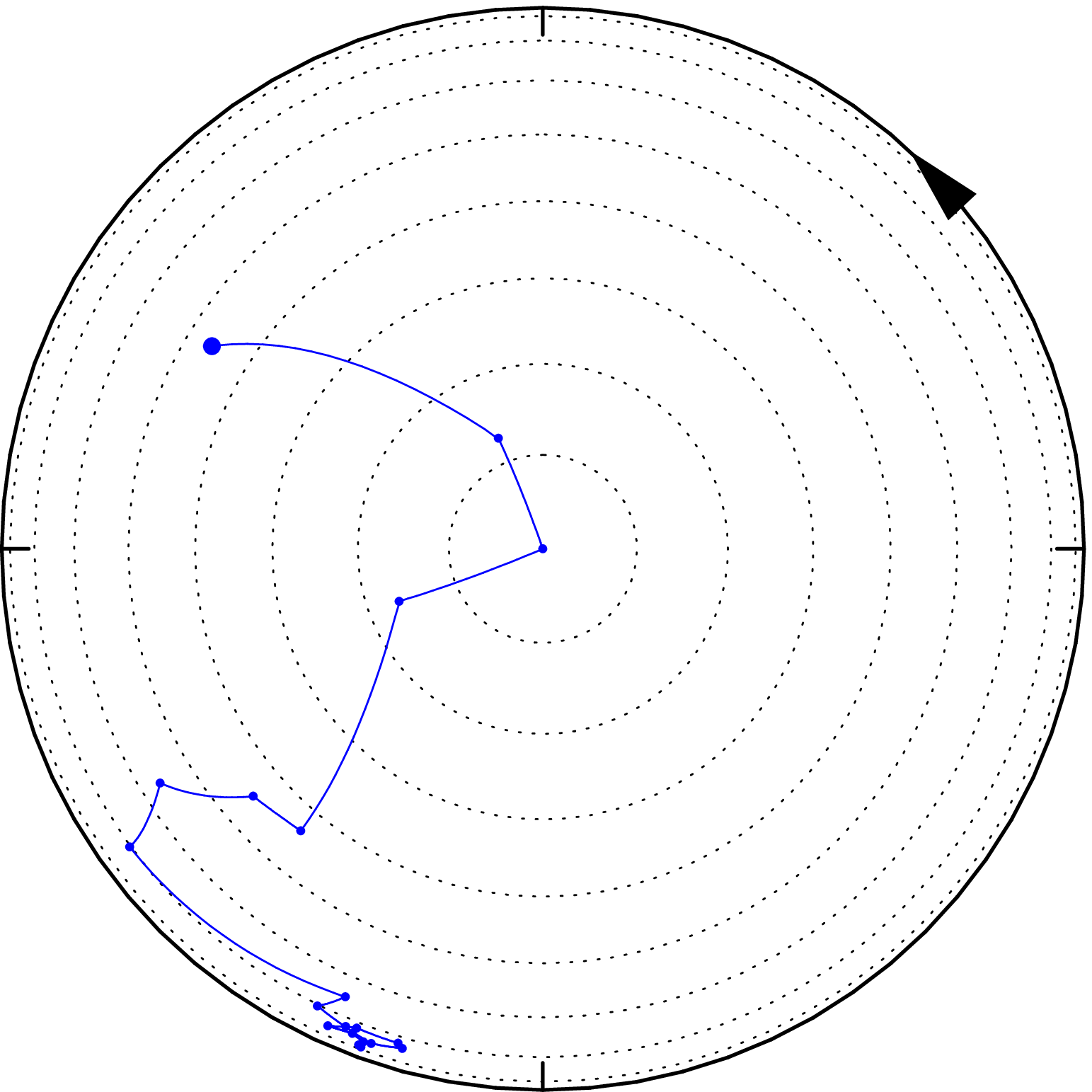}
\includegraphics[width=0.3\textwidth]{xx_08b.eps}
\includegraphics[width=0.34\textwidth]{xx_08c.eps}
\caption{
Tip-LON plots for the \ion{H}{i} disk of NGC~2685. Left and right
panel: tip-LON plots (see \citealt{Briggs90} and Paper II), the filled
circles are the projection of the tip of the normal vectors of the
rings at different radii onto a reference plane; the centre of the
galaxy is denoted by a larger circle. The circles are drawn at
intervals of $10\deg$. Left panel: The reference plane is the ring at
a radius of $36\arcsec$, the arrow indicating the direction of
rotation. As NGC~2685 does not contain an inner flat disk, this ring
was chosen to be representative to check for the third rule of
\citet[][]{Briggs90}. Middle panel: linear representation of the
tip-LON plot shown in the left panel. Right panel: the reference plane
is chosen such that two straight LONs are roughly visible, resembling
the fourth rule of \citet[][]{Briggs90}. At larger radii the dots are
clustered around the same position, indicating an outer planar disk.
}
\label{Fig_08}
\end{center}
\end{figure*}
%
%
We used equation 11 in Sak00 to estimate the total
extinction-corrected I-band magnitude of the galaxy. To correct for
inclination, the \ion{H}{i} 50\% line width was derived by changing
all modelled inclinations to $90\deg$ and determining a line width of
$320\pm23\,{\rm km}\,{\rm s}^{-1}$ by artificially observing NGC~2685
as if observing an edge-on disk. The apparent I-band magnitude of
$9\fm 9\pm 0\fm 1$ (see Table~\ref{Tab_1}) was taken from
\citet[][]{Prugniel98} as listed in \citet[][]{LEDA}. A (marginal)
k-correction was applied and the magnitude was corrected for Galactic
(0\fm 121, NED, following \citealt{Schlegel98}) and internal
extinction. For the latter, we use a 20\%-linewidth of $306\pm17\,{\rm
km}\,{\rm s}^{-1}$ (not corrected for inclination), derived from our
model data cube, which compares well to $298\,{\rm km}\,{\rm s}^{-1}$
published by \citeauthor[][]{Driel00} (Table 2, from
\citealt{Richter94}, Table 1), and the logarithmic axis ratio
$\log(a/b) = 0.27\pm 0.06$ \citep[LEDA, ][]{LEDA} to apply formula 2
\citep[][]{Tully98} in Sak00.  This yields a Tully-Fisher distance of
$15.2\pm3.8\,{\rm Mpc}$. The large error mainly arises from the
conservative errors in the 50\%-line width and the axis ratio
$\log(a/b)$ that is used to estimate dust extinction. We find this
result to be well in agreement with a distance estimate using a Hubble
constant of $72\,{\rm km}\,{\rm s}^{-1}\,{\rm Mpc}^{-1}$ yielding
distances of $13.9\,{\rm Mpc}$ (corrected the solar motion relative to
the CMB) and $13.4\,{\rm Mpc}$ (corrected for the solar motion
relative to the Local Group) with the inherent uncertainty of the
peculiar motion of the galaxy.  Therefore, we consider the usage of
the Tully-Fisher relation a valid method to estimate the distance to
NGC 2685.
%
%
\begin{figure}[htbp]
\begin{center}
\includegraphics[angle=270,width=0.48\textwidth]{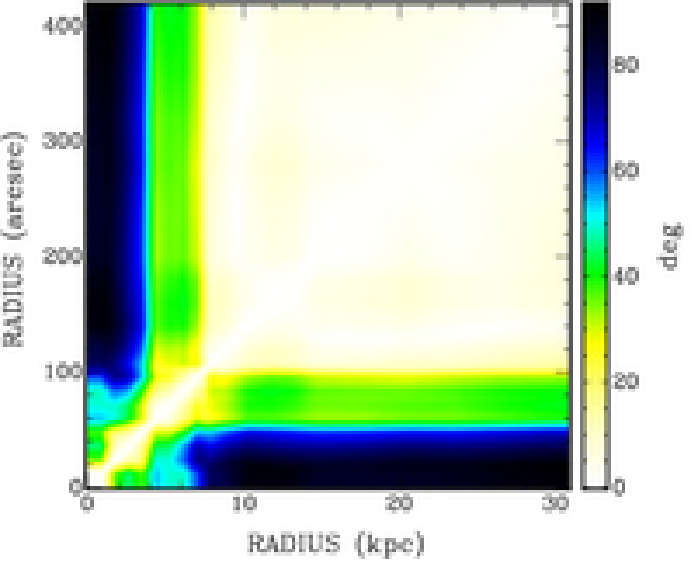}
\includegraphics[angle=270,width=0.3\textwidth]{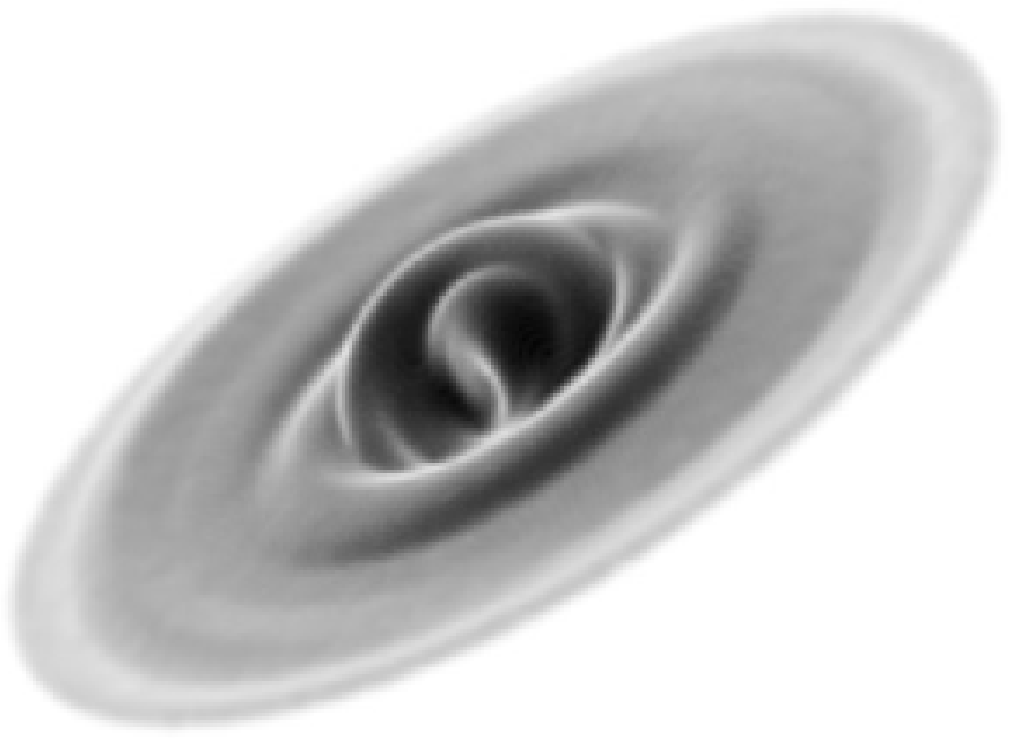}
\caption{
Top panel: ``tiltogram'' of NGC~2685. This is a pixel map showing the
mutual inclination of rings at different radii (see Paper II). Bottom
panel: three-dimensional model of the \ion{H}{i} disk of NGC~2685
according to the best-fit tilted-ring model. The model is rotated and
artificially made opaque to show again the presence of an outer planar
disk.
}
\label{Fig_09}
\end{center}
\end{figure}
%
%
%
%
\section{The structure of the warp}
\label{Sect_6}
The tilted-ring model shown in Fig.~\ref{Fig_06} is visualised in
various forms in Figs.~\ref{Fig_05}, \ref{Fig_08}, and
Fig.~\ref{Fig_09}. Our findings again support our picture drawn in
Sect.~{\ref{Sect_4}}; a sudden change in orientation, as would be
expected in case of the presence a distinct polar structure, should be
reflected in our parametrisation resulting from a free fit. In
contrast, we find that the best-fit model is similar to the warp toy
model shown in Sect.~\ref{Sect_4}. As is the case for the spiral
galaxy NGC~3718 \citep{Schwarz85}, the gaseous disk of NGC~2685
changes its orientation continuously from the innermost radii onward,
and then becomes remarkably flat towards large radii. For the
innermost $12\arcsec -36\arcsec$, however, no reliable information can
be derived from the data, as obviously the gas becomes ionised towards
the centre of the galaxy.

Towards the centre the \ion{H}{i} disk changes its orientation
continuously. Therefore, it has a different quality compared to warps
in common disks of spiral galaxies, which show an inner flat part, the
bending occurring at larger radii \citep[][]{Briggs90}. It is,
however, remarkable that nevertheless some typical warp features seem
to be reproduced (for a brief review on the properties of \ion{H}{i}
warps see Paper II). Figure~\ref{Fig_08} shows tip-LON plots
\citep[][]{Briggs90}, a projection of the spin normal vectors of the
disk onto a reference plane. In the left-hand panel this plane is
chosen to be the orbital plane at a radius of $36^{\prime\prime}$, the
first radius at which we consider the parametrisation being
reliable. Our analysis indicates that -- in this reference frame --
quite some portion of the inner disk shares a common line-of-nodes:
with increasing radius, in the chosen projection, the LON-angle is
constant within the errors (see Fig.~\ref{Fig_08}, middle panel). For
spiral galaxies the same feature is observed when choosing the flat
inner disk as a reference plane and is indicative of co-precession in
the inner warped disk \citep[][]{Briggs90}. We emphasise that this
feature depends on the choice of the reference plane: the same feature
would not be observed choosing the orbital plane of the outer flat
disk as reference. A rapid change of the direction of the LON occurs
at a radius of $100\,\arcsec$. At larger radii ($> 120\,\arcsec$), the
data points fall into the same region, such that in the outer region
they share the same LON angle again. Consistent with observations of
normal warps, the LON advances in the direction of galactic rotation
(\citealt{Briggs90}, Paper II), indicating an outwards decreasing
torque \citep[][]{Shen06}. The right panel in Fig.~\ref{Fig_08}
illustrates that also for NGC~2685 a reference plane can be found that
divides the \ion{H}{i} disk into two parts, connected by a small
transition region, with a comparably constant LON angle
(\citealt{Briggs90}, Paper II). The implication made for warped disk
galaxies is that the disk is divided in two connected regions, each of
which is in a state of co-precession \citep[][]{Revaz01}.  The top
panel of Fig.~\ref{Fig_09} shows the mutual inclination of the spin
vectors of the disk. Beyond a radius of $100\arcsec$, the values in
this diagram are consistent with being zero, meaning that NGC~2685
possesses a flat outer disk, as illustrated in an opaque 3d-model
based on our best-fit parametrisation shown in the bottom panel
Fig.~\ref{Fig_09}. This is also evident from the clustering of the
data points at larger radii in the tip-LON plots of Fig.~\ref{Fig_08}
and the fact that the orientational parameters of the tilted-ring
parametrisation (Fig.~\ref{Fig_06}) do not change within the error
bars. The occurrence of the flat outer disk is a feature that has been
previously found to be present in some warped disk galaxies (Paper II
and references therein).

Since the warped structure of the gas disk indicates that the galaxy
is not a typical polar-ring galaxy as defined by
\citet[][]{Whitmore90}, the question arises to which degree a polar
orbit is reached at all in this system, either with respect to the
outer disk or with respect to the central stellar
body. Figure~\ref{Fig_08} shows that at least for radii $\geq
36\arcsec$, a polar orientation of outer and inner \ion{H}{i} disk is
not compatible with our model. The outer disk, assumed to be flat, is
inclined by $\approx 75\deg$ with respect to the disk at a radius of
$36\arcsec$. This is remarkably consistent with the same value found
by \citet[][]{Sarzi06} for the mutual inclination of the central
stellar body and ionised gas disk of NGC~2685 \citep[see
also][]{Nicholson87}. Our analysis disfavours the hypothesis of a true
polar structure at radii $\geq 36\arcsec$, at smaller radii our model
becoming unreliable. However, the results of \citet[][]{Sarzi06} can
be interpreted as evidence against a polar orientation in the central,
ionised part of the gaseous disk.
%
%
\section{Non-circular motions}
\label{Sect_7}
The previous sections showed that the kinematics of NGC~2685, as they
appear in projection, can be well represented by the tilted-ring
model. However, beacause of the peculiar structure of NGC~2685, the
kinematics of the gas disk of the galaxy is expected to show
deviations from the assumed symmetry. The orbits of any tracer
material are expected to deviate from circularity. Given the lack of
appropriate analysis software to deal with non-circular motions in
extremely warped cases like NGC~2685, we have to restrict ourselves to
a brief discussion at a phenomenological level.

\begin{figure}[htbp]
\begin{center}
\resizebox{\hsize}{!}{\includegraphics{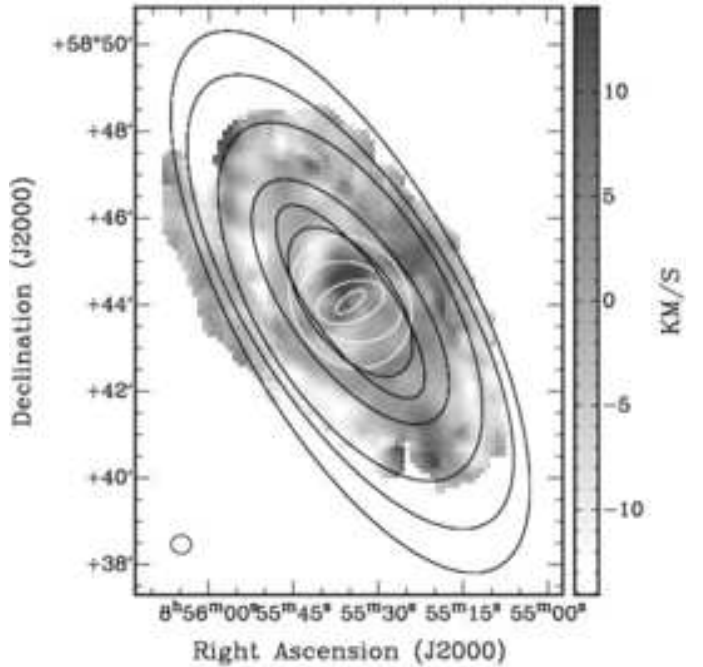}}
\caption{
Difference- or residual velocity map derived by subtracting the model
first-moment map from the original first-moment map. Overlaid are the
projected rings of the best-fit model at radii
$12\arcsec,36\arcsec,60\arcsec,84\arcsec,108\arcsec,140\arcsec,
180\arcsec,240\arcsec,320\arcsec,$ and $400\arcsec$. Apart from the
antisymmetric feature between a radius of $30\arcsec$ and $70\arcsec$,
no significant deviations of modelled and measured velocity field are
seen.
}
\label{Fig_10}
\end{center}
\end{figure}

Figure~\ref{Fig_10} shows the residual obtained when subtracting the
first-moment velocity maps derived from the low-resolution original-
and the model data cube for NGC~2685. In such a map, systematic
patterns indicate a mismatch of the model and measurement. Such a
mismatch can be due to i) a wrong fit, ii) an incompatibility of
morphology and/or kinematics of the observed structure with the
assumed tilted-ring symmetry, or iii) an erroneous calculation of the
weighted mean recession velocity in the noisy observed data cube. We
cannot provide any independent tests for the accuracy of our fit
method. Hence we cannot exclude that residuals in velocity arise from
a wrong fit (if they are of first-order harmonic nature). However,
extensive tests (see Paper I) suggest that for the given geometry of
the \ion{H}{i} disk, such errors are unlikely (for radii $>\,1.5-2$
synthetic beams) to show up in the residual velocity field.
%
%
%
%
The single obvious feature in the residual map is located at the tip
of the inner stellar body, roughly within a radial range between
$48\arcsec$ and $60\arcsec$, and is of antisymmetric nature. An
interpretation can, however, only be of a speculative nature: we use a
primitive approach as NGC~2685 is 
probably
a dynamically complex
case. Using epicyclic theory, \citet[][]{Franx94} connect an
antisymmetric deviation from circular symmetry in the radial
velocities to a symmetric distortion in the potential, hence a bar or
an elliptic potential. However, the basic assumption made therein is
that the disk is flat, which is obviously not the case for
NGC~2685. Another, straightforward, suggestion would be that we
witness (vertical) precession, which would cause a first-order
harmonic distortion in radial velocity along a ring. With the
(symmetrically) warped nature of the disk, precession does not seem
unlikely. We emphasise, however, that the observed deviation of
observed and synthetic velocity field can also be caused by a purely
morphological asymmetry in the tracer material. Given the line-widths
at the radii concerned (see Fig.~\ref{Fig_07}), over which the
velocities are calculated as an intensity-weighted mean, this is also
a possible option. Since the feature seen is very localised, we prefer
this interpretation.

Except for the very localised axisymmetric feature in the vicinity of
the central stellar constituent, the residuals are low compared to the
rotation velocity. Also the direct comparison of our model with the
data (Figs.~\ref{Fig_07},~\ref{Fig_A1}) shows that we reproduce the
data in great detail, without obvious systematic deviations.  In the
presence of large deviations from the circularity of orbits, those
would usually be expected to show up
\citep[][]{Franx94,Schoenmakers97}.  Hence, in the scope of our model,
we find no obvious evidence for a global deviation from the
tilted-ring symmetry and, with that, for a significant global
non-circularity of orbits.

Especially the suggestion that the outer projected ring could be a
real accumulation of gas and stars at an outer Lindblad resonance of a
bar \citep[][]{Schinnerer02} does not fit our results. While a change
of position angle and inclination in a tilted-ring model can be
indicative for a bar
\citep[][]{Schinnerer02}, streaming motions would leave a distinct
large-scale imprint on the residual velocity field shown in
Fig.~\ref{Fig_10}, which is not seen \citep[see also][]{Erwin03}.

Our fitting results would certainly be improved including an
appropriate treatment of noncircular motions, like allowing for
azimuthal variations of velocities along the sight 
\citep[][]{Franx94,Schoenmakers97}. 
On the other hand, our simplified
treatment indicates that there is probably not much room for a large
amplitude of harmonic terms indicative for non-circular motions,
provided there is no significantly different parametrisation that
reproduces the data equally well. In Sect.~\ref{Sect_4}, we discuss
that at least a two-ringed solution seems improbable, but we are not
able, technically and principally, to search the complete parameter
space to exclude the existence of such a solution.
%
%
\section{Dynamical considerations}
\label{Sect_8}
While polar-ring galaxies are not frequent (\citealt{Whitmore90}
calculate a detection rate of roughly 0.5\% among all nearby,
lenticular galaxies, and estimate a fraction of 4.5\% of all nearby,
lenticulars to have a polar ring), they are subject to quite some
interest. Their kinematics contains information about the
three-dimensional shape of the overall potential, since the orbits of
the observed tracers (stars and gas) do not share a common plane of
motion. The same accounts in principle for a galaxy with an extreme
warp in a low-surface-brightness disk in co-existence with a rotating,
central stellar body. The main purpose of the paper is to present our
observations and the consequent kinematical description of NGC~2685,
while a detailed dynamical treatment under consideration of further
data will follow in a different paper.

Here, we briefly discuss some implications of our tilted-ring model
concerning the dynamics of the object.
%
%
\subsection{The age of the low-surface brightness disk}
\label{Sect_8.1}
Our analysis indicates the presence of a relaxed \ion{H}{i} disk at
large radii, rotating regularly with a rotational period of $1.3\,\rm
Gyr$ (see Table~\ref{Tab_1}). The coherence in the \ion{H}{i} disk of
NGC~2685 suggests that its major portion has been acquired in a single
event in the evolution of NGC~2685. Due to the fact that a settling
onto regular orbits requires a few orbital periods, this implies an
age of several Gyr of the warped \ion{H}{i} disk of NGC~2685. In line
with this is the observation that the orientation of the outer
\ion{H}{i} (- and stellar) disk of NGC~2685 has the same orientation
as the inner lenticular stellar body, suggesting a connection of the
outer gaseous and stellar disk \citep[][]{Morganti06}. Following that
notion, \citet[][]{Morganti06} proposed that the warp of the inner
low-luminosity disk formed in a later event.

Optical observations, however, suggest that the warp is a long-lived
configuration. \citet[][]{Peletier93} estimated the age of the stellar
population in the helical structure to lie between 2 and 5~Gyr,
depending on metallicity. Following \citet[][]{Eskridge97}, who found
a solar abundance, the age of the stars in the helical structure is
about 2~Gyr. The most recent estimates by \citet[][]{Karataeva04}
point to a low metallicity typical for a dIrr or an LSB galaxy, which
corresponds to a larger age of the helix, possibly reaching the age of
the outer disk inferred from the rotation period. A further argument
for the longevity of the warped structure comes from the analysis of
our model parameters, that suggest a co-precession within the helical
structure, as indicated by the straight LON angle in the region of the
inner warp (Sect.~\ref{Sect_6}, Fig.~\ref{Fig_08}); the signature of a
transient warp would be a gradient in the LON angle due to a rapid
wind-up through differential precession.
%
%
\subsection{The kinematical model and the shape of the potential}
\label{Sect_8.2}
The low-surface-brightness component in NGC~2685 tends towards a polar
orientation at small radii, while at large radii the disk is aligned
with the central stellar body. The age estimates for the helical
structure point towards the fact that this feature is long-lived (see
previous section). Without being conclusive, we discuss possibilities
to explain the warped low-surface-brightness component as a long-lived
feature.

It had already been pointed out by \citet[][]{Simonson83} that NGC~2685
could contain an extremely warped component stabilised under the
influence of a prolate, tumbling potential. This was motivated by
\citet[][]{Tohline82} and \citet[][]{Albada82}, who found that under
the influence of such a potential the gas would settle into a stable
extremely warped plane. The tumbling of the potential would be
necessary to prevent the definition of two separate planes in which
orbits stable against differential precession are possible \citep[see
e.g.][]{Whitmore90,Reshetnikov94}, ultimately leading to a break-up of
the warp. Since the central stellar object at that time was not
clearly identified as a ``pancake'', the alternative, an elliptical
galaxy tumbling with a rotation axis orientated towards the pole of
the outer disk could explain the shape of the warp \citep[see
also][]{Varnas90,Peletier93,Bekki02}. While the nature of the central
object as a lenticular galaxy has been clarified, the possibility
remains that nevertheless the overall potential is triaxial and
tumbling, which would lead to the observed structure. Consequently,
the central lenticular galaxy would be subject to the same deviation
from axisymmetry in the potential. Therefore, its kinematics should
show accordingly oval distortions. We remark that such deviation is
not reported in the literature despite the existence of
high-resolution stellar absorption line kinematics
\citep[][]{Emsellem04}. The absorption-line velocity-field
published by \citet[][]{Emsellem04} does not show obvious signatures
of deviations from axial symmetry indicative for an oval distortion of
the potential out to a radius of $\approx 40\arcsec$ or about one third of
the region where we model the \ion{H}{i} disk to be severely
warped. Since therefore the potential is traced in the optical on a smaller scale than the extent of the warped region,
the lack of such obvious signatures is indicative, but it does not
exclude the possibility of a tumbling, triaxial potential at larger
radii.

A second possibility is that the inner warped structure is stabilised
via self-gravity under the influence of an oblate potential, aligned
with the central stellar body. In a simplified analysis (consisting of
two heavy rings, or ``wires'' under the influence of an oblate, scale
free logarithmic potential), \citet[][]{Sparke86} showed that
neutrally stable configurations exist, where the inner portion of a
self-gravitating disk lies close to polar, while the outer disk spin
axis aligns with the symmetry axis of the potential \citep[comparable
to Type II warps as defined by][]{Sparke88}. However, the assumption
of the presence of such a modified-tilt mode \citep[][]{Sparke88}, in
which the disk would be in a state of co-precession throughout is in
conflict with our observations (Sect.~\ref{Sect_6},
Fig.~\ref{Fig_08}).

The time-development of heavy disks under the influence of aspherical
potentials was investigated by \citet[][]{Sparke86} (oblate) and
\citet[][]{Arnaboldi94} (prolate, figure-rotating potentials). Within
the parameter range investigated, they observed that in case of a
large change in disk orientation (not only a slight warp), a break-up
of the disk into multiple regions occurs, which themselves show
co-precession. A global mode of co-precession is not reached. While
this is reminiscent of the rapid change in the LON angle at a radius
$r\approx 100\arcsec$ (Fig.~\ref{Fig_08}), it is questionable if the
coherency of the disk orientation in NGC~2685 is reproducible in a
model setup of a heavy disk in a distinctly aspherical potential. We
emphasise, however, that in the studies discussed here
\citep[][]{Sparke86,Sparke88,Arnaboldi94}, the parameter space was not
searched to fit an extremely warped disk, since the goal was to
resemble true polar ring galaxies with slight warps in their disks or
stable modes.

If we assume that the low-surface brightness disk of NGC~2685 has a
negligible mass compared to the common potential of the bright S0
stellar body and the dark-matter halo, we can estimate the flattening
q of the potential under the assumption of free precession in a scale
free logarithmic potential aligned with the main stellar body and
hence roughly with the outer \ion{H}{i} disk \citep[][]{Morganti06},
giving an upper limit. In the approximation of a nearly spherical,
logarithmic, scale free potential, the disk precesses about the pole
at a precession rate given by \citep[][]{Sparke86,Cox06}
\begin{equation}
\dot{\phi_i} = -\frac{3V_i}{4r_i}\eta \cos \theta_i\qquad {\rm ,}
\end{equation}
where $\theta_i$ is the inclination of the spin vector of the disk at
radius $r_i$ with respect to the symmetry axis of the potential, $V_i$
is the rotation velocity at $r_i$, and $\eta=1-b/a$ is the ellipticity
of the potential with axis ratio $b/a$, $a$ being the long axis. If we
assume the potential to be aligned with the outer disk, we can
calculate from our tilted-ring model the inclination of the disk spin
vectors with respect to the symmetry axis of the potential. With
$\phi_i$ being the angle enclosed by an arbitrary zero-point and the
projection of the spin normal vector at radius $r_i$ onto the symmetry
plane of the potential (the LON angle in a Briggs-plot with the outer
disk as reference plane; in this reference plane, the LON is not
straight), we can calculate the ellipticity employing above estimates
$t_{\rm warp}=3.5\pm1.5,\ \rm Gyr$ for the age of the warp feature:
\begin{equation}
\eta = \frac{4}{3}(\phi_i-\phi_j)(t_{\rm warp}(V_i/r_i \cos \theta_i -V_j/r_j \cos \theta_j))^{-1}
\end{equation}
We define the symmetry plane as the orbital plane given by the mean of
the orientational parameters between radii of $160\arcsec$ and
$360\arcsec$. Building a weighted mean of the potential ellipticities
calculated for every pair of radii between $36\arcsec$ and $96\arcsec$
yields $\eta = 0.011 \pm 0.009$: under the assumption of free
precession under the influence of a flattened potential aligned with
the central S0 stellar body, the potential of NGC~2685 in the reach of
the warp is close to spherical. Given the high orbital frequencies
within the radial range considered, the coherency of the disk, and the
high assumed life time of the warp, this result is expected.

The assumption of the presence of at most slight deviations from
sphericity of the overall potential is supported by the absence of
obvious global features in the residual velocity map
(Sect.~\ref{Sect_7}). Furthermore, we find that the measured rotation
curve does not show any peculiarities compared to a rotation curve
measured for a non-warped early-type galaxy. In the case of a halo
flattening towards the polar orientation as suggested for polar ring
galaxies by \citet[][]{Iodice03}, one would expect a distinct drop in
the measured rotation velocity within the warped region. The slight
decrease in rotation velocity towards larger radii and the shallow
maximum in a radial range of $60\arcsec$ to $100\arcsec$ is a common
feature observed for (non-warped) early type galaxies
\citep[e.g.][]{Noordermeer07}. The shape of the rotation curve is
therefore not indicative of a flattening of the potential.

In the same notion, comparing our distance estimate from the
Tully-Fisher relation to the Hubble Distance, we do not find a
significant offset from the Tully-Fisher relation as reported by
\citet[][NGC~2685 being a galaxy in their sample]{Iodice03}: NGC~2685
does not appear significantly dimmer than suggested by the \ion{H}{i}
line width (see Sect.~\ref{Sect_5}).  In a statistical treatment of
several polar ring galaxies, \citet[][]{Iodice03} interprete such an
offset from the Tully-Fisher relation as indication for a deformation
of the potential. Thus, we conclude that also the peak amplitude of
the rotation curve of NGC~2685 (determining the \ion{H}{i} line width)
is not indicative of a flattening of the potential.
%
%
\section{The environment of the Spindle}
\label{Sect_9}
\begin{figure}[htbp]
\resizebox{\hsize}{!}{\includegraphics[angle=270]{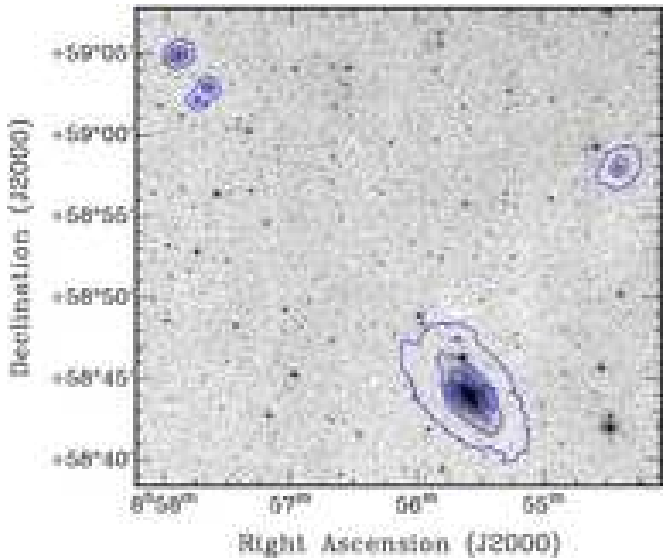}}
\caption{
\ion{H}{i} total-intensity maps of the observed field overlaid on a
DSS (red) image. From top to bottom: UGC~4683, two unknown possibly
merging objects, PGC~25002, NGC~2685. The contours represent the $1.5,
40, 85, 145 \tth$ levels. Note that the \ion{H}{i} map is primary-beam
corrected.
}
\label{Fig_11}
\end{figure}

%
%
\begin{table*}[htbp]
\caption{
Basic properties of the detected dwarf galaxies in the vicinity of
NGC~2685. For all objects the same distance of NGC~2685
(Table~\ref{Tab_1}) was assumed. The data show that these galaxies
probably cannot account for the ring in NGC~2685 and are dynamically
unimportant.}
\label{Tab_4}
\begin{center}
\begin{tabular}{llrrr}
\hline
\hline                    
 Descr.                                                                                     & Par.                       & PGC 25002                  & UGC 4683                    & unknown                     \\
\hline                    
 Classification of galaxy (\citealt[][]{Vaucouleurs91})                                     & Type                       & IB(rs)m                    & dI                          & unknown                     \\
 Right Ascension (J2000) (NED)                                                              & RA (J2000)                 & $ \hmsm{8}{54}{22}{50.0} $ & $ \hmsm{8}{57}{54}{37.0}  $ & $ \hmsm{8}{57}{42}{51.0}  $ \\
 Declination (J2000) (NED)                                                                  & Dec (J2000)                & $ \dmsm{58}{58}{4}{0.00} $ & $ \dmsm{59}{4}{57}{7.00}  $ & $ \dmsm{59}{2}{28}{85.00} $ \\
 Optical heliocentric systemic velocity (${\rm km}\,{\rm s}^{-1}$)                          & $ V_{\rm sys} $            & $  1017.2 \,\pm\,  8.2   $ & $  927.0 \,\pm\,  8.2     $ & $ 1187.0   \,\pm\,  8.2   $ \\
 Apparent B-band magnitude ($\rm mag$, \citealt[][]{Paturel89})                             & $ m_{\rm B} $              & $    16.4                $ & $   14.8                  $ &  very low                   \\
Total \ion{H}{i} flux (${\rm Jy}\,{\rm km}\,{\rm s}^{-1}$)                                  & $ F_{\rm \ion{H}{i}} $     & $     3.3 \,\pm\,  0.3   $ & $    2.2 \,\pm\,  0.2     $ & $    2.1   \,\pm\,  0.2   $ \\
\ion{H}{i} mass ($10^7\,{\rm M}_\odot$)                                                     & $ M_{\rm \ion{H}{i}} $     & $    17.8 \,\pm\,  1.9   $ & $   12.2 \,\pm\,  1.3     $ & $   11.3   \,\pm\,  1.2   $ \\
 Projected distance of object from NGC~2685 ($\,\arcsec$)                                   & $ \rm d_{2685} $           & $  1016   \,\pm\, 28     $ & $ 1651   \,\pm\, 28       $ & $ 1484     \,\pm\, 28     $ \\
  Distance of object from NGC~2685 ($\rm kpc$)                                              & $ \rm D_{2685} $           & $    75   \,\pm\, 20     $ & $  122   \,\pm\, 31       $ & $  109     \,\pm\, 28     $ \\
   Profile width at 20 percent of the peak flux density (${\rm km}\,{\rm s}^{-1}$)          & $ w_{20} $                 & $    95.0 \,\pm\,  9.6   $ & $   64.0 \,\pm\,  6.5     $ & $   72.5   \,\pm\,  7.2   $ \\
  Inclination ($\,\deg$)                                                                    & $ i $                      & $    29.2                $ & $   31.6                  $ & $   90.0 $                  \\
 Dynamical mass ($10^8{\rm M}_\odot$)                                                       & $ M_{\rm dyn} $            & $ > 137 \,\pm\, 53       $ & $ > 32 \,\pm\, 18         $ & $ >    8.7 \,\pm\,  6.0   $ \\
\hline
\end{tabular}
\end{center}
\end{table*}
%
%
In \ion{H}{i}, NGC~2685 and four additional, small objects were
detected (Fig.~\ref{Fig_11}). This means that NGC~2685 is probably the
dominating member of a small loose group of galaxies; hence, the
question arises whether the peculiar structure of NGC~2685 might be
connected to a past interaction with (one of) its companions. Two
\ion{H}{i} detections can be associated with the known dwarf galaxies
UGC~4683 and PGC~25002, while two objects were detected that are
barely, if at all, visible in the DSS. Their projected separation is
very small ($\approx65\arcsec\,\hat{=}\,4.8\,{\rm kpc}$) and a faint
\ion{H}{i} bridge between the two gas clouds is present. However,
their kinematical axes are misaligned, indicating that these are two
individual, possibly interacting dwarf galaxies with an extremely low
surface brightness. Figure~\ref{Fig_11} shows an overlay of the total
\ion{H}{i} map on a DSS image containing all objects detected in
\ion{H}{i}.
%
%
%
%
Table~\ref{Tab_4} lists the properties of the \ion{H}{i} detections
and associated galaxies. The \ion{H}{i} masses are derived under the
assumption that each object is at the same distance as NGC~2685
($15.2\,\rm kpc$, see Table~\ref{Tab_1}). The diameters needed to
calculate their dynamical masses are estimated from the moment
maps. The integrated \ion{H}{i} spectra of the galaxies are used to
calculate the $20\%$ peak line width and the systemic velocities of
the objects. Then, the (lower limit to the) dynamical mass is derived
from these line widths and the diameters of the targets. In case of
the two unknown objects, only rough estimates can be made. They are
approximately of the same size and have probably the same rotation
amplitude. The \ion{H}{i} flux, systemic velocity and linewidth are
calculated from the spectrum of both objects, while the dynamical mass
was derived taking into account the extent of only one of the objects.
Table~\ref{Tab_4} shows that none of the possible companion galaxies
is large enough to be important for the dynamics of NGC~2685. The most
massive companion is PGC~25002 with about one tenth of the (dynamical)
mass of NGC~2685; its projected distance to NGC~2685 is $75\,\rm
kpc$. The mass-ratio of PGC~25002 and NGC~2685 is roughly the same as
that of the Milky Way and the Large Magellanic Cloud, which is
currently at its pericentric passage and depositing quite some gas on
a polar orbit about the Milky Way albeit at a large distance from the
Galactic centre. The position of PGC~25002 lies roughly in the
direction of the projected minor axis of the central stellar body of
NGC~2685, such that it might be speculated that this object deposited
some amount of gas into a polar orbit about NGC~2685. However, the
rather high ratio of gas mass to B-band luminosity of $1.8\pm 0.2
\,{M_{\rm {\ion{H}{i}}}}/{L_{\rm B}}$ for PGC~25002 (estimated from
Table~\ref{Tab_4}) indicates that this galaxy has not lost any major
fraction of its gas. Consequently a comparison of the total gas masses
of NGC~2685 and PGC~25002 (10:1) makes it questionable as a donor for
the complete amount of neutral gas in NGC~2685.  This conclusion
remains if we consider only the gas mass contained in the inner warped
structure, assuming that only the gas in the warped structure has been
acquired in an accretion event: from our model we estimate a gas mass
of $5.2\cdot 10^8 {M}_\odot$ inside radii $\leq 96^{\prime\prime}$,
which is a third of the total \ion{H}{i} mass of NGC~2685.

UGC~4683 has a ratio of gas mass to B-band luminosity of $0.28\pm 0.03
\,{M_{\rm {\ion{H}{i}}}}/{L_{\rm B}}$. Under the assumption that the
\ion{H}{i} in NGC~2685 comes from this galaxy, an initial ratio of gas
mass to B-band luminosity of about $4\,{M_{\rm {\ion{H}{i}}}}/{L_{\rm
B}}$ before the encounter with NGC~2685 would be required. In this
case, because of the position of the galaxy with respect to NGC~2685,
UGC~4683 must have donated its gas on an orbit well aligned with the
outer gas disk and hence to the inner lenticular stellar body. Also
here it is unlikely that the complete gas content of NGC~2685 was
deposited by UGC~4683.

The optical observations show a very faint elliptical stellar system
located at the SW side of NGC~2685 (see Fig.~\ref{Fig_01}, upper left
panel). It is, however, unclear at which distance this object is with
respect to NGC~2685, since it is not detected in \ion{H}{i}. There is
no evidence for any interaction as no \ion{H}{i} feature in the gas
belonging to NGC~2685 seems to be connected with the object.

We therefore conclude that the peculiar structure of NGC~2685 is
probably not caused by an interaction with its direct neighbours. It
is, however, clear that with a lifetime of the helix of more than 2
Gyr a possible donor galaxy that had an encounter with NGC~2685 can
now be located at a distance of Megaparsecs. Hence, the lack of
possible donors in the field-of-view of our \ion{H}{i} observations
does not exclude the possibility of a gas accretion from a bypassing
galaxy as a possible origin for the peculiar structure of NGC~2685.
%
%
\section{Summary and discussion}
\label{Sect_10}
We analyse deep WSRT HI synthesis observations and deep INT WFC
observations in the i$^\prime$ band of the Spindle Galaxy
NGC~2685. Our analysis includes a tilted-ring parametrisation of the
\ion{H}{i} disk. Our analysis 
supports the following
hypotheses:
\begin{itemize}
\item{
NGC~2685 possesses two separate components, one being the central S0
stellar body. The other is a coherent large-scale warped disk
consisting of gas, stars and dust. The appearance of a two-ringed
structure \citep[][]{Shane77,Shane80,Mahon92,Schinnerer02} 
is due to
projection effects. This makes scenarios in which the rings,
respectively the disk, of NGC~2685 have been acquired in two separate
accretion events unlikely.
}
\item{
the large-scale disk shares many properties of a small spiral-galaxy
disk of low surface brightness, containing molecular-, neutral-, and
ionised gas and stars. The helical dust-lane structure appearing in
optical images can be identified with spiral arms that are matched by
density peaks in the neutral gas component. The spiral arms also occur
on the far side of the galaxy in high resolution \ion{H}{i} maps.  The
\ion{H}{i} surface density profile resembles that of any spiral
galaxy.
}
\item{
our analysis indicates the presence of a relaxed \ion{H}{i} disk at
large radii, rotating regularly with a rotational period of 1.3 Gyr
(see Table~\ref{Tab_1}). The age of the stellar population in the
helical structure is between 2 and 5 Gyrs
\citep[][]{Peletier93,Eskridge97,Karataeva04}. Since after an
accretion event several rotational periods are necessary for the gas
to settle on regular orbits, we conclude that the age of the gas disk
is probably significantly larger than 2 Gyr. This is consistent with
the findings of \citet[][]{Karataeva04}. In contrast to
\citet[][]{Eskridge97}, they found a metallicity typical for a dIrr or
LSB galaxy with the consequence of a higher stellar age
\citep[][]{Peletier93}. This suggests that the large-scale disk of
NGC~2685, part of which is the helical structure, has not been
acquired recently, but is of intermediate age.
}
\item{
The warp in the large-scale disk shares important properties with
warps of lower amplitude in spiral galaxies. The disk is divided into
two regimes each of which shares a common line-of-nodes
\citep[][]{Briggs90}. Hence, the disk is co-precessing (or not
precessing) in both regimes. The disk is coherent, hence both regimes
are connected by a small transition region. In the transition region
the nodal line progresses in the direction of rotation, as was
previously found by \citet[][]{Briggs90} for other galaxies. The outer
disk of NGC~2685 is flat out to a radius of about $420\arcsec\,
\hat{=} \, 31\, \rm kpc$.
}
\item{
For radii $\geq 36\arcsec$, where we consider our model reliable, the
maximal mutual inclination of two spin vectors is $\approx75\deg$ and
no polar orientation w.r.t. to the central bright stellar body is
reached.  Optical analyses indicate that this also holds for the
ionised gas disk at radii $\leq 36\arcsec$ \citep[][]{Sarzi06}.
}
\item
{
Within the scope of our measurements and our analysis, 
we do not find obvious indications for a large deviation of the
potential from sphericity. The absence of global residuals when
subtracting the modelled velocity field from the observed ones and the
fact that the rotation curve does not show any peculiarities compared
to rotation curves of other early-type galaxies rather indicates that
the potential is close to spherical. This is consistent with an
estimate of the ellipticity of the potential in the limit of free
precession ($\eta = 0.011 \pm 0.009 $), providing an upper limit for
the ellipticity.
}
\item
{
Because of their low \ion{H}{i} mass compared to that of NGC~2685,
and the lack of an \ion{H}{i}-deficiency, companions of NGC~2685
detected in \ion{H}{i} can be excluded as donors to create the entire
coherent \ion{H}{i} disk. If an interaction with a present companion
galaxy created the warped structure of the large-scale disk, the
interaction was mainly of tidal nature.
}
\end{itemize}
As stated in Sect.~\ref{Sect_4}, cases similar to NGC~2685 are known
and the question arises how such a structure can form. The fact that
the colours of the inner stellar body and the helical structure differ
\citep[][]{Peletier93} leaves little doubt that the inner warp and
segregation of two rotating systems is the result of a more recent
interaction rather than it has formed at early times.

As it is the case for polar ring galaxies
\citep[e.g.][]{Bekki97,Bournaud03}, two scenarios come into
question. The first is that NGC~2685 is the result of a merger
event. Simulations of wet, dissipative mergers of spiral galaxies have
shown that indeed extremely warped structures of resulting gas disks
as reported here can show up as a result of a merger \citep[][Fig.~2
therein showing a case with a gas structure very similar to that of
NGC~2685]{Barnes02,Naab06}. \citet[][]{Barnes02} shows that extended,
regular gas disks (containing up to about half of the gas of the
progenitor galaxies) can build up after a major merger that would
resemble the outer \ion{H}{i} disk of NGC~2685, while towards the
centre a highly inclined disk turns up, in some cases connected to the
outer disk. In contrast to \citet[][]{Naab06}, \citet[][]{Barnes02}
finds extremely warped gas remnants of a merger to turn up also in the
case of unequal (3:1) mass mergers. This is of relevance insofar as
simulations
\citep[][]{Naab99,Naab03} show that generally equal-mass mergers tend
to result in slowly rotating, boxy stellar remnants, while the central
stellar body of NGC~2685, despite showing a de Vaucouleurs profile in
surface-brightness \citep[][]{Peletier93}, is clearly a fast rotating,
flattened system \citep[][]{Emsellem04}, as is produced in simulations
with preference in unequal-mass mergers.

A second scenario could be that the central warp in NGC 2685 is the
result of gas accretion, the gas being acquired either in the course
of a tidal interaction with a (now distant, Sect.~\ref{Sect_9}) donor
galaxy \citep[][]{Schweizer83,Reshetnikov97,Bournaud03}, or through
direct infall of cold, primordial gas \citep[][]{Maccio06}.  In the
latter scenario, the (cold) accreted gas falls onto an inner part of a
pre-existing spiral galaxy, is stopped in the gas disk and mixes its
angular momentum with the original gas disk only in the region of the
resulting warp. Depending on the ratio of angular momentum carried by
the infalling material and the disk at differing radii, in this way a
warped gas structure might build up. A similar scenario has been
successfully simulated by \citet[][]{Bournaud03} to explain the
co-existence of an inner polar and an outer equatorial ring. The
important difference of their model predictions and our simple model
for the gas structure in NGC~2685 is that in their simulations a
coherent disk and a warp of large amplitude never turns up. This is
due to the fact that within their setup the gas on an inclined orbit
with respect to the progenitor disk is subject to a torque strong
enough to prevent the formation of warped structure with partly low
inclination with respect to the equatorial plane. Only gas on highly
inclined orbits is stable against differential precession. This
situation might change if the torque imposed on the mixture of
infalling- and disk material is attenuated by choosing a different
setup, e.g. a more dominating spherical DM halo potential. In favour
of the accretion scenario could be the excess of \ion{H}{i} surface
brightness (Fig.~\ref{Fig_06}) in the region of the inner warp.  The
shape of the \ion{H}{i} surface-brightness profile is not unusual
compared to a spiral galaxy. A concern against the accretion scenario
is the light profile of the central stellar
body. \citet[][]{Peletier93} showed that it follows a de
Vaucouleurs-law, while an exponential profile would be expected if the
central stellar body would form from a spiral galaxy stripped from the
gas in the inner regions. Furthermore, the accretion scenario would
require fine-tuning, since the structure of the gas disk of NGC~2685
is highly symmetric. An accretion event would likely also impose
linear momentum on the progenitor disk, which would then probably
result in a more asymmetric warp and a lopsided kinematics and gas
distribution. The task would be to find a scenario where the transfer
of linear momentum is small or where the resulting displacement of
inner, warped disk and outer disk vanishes quickly
\citep[see e.g. ][on the topic of warp formation through gas
infall]{Sanchez-Salcedo05}.

A possibility to distinguish between the two tentative scenarios might
be to consider photometric observations. Obviously, a stellar
component has formed within the gaseous, warped disk of NGC~2685. In
the merging scenario, the gas disk grows from inside out
\citep[][]{Barnes02}, after a rapid initial transport of gas to the
centre. Therefore, it is to be expected that the age of the
subsequently formed stars becomes younger towards larger radii. Hence,
if at all, a colour gradient towards a bluer population with
increasing radius is expected in the merging scenario. In the
accretion scenario, since the bulk of the new star forming material
should be placed in the centre of the galaxy, a transition in colour
from the inner, warped, region, to the outer remnant of the progenitor
disk is expected to occur from bluer to redder colours.  Considering
the colour maps published by \citet[][]{Erwin02}, it is immediatily
evident that this distinction cannot be made easily. In fact, the
inner helical structure appears reddened with respect to the outer
projected ring, but this may well be due to reddening by dust and a
mixing of the light from the low-surface-brightness disk and the
central stellar body on the line-of-sight. According to our
tilted-ring model (Fig.~\ref{Fig_05}), the material along the
projected outer ring is a mixture of material at smaller and larger
radii, while on the apparent minor axis it is dominated by material at
smaller radii and on the apparent major axis it is dominated by
material at larger radii. Thus, one can inspect the projected ring to
search for a change in colour along the projected ring. While the
colour changes indeed along the projected outer ring, no clear trend
can be made out in the colour maps published by \citet[][]{Erwin02}.

We cannot be conclusive about the origin of the structure of
NGC~2685. However, due to the facts that the inner stellar body does
not follow an exponential light profile, and that warped gas
structures do turn up in simulations of wet mergers, we give our
preference to the hypothesis that NGC~2685 is the result of a merger
of two disk galaxies.

NGC~2685 remains an interesting object, mainly because of the
co-existence of two rotating disks of extreme symmetry at differing
orientation. Exploiting the kinematics of both the stars of the
central lenticular stellar body and the gas disk should lead to
further insight in the 3D structure of the galaxy.

\begin{acknowledgements}

We thank the first, anonymous referee for initiating the discussion
about the origin of the intrinsic structure of NGC~2685 in
Sect.~\ref{Sect_10}, and for animating us to compare our results
concerning the Tully Fisher relation to the work by Enrichetta Iodice
and collaborators.

We like to thank the editor, Fran\c{c}oise Combes, for finishing the refereeing process and for giving good advise.

The underlying PhD thesis project was partly financed by the Deutsche
Forschungsgemeinschaft in the framework of the Graduiertenkolleg 787
``Galaxy Groups as Laboratories for Baryonic and Dark Matter''.

The underlying PhD thesis project was also partly financed by the
University of Bonn in the framework of the ``Research Group Bonn: Dark
Matter and Dark Energy''.

GJ likes to emphasise that the NGC~2685-project originates from a
summer studentship at ASTRON (Dwingeloo, The Netherlands).

The Westerbork Synthesis Radio Telescope is operated by the ASTRON
(Netherlands Foundation for Research in Astronomy) with support from
the Netherlands Foundation for Scientific Research (NWO).

The Isaac Newton Telescope is operated on the island of La Palma by
the Isaac Newton Group in the Spanish Observatorio del Roque de los
Muchachos of the Instituto de Astrof\'isica de Canarias.

This work makes use of a digitised image of the Second Palomar
Observatory Sky Survey (POSS-II) obtained through the Digitized Sky
Survey (DSS) websites. The Digitized Sky Surveys were produced at the
Space Telescope Science Institute under U.S. Government grant NAG
W-2166.

This research has made use of the NASA/IPAC Extragalactic Database
(NED) which is operated by the Jet Propulsion Laboratory, California
Institute of Technology, under contract with the National Aeronautics
and Space Administration\footnote{The NASA/IPAC Extragalactic Database
(NED) is operated by the Jet Propulsion Laboratory, California
Institute of Technology, under contract with the National Aeronautics
and Space Administration.}.

We acknowledge the usage of the HyperLeda database (http://leda.univ-lyon1.fr).
\end{acknowledgements}

\bibliographystyle{aa}
\bibliography{n2685_aa}

\begin{thebibliography}{69}
\expandafter\ifx\csname natexlab\endcsname\relax\def\natexlab#1{#1}\fi

\bibitem[{{Arnaboldi} \& {Galletta}(1993)}]{Arnaboldi93}
{Arnaboldi}, M. \& {Galletta}, G. 1993, \aap, 268, 411

\bibitem[{{Arnaboldi} \& {Sparke}(1994)}]{Arnaboldi94}
{Arnaboldi}, M. \& {Sparke}, L.~S. 1994, \aj, 107, 958

\bibitem[{{Barnes}(2002)}]{Barnes02}
{Barnes}, J.~E. 2002, \mnras, 333, 481

\bibitem[{{Bekki}(1997)}]{Bekki97}
{Bekki}, K. 1997, \apjl, 490, L37

\bibitem[{{Bekki} \& {Freeman}(2002)}]{Bekki02}
{Bekki}, K. \& {Freeman}, K.~C. 2002, \apjl, 574, L21

\bibitem[{{Bournaud} \& {Combes}(2003)}]{Bournaud03}
{Bournaud}, F. \& {Combes}, F. 2003, \aap, 401, 817

\bibitem[{{Briggs}(1990)}]{Briggs90}
{Briggs}, F.~H. 1990, \apj, 352, 15

\bibitem[{{Cox} {et~al.}(2006){Cox}, {Sparke}, \& {van Moorsel}}]{Cox06}
{Cox}, A.~L., {Sparke}, L.~S., \& {van Moorsel}, G. 2006, \aj, 131, 828

\bibitem[{{de Vaucouleurs} {et~al.}(1991){de Vaucouleurs}, {de Vaucouleurs},
  {Corwin}, {Buta}, {Paturel}, \& {Fouque}}]{Vaucouleurs91}
{de Vaucouleurs}, G., {de Vaucouleurs}, A., {Corwin}, H.~G., {et~al.} 1991,
  {Third Reference Catalogue of Bright Galaxies} (Berlin Heidelberg New York:
  Springer)

\bibitem[{{Emsellem} {et~al.}(2004){Emsellem}, {Cappellari}, {Peletier},
  {McDermid}, {Bacon}, {Bureau}, {Copin}, {Davies}, {Krajnovi{\'c}},
  {Kuntschner}, {Miller}, \& {de Zeeuw}}]{Emsellem04}
{Emsellem}, E., {Cappellari}, M., {Peletier}, R.~F., {et~al.} 2004, \mnras,
  352, 721

\bibitem[{{Erben} {et~al.}(2005){Erben}, {Schirmer}, {Dietrich}, {Cordes},
  {Haberzettl}, {Hetterscheidt}, {Hildebrandt}, {Schmithuesen}, {Schneider},
  {Simon}, {Deul}, {Hook}, {Kaiser}, {Radovich}, {Benoist}, {Nonino}, {Olsen},
  {Prandoni}, {Wichmann}, {Zaggia}, {Bomans}, {Dettmar}, \&
  {Miralles}}]{Erben05}
{Erben}, T., {Schirmer}, M., {Dietrich}, J.~P., {et~al.} 2005, Astronomische
  Nachrichten, 326, 432

\bibitem[{{Erwin} \& {Sparke}(2002)}]{Erwin02}
{Erwin}, P. \& {Sparke}, L.~S. 2002, \aj, 124, 65

\bibitem[{{Erwin} \& {Sparke}(2003)}]{Erwin03}
{Erwin}, P. \& {Sparke}, L.~S. 2003, \apjs, 146, 299

\bibitem[{{Eskridge} \& {Pogge}(1997)}]{Eskridge97}
{Eskridge}, P.~B. \& {Pogge}, R.~W. 1997, \apj, 486, 259

\bibitem[{{Franx} {et~al.}(1994){Franx}, {van Gorkom}, \& {de Zeeuw}}]{Franx94}
{Franx}, M., {van Gorkom}, J.~H., \& {de Zeeuw}, T. 1994, \apj, 436, 642

\bibitem[{{Iodice} {et~al.}(2003){Iodice}, {Arnaboldi}, {Bournaud}, {Combes},
  {Sparke}, {van Driel}, \& {Capaccioli}}]{Iodice03}
{Iodice}, E., {Arnaboldi}, M., {Bournaud}, F., {et~al.} 2003, \apj, 585, 730

\bibitem[{{J\'ozsa}(2006)}]{Jozsa06}
{J\'ozsa}, G.~I.~G. 2006, PhD thesis, Univ. Bonn, available at
  http://hss.ulb.uni-bonn.de/diss\_online/math\_nat\_fak/
  2006/jozsa\_gyula/index.htm

\bibitem[{{J\'ozsa}(2007)}]{Jozsa07b}
{J\'ozsa}, G.~I.~G. 2007, \aap, ~(Paper II), in press

\bibitem[{{J\'ozsa} {et~al.}(2007){J\'ozsa}, {Kenn}, Klein, \&
  {Oosterloo}}]{Jozsa07a}
{J\'ozsa}, G.~I.~G., {Kenn}, F., Klein, U., \& {Oosterloo}, T.~A. 2007, \aap,
  ~(Paper 1), in press

\bibitem[{{Karataeva} {et~al.}(2004){Karataeva}, {Drozdovsky}, {Hagen-Thorn},
  {Yakovleva}, {Tikhonov}, \& {Galazutdinova}}]{Karataeva04}
{Karataeva}, G.~M., {Drozdovsky}, I.~O., {Hagen-Thorn}, V.~A., {et~al.} 2004,
  \aj, 127, 789

\bibitem[{{Macci{\`o}} {et~al.}(2006){Macci{\`o}}, {Moore}, \&
  {Stadel}}]{Maccio06}
{Macci{\`o}}, A.~V., {Moore}, B., \& {Stadel}, J. 2006, \apjl, 636, L25

\bibitem[{{Mahon}(1992)}]{Mahon92}
{Mahon}, M.~E. 1992, Bulletin of the American Astronomical Society, 24, 1267

\bibitem[{{McMahon} {et~al.}(2001){McMahon}, {Walton}, {Irwin}, {Lewis},
  {Bunclark}, \& {Jones}}]{McMahon01}
{McMahon}, R.~G., {Walton}, N.~A., {Irwin}, M.~J., {et~al.} 2001, New Astronomy
  Review, 45, 97

\bibitem[{{Monet}(1998)}]{Monet98}
{Monet}, D.~G. 1998, in Bulletin of the American Astronomical Society, Vol.~30,
  Bulletin of the American Astronomical Society, 1427--1428

\bibitem[{{Morganti} {et~al.}(2006){Morganti}, {de Zeeuw}, {Oosterloo},
  {McDermid}, {Krajnovi{\'c}}, {Cappellari}, {Kenn}, {Weijmans}, \&
  {Sarzi}}]{Morganti06}
{Morganti}, R., {de Zeeuw}, P.~T., {Oosterloo}, T.~A., {et~al.} 2006, \mnras,
  371, 157

\bibitem[{{Morganti} {et~al.}(1997){Morganti}, {Sadler}, {Oosterloo},
  {Pizzella}, \& {Bertola}}]{Morganti97}
{Morganti}, R., {Sadler}, E.~M., {Oosterloo}, T., {Pizzella}, A., \& {Bertola},
  F. 1997, \aj, 113, 937

\bibitem[{{Naab} \& {Burkert}(2003)}]{Naab03}
{Naab}, T. \& {Burkert}, A. 2003, \apj, 597, 893

\bibitem[{{Naab} {et~al.}(1999){Naab}, {Burkert}, \& {Hernquist}}]{Naab99}
{Naab}, T., {Burkert}, A., \& {Hernquist}, L. 1999, \apjl, 523, L133

\bibitem[{{Naab} {et~al.}(2006){Naab}, {Jesseit}, \& {Burkert}}]{Naab06}
{Naab}, T., {Jesseit}, R., \& {Burkert}, A. 2006, \mnras, 372, 839

\bibitem[{{Nicholson} {et~al.}(1987){Nicholson}, {Taylor}, {Sparks}, \&
  {Bland}}]{Nicholson87}
{Nicholson}, R.~A., {Taylor}, K., {Sparks}, W.~B., \& {Bland}, J. 1987, in IAU
  Symposium, Vol. 127, Structure and Dynamics of Elliptical Galaxies, ed. P.~T.
  {de Zeeuw}, 415--+

\bibitem[{{Noordermeer} {et~al.}(2007){Noordermeer}, {van der Hulst},
  {Sancisi}, {Swaters}, \& {van Albada}}]{Noordermeer07}
{Noordermeer}, E., {van der Hulst}, J.~M., {Sancisi}, R., {Swaters}, R.~S., \&
  {van Albada}, T.~S. 2007, \mnras, 376, 1513

\bibitem[{{Oosterloo} {et~al.}(2007){Oosterloo}, {Morganti}, {Sadler}, {van der
  Hulst}, \& {Serra}}]{Oosterloo07}
{Oosterloo}, T.~A., {Morganti}, R., {Sadler}, E.~M., {van der Hulst}, T., \&
  {Serra}, P. 2007, \aap, 465, 787

\bibitem[{{Oosterloo} {et~al.}(2002){Oosterloo}, {Morganti}, {Sadler},
  {Vergani}, \& {Caldwell}}]{Oosterloo02}
{Oosterloo}, T.~A., {Morganti}, R., {Sadler}, E.~M., {Vergani}, D., \&
  {Caldwell}, N. 2002, \aj, 123, 729

\bibitem[{{Paturel}(1989)}]{Paturel89}
{Paturel}, G. 1989, {Catalogue of principal galaxies (PGC)} (Lyon: Base de
  Donnees Extragalactiques, Observatoire de Lyon)

\bibitem[{{Paturel} {et~al.}(2003){Paturel}, {Petit}, {Prugniel}, {Theureau},
  {Rousseau}, {Brouty}, {Dubois}, \& {Cambr{\'e}sy}}]{LEDA}
{Paturel}, G., {Petit}, C., {Prugniel}, P., {et~al.} 2003, \aap, 412, 45

\bibitem[{{Peletier} \& {Christodoulou}(1993)}]{Peletier93}
{Peletier}, R.~F. \& {Christodoulou}, D.~M. 1993, \aj, 105, 1378

\bibitem[{{Prugniel} \& {Heraudeau}(1998)}]{Prugniel98}
{Prugniel}, P. \& {Heraudeau}, P. 1998, \aaps, 128, 299

\bibitem[{{Reshetnikov} \& {Sotnikova}(1997)}]{Reshetnikov97}
{Reshetnikov}, V. \& {Sotnikova}, N. 1997, \aap, 325, 933

\bibitem[{{Reshetnikov} {et~al.}(1994){Reshetnikov}, {Hagen-Thorn}, \&
  {Yakovleva}}]{Reshetnikov94}
{Reshetnikov}, V.~P., {Hagen-Thorn}, V.~A., \& {Yakovleva}, V.~A. 1994, \aap,
  290, 693

\bibitem[{{Revaz} \& {Pfenniger}(2001)}]{Revaz01}
{Revaz}, Y. \& {Pfenniger}, D. 2001, \aap, 372, 784

\bibitem[{{Richter} {et~al.}(1994){Richter}, {Sackett}, \&
  {Sparke}}]{Richter94}
{Richter}, O.-G., {Sackett}, P.~D., \& {Sparke}, L.~S. 1994, \aj, 107, 99

\bibitem[{{Sakai} {et~al.}(2000){Sakai}, {Mould}, {Hughes}, {Huchra}, {Macri},
  {Kennicutt}, {Gibson}, {Ferrarese}, {Freedman}, {Han}, {Ford}, {Graham},
  {Illingworth}, {Kelson}, {Madore}, {Sebo}, {Silbermann}, \&
  {Stetson}}]{Sakai00}
{Sakai}, S., {Mould}, J.~R., {Hughes}, S.~M.~G., {et~al.} 2000, \apj, 529, 698

\bibitem[{{S{\'a}nchez-Salcedo}(2006)}]{Sanchez-Salcedo05}
{S{\'a}nchez-Salcedo}, F.~J. 2006, \mnras, 365, 555

\bibitem[{{Sandage}(1961)}]{Sandage61}
{Sandage}, A. 1961, {The Hubble atlas of galaxies} (Washington: Carnegie
  Institution)

\bibitem[{{Sarzi} {et~al.}(2006){Sarzi}, {Falc{\'o}n-Barroso}, {Davies},
  {Bacon}, {Bureau}, {Cappellari}, {de Zeeuw}, {Emsellem}, {Fathi},
  {Krajnovi{\'c}}, {Kuntschner}, {McDermid}, \& {Peletier}}]{Sarzi06}
{Sarzi}, M., {Falc{\'o}n-Barroso}, J., {Davies}, R.~L., {et~al.} 2006, \mnras,
  366, 1151

\bibitem[{{Sault} {et~al.}(1995){Sault}, {Teuben}, \& {Wright}}]{Sault95}
{Sault}, R.~J., {Teuben}, P.~J., \& {Wright}, M.~C.~H. 1995, in Astronomical
  Data Analysis Software and Systems IV, ASP Conf. Ser., 77 (San Francisco:
  ASP), 433

\bibitem[{{Schechter} \& {Gunn}(1978)}]{Schechter78}
{Schechter}, P.~L. \& {Gunn}, J.~E. 1978, \aj, 83, 1360

\bibitem[{{Schinnerer} \& {Scoville}(2002)}]{Schinnerer02}
{Schinnerer}, E. \& {Scoville}, N. 2002, \apjl, 577, L103

\bibitem[{{Schirmer} {et~al.}(2003){Schirmer}, {Erben}, {Schneider},
  {Pietrzynski}, {Gieren}, {Carpano}, {Micol}, \& {Pierfederici}}]{Schirmer03}
{Schirmer}, M., {Erben}, T., {Schneider}, P., {et~al.} 2003, \aap, 407, 869

\bibitem[{{Schlegel} {et~al.}(1998){Schlegel}, {Finkbeiner}, \&
  {Davis}}]{Schlegel98}
{Schlegel}, D.~J., {Finkbeiner}, D.~P., \& {Davis}, M. 1998, \apj, 500, 525

\bibitem[{{Schoenmakers} {et~al.}(1997){Schoenmakers}, {Franx}, \& {de
  Zeeuw}}]{Schoenmakers97}
{Schoenmakers}, R.~H.~M., {Franx}, M., \& {de Zeeuw}, P.~T. 1997, \mnras, 292,
  349

\bibitem[{{Schwarz}(1985)}]{Schwarz85}
{Schwarz}, U.~J. 1985, \aap, 142, 273

\bibitem[{{Schweizer} {et~al.}(1983){Schweizer}, {Whitmore}, \&
  {Rubin}}]{Schweizer83}
{Schweizer}, F., {Whitmore}, B.~C., \& {Rubin}, V.~C. 1983, \aj, 88, 909

\bibitem[{{Serra} {et~al.}(2006){Serra}, {Trager}, {van der Hulst},
  {Oosterloo}, \& {Morganti}}]{Serra06}
{Serra}, P., {Trager}, S.~C., {van der Hulst}, J.~M., {Oosterloo}, T.~A., \&
  {Morganti}, R. 2006, \aap, 453, 493

\bibitem[{{Shane}(1977)}]{Shane77}
{Shane}, W.~W. 1977, \baas, 9, 362

\bibitem[{{Shane}(1980)}]{Shane80}
{Shane}, W.~W. 1980, \aap, 82, 314

\bibitem[{{Shen} \& {Sellwood}(2006)}]{Shen06}
{Shen}, J. \& {Sellwood}, J.~A. 2006, \mnras, 370, 2

\bibitem[{{Simonson} \& {Tohline}(1983)}]{Simonson83}
{Simonson}, G.~F. \& {Tohline}, J.~E. 1983, \apj, 268, 638

\bibitem[{{Sparke}(1986)}]{Sparke86}
{Sparke}, L.~S. 1986, \mnras, 219, 657

\bibitem[{{Sparke} \& {Casertano}(1988)}]{Sparke88}
{Sparke}, L.~S. \& {Casertano}, S. 1988, \mnras, 234, 873

\bibitem[{{Sparke} {et~al.}(2008){Sparke}, {van Moorsel}, {Erwin}, \&
  {Wehner}}]{Sparke08}
{Sparke}, L.~S., {van Moorsel}, G., {Erwin}, P., \& {Wehner}, E.~M.~H. 2008,
  \aj, 135, 99

\bibitem[{{Taniguchi} {et~al.}(1990){Taniguchi}, {Sofue}, {Wakamatsu}, \&
  {Nakai}}]{Taniguchi90}
{Taniguchi}, Y., {Sofue}, Y., {Wakamatsu}, K.-I., \& {Nakai}, N. 1990, \aj,
  100, 1086

\bibitem[{{Tohline} \& {Durisen}(1982)}]{Tohline82}
{Tohline}, J.~E. \& {Durisen}, R.~H. 1982, \apj, 257, 94

\bibitem[{{Tully} {et~al.}(1998){Tully}, {Pierce}, {Huang}, {Saunders},
  {Verheijen}, \& {Witchalls}}]{Tully98}
{Tully}, R.~B., {Pierce}, M.~J., {Huang}, J.-S., {et~al.} 1998, \aj, 115, 2264

\bibitem[{{Ulrich}(1975)}]{Ulrich75}
{Ulrich}, M.-H. 1975, \pasp, 87, 965

\bibitem[{{van Albada} {et~al.}(1982){van Albada}, {Kotanyi}, \&
  {Schwarzschild}}]{Albada82}
{van Albada}, T.~S., {Kotanyi}, C.~G., \& {Schwarzschild}, M. 1982, \mnras,
  198, 303

\bibitem[{{van Driel} {et~al.}(2000){van Driel}, {Arnaboldi}, {Combes}, \&
  {Sparke}}]{Driel00}
{van Driel}, W., {Arnaboldi}, M., {Combes}, F., \& {Sparke}, L.~S. 2000, \aaps,
  141, 385

\bibitem[{{Varnas}(1990)}]{Varnas90}
{Varnas}, S.~R. 1990, \mnras, 247, 674

\bibitem[{{Whitmore} {et~al.}(1990){Whitmore}, {Lucas}, {McElroy},
  {Steiman-Cameron}, {Sackett}, \& {Olling}}]{Whitmore90}
{Whitmore}, B.~C., {Lucas}, R.~A., {McElroy}, D.~B., {et~al.} 1990, \aj, 100,
  1489

\end{thebibliography}
\appendix
\newpage
\section{Data cubes and moment maps}
\label{Sect_4.8}
%
%
\begin{figure*}[htbp]
\begin{center}
\includegraphics[width=0.9\textwidth]{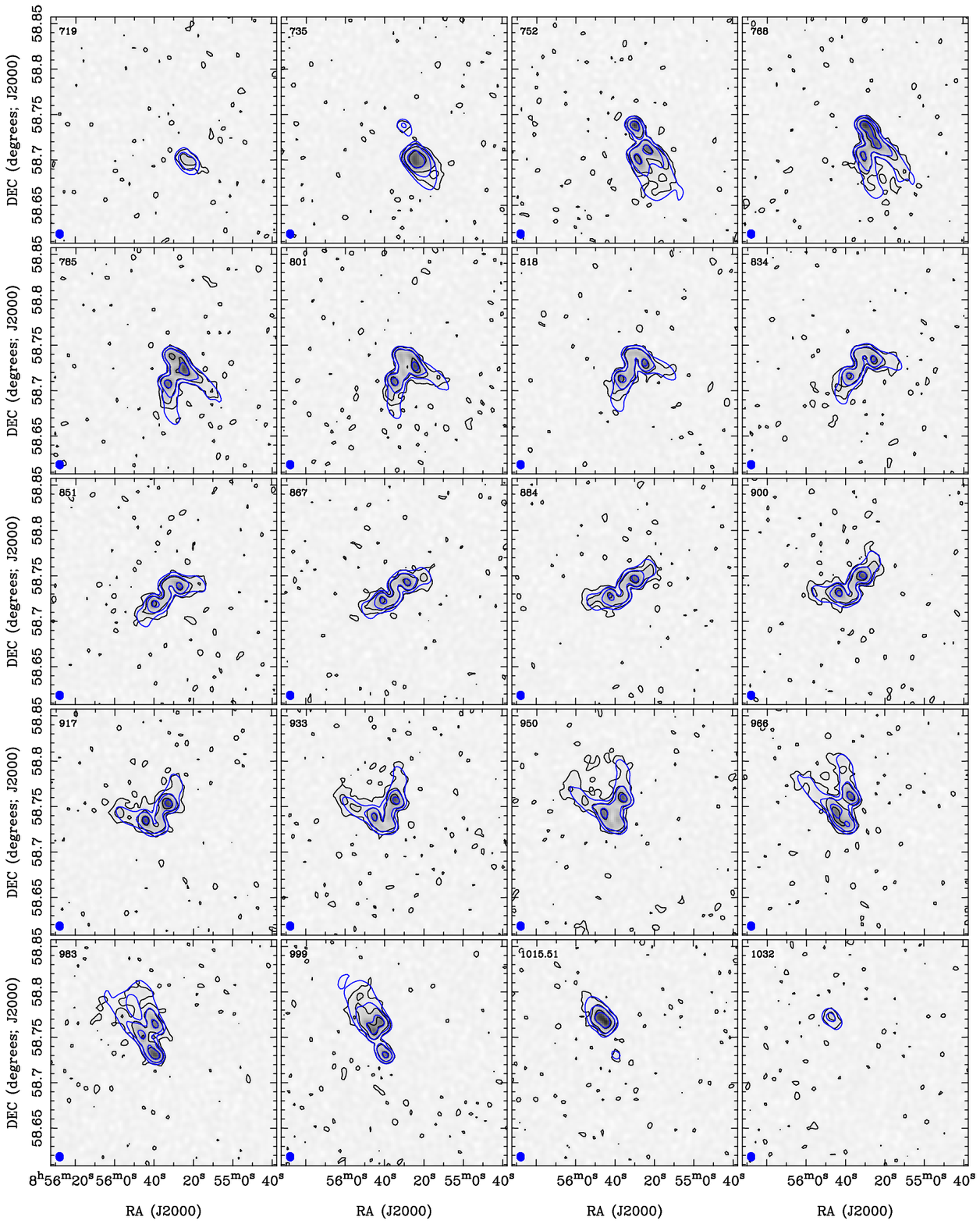}
\caption{
Selected images from the low-resolution data cube (channel width of
$4.12\,{\rm km}\,{\rm s}^{-1}$, Robust weighting 0.4, baselines of
length $<6.4\,{\rm k\lambda}$ used) overlaid with contours of the
original dataset (black) and the final model data cube (grey, blue in
online-version). The numbers on the upper left give the heliocentric
radio velocity in ${\rm km}\,{\rm s}^{-1}$. The dot in the lower left
corner represents the clean beam. The contours represent the $0.5, 2,
8 \mJyb$ levels.  }
\label{Fig_A1}
\end{center}
\end{figure*}
%
%
\begin{figure*}[htbp]
\begin{center}
\includegraphics[angle=270,width=0.9\textwidth]{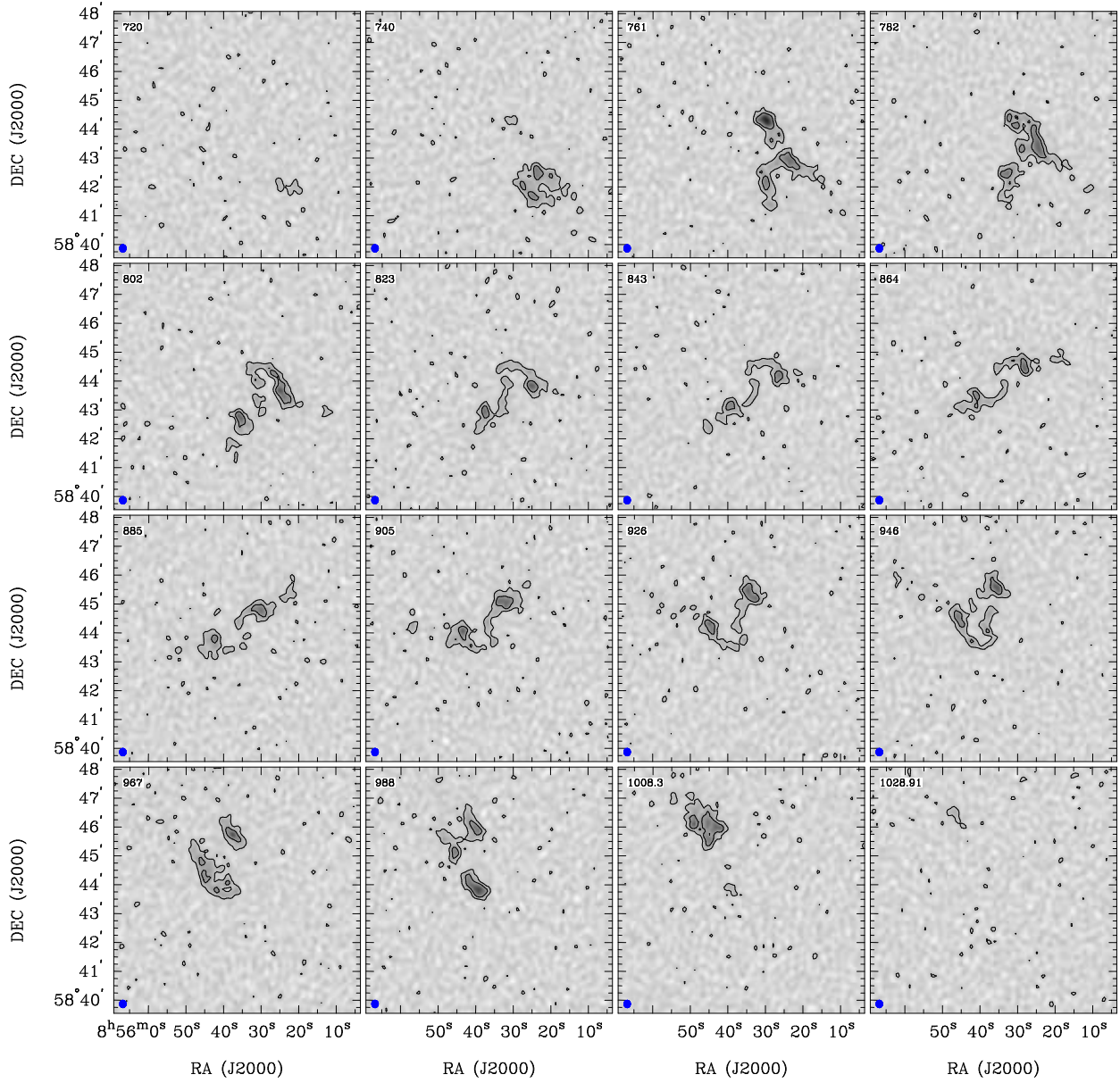}
\caption{
Selected images from the high-resolution data cube (channel width
of $2.06\,{\rm km}\,{\rm s}^{-1}$, uniform weighting, all visibilities
regarded). The numbers on the upper left give the heliocentric radio
velocity in ${\rm km}\,{\rm s}^{-1}$ . The dot in the lower left
corner represents the clean beam. The contours represent the $0.75, 3,
4.5 \mJyb$ levels.
}
\label{Fig_A2}
\end{center}
\end{figure*}
%
%
\begin{figure*}[hbp]
  \begin{center}
  \includegraphics[angle=270,width=8.5cm]{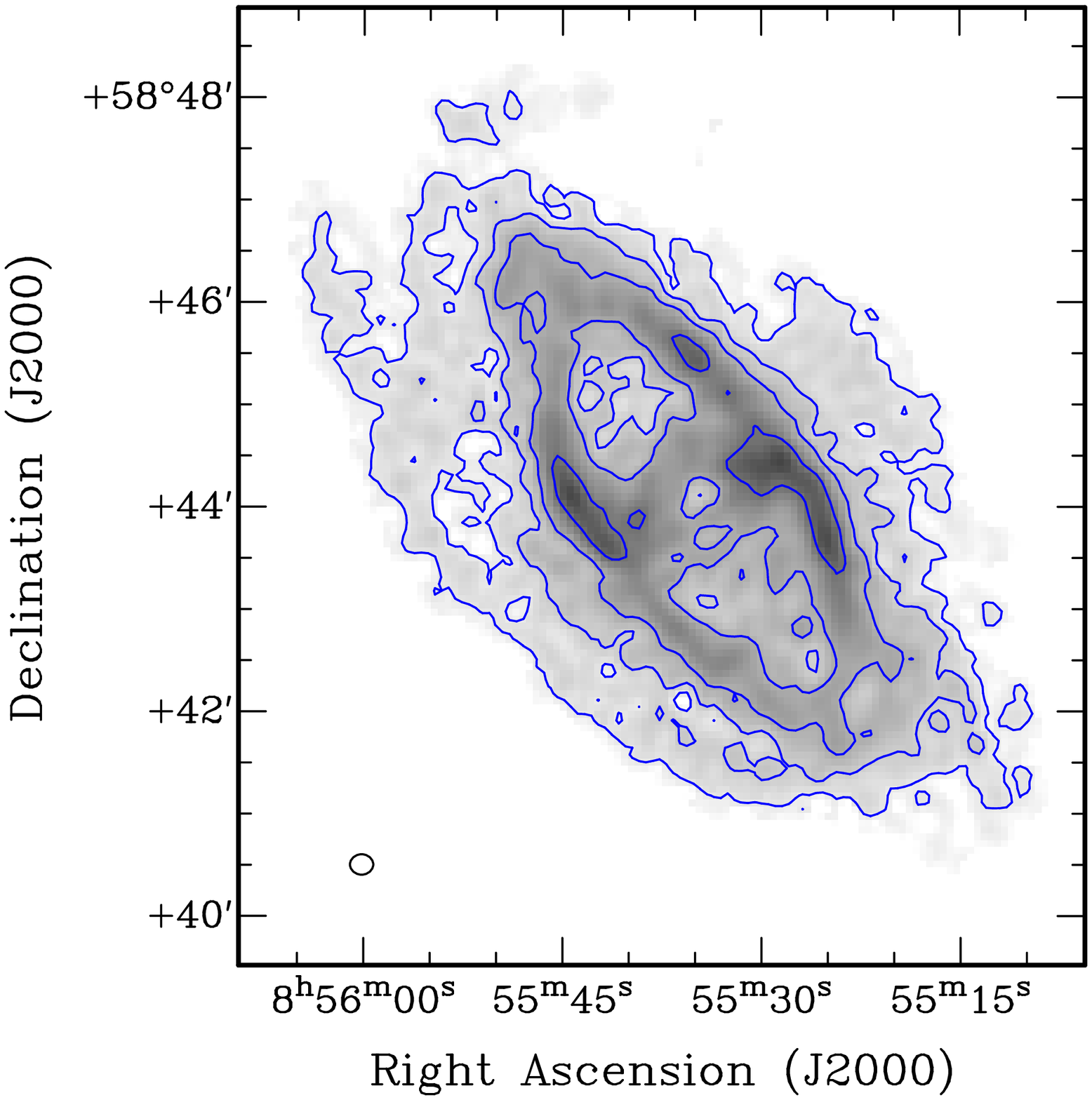}
  \includegraphics[angle=270,width=8.5cm]{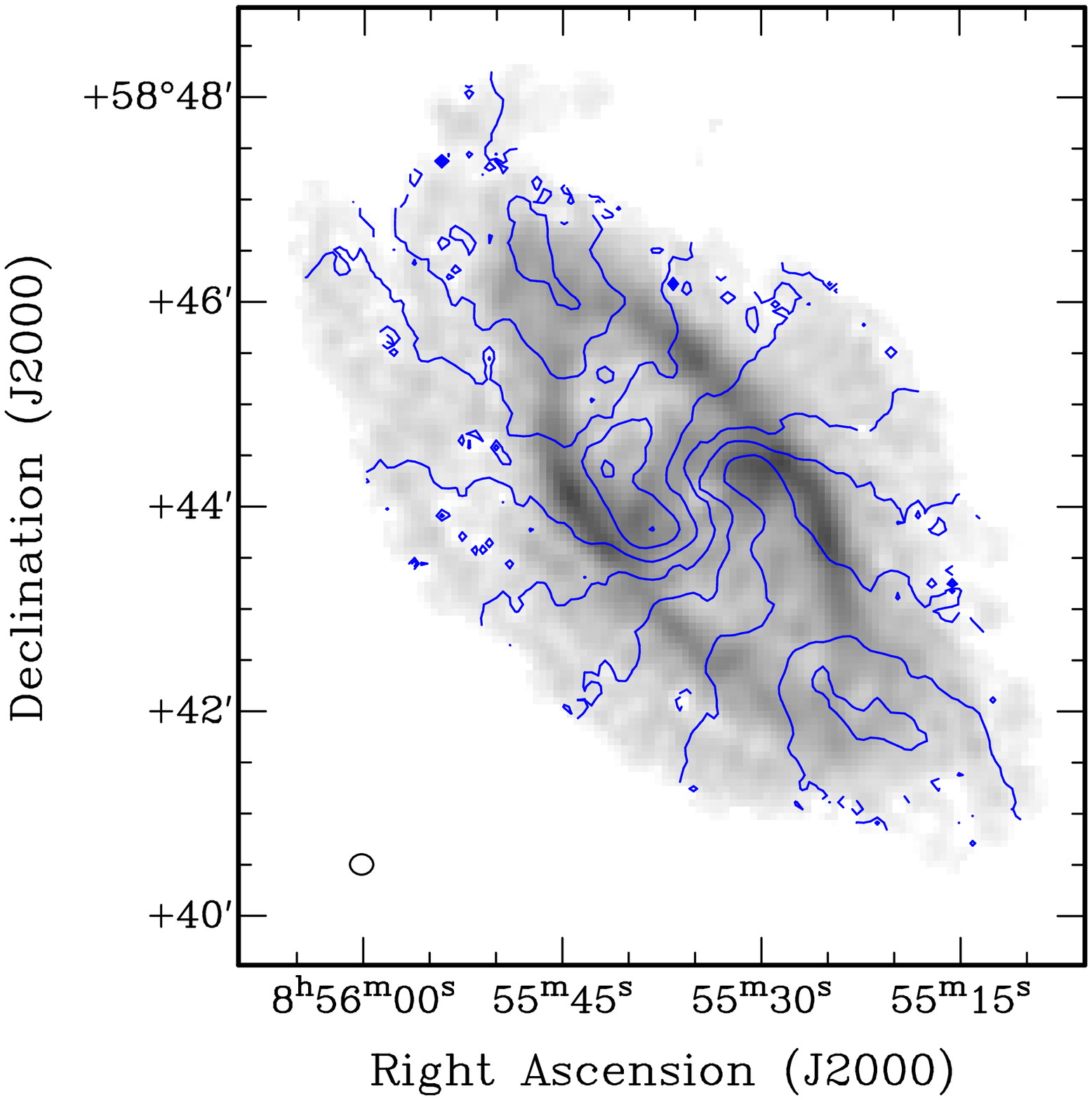}
\caption{
Total-intensity map (left) and first-moment velocity field (right)
derived from the high-resolution data cube overlaid on a total
intensity grey-scale map. The ellipse in the lower left corner
represents the clean beam. The contours represent the $5,20,50,140
\tth$ levels and the $V_{\rm sys} \pm 0,40,80,120,140\,{\rm km}\,{\rm
s}^{-1}$ levels respectively (see Table~\ref{Tab_2}).
}
\label{Fig_A3}
  \end{center}
\end{figure*}
%
%
\clearpage
\section{Results from spectral fits}
\begin{table*}[htbp]
\begin{center}
\caption{
Velocities and masses of gas detected at ``anomalous'' velocities as
derived from fitting a double-Gaussian to several profiles of the data
cube.
{\bf (1)}  Position, right ascension (J2000) of profile.
{\bf (2)}  Position, declination (J2000) of profile.
{\bf (3)} Difference between peak position belonging to inner
component and systemic velocity: assumed projected rotation velocity
(${\rm km}\,{\rm s}^{-1}$).
{\bf (4)} Difference between peak position belonging to inner
component and far side 20 percent peak position of residual: maximal
excess motion with respect to inner component (${\rm km}\,{\rm
s}^{-1}$).
{\bf (5)} Difference between peak position belonging to inner
component and bulk velocity of residual: bulk excess motion with
respect to inner component (${\rm km}\,{\rm s}^{-1}$).
{\bf (6)} Difference between peak position belonging to outer
component and far side 20 percent peak position of residual: maximal
excess motion with respect to inner component (${\rm km}\,{\rm
s}^{-1}$).
{\bf (7)} Difference between peak position belonging to outer
component and bulk velocity of residual: bulk excess motion with
respect to inner component (${\rm km}\,{\rm s}^{-1}$).
{\bf (8)} Total flux of residual within the boundaries set by 20
percent of the peak intensity (${\rm Jy}\,{\rm km}\,{\rm s}^{-1}$).
{\bf (9)} Total flux derived from Gaussian fit to the inner component
(${\rm Jy}\,{\rm km}\,{\rm s}^{-1}$).
{\bf (10)} Total flux derived from Gaussian fit to the outer component
(${\rm Jy}\,{\rm km}\,{\rm s}^{-1}$).
}
\label{Tab_3a}
\begin{tabular}{r r r r r r r r r r}
\hline
\hline

$ \rm RA $ & $ \rm Dec $ & $\Delta V_{\rm i,sys} $ & $\Delta V_{\rm i,max} $ & $\Delta V_{\rm i,bulk} $ & $\Delta V_{\rm o,max} $ & $\Delta V_{\rm o,bulk} $ & $F_{\rm {\ion{H}{i}},e}$ & $F_{\rm {\ion{H}{i}},i}$ & $F_{\rm {\ion{H}{i}},o}$ \\ 
(1)           & (2)           & (3)                 & (4)               & (5)                     & (6)                         & (7)                                 & (8)      & (9)              & (10)          \\
\hline
$ \hmsm{08}{55}{29}{204} $ & $ \dmsm{58}{44}{23}{98} $ & $ 110.0  $ & $  71.5  $ & $  44.7  $ & $  82.8  $ & $  56.0  $ & $  0.21  $ & $  0.60  $ & $  0.34  $ \\
$ \hmsm{08}{55}{39}{994} $ & $ \dmsm{58}{43}{41}{99} $ & $ 112.3  $ & $  97.2  $ & $  58.1  $ & $  94.4  $ & $  55.2  $ & $  0.23  $ & $  0.47  $ & $  0.18  $ \\
$ \hmsm{08}{55}{29}{204} $ & $ \dmsm{58}{44}{30}{98} $ & $ 109.6  $ & $  75.2  $ & $  48.4  $ & $  82.1  $ & $  55.3  $ & $  0.21  $ & $  0.42  $ & $  0.47  $ \\
$ \hmsm{08}{55}{39}{993} $ & $ \dmsm{58}{43}{34}{99} $ & $ 114.0  $ & $  98.9  $ & $  57.7  $ & $ 102.2  $ & $  61.0  $ & $  0.29  $ & $  0.26  $ & $  0.26  $ \\
$ \hmsm{08}{55}{29}{204} $ & $ \dmsm{58}{44}{37}{99} $ & $ 109.9  $ & $  79.6  $ & $  48.7  $ & $  90.5  $ & $  59.6  $ & $  0.18  $ & $  0.22  $ & $  0.56  $ \\
$ \hmsm{08}{55}{39}{993} $ & $ \dmsm{58}{43}{27}{98} $ & $ 113.8  $ & $  98.7  $ & $  55.4  $ & $ 109.9  $ & $  66.6  $ & $  0.25  $ & $  0.12  $ & $  0.36  $ \\
$ \hmsm{08}{55}{28}{306} $ & $ \dmsm{58}{44}{23}{98} $ & $ 107.8  $ & $  69.3  $ & $  44.6  $ & $  74.6  $ & $  49.9  $ & $  0.21  $ & $  0.46  $ & $  0.42  $ \\
$ \hmsm{08}{55}{40}{893} $ & $ \dmsm{58}{43}{41}{99} $ & $ 113.1  $ & $  98.0  $ & $  56.8  $ & $  96.4  $ & $  55.2  $ & $  0.35  $ & $  0.32  $ & $  0.26  $ \\
$ \hmsm{08}{55}{28}{304} $ & $ \dmsm{58}{44}{30}{98} $ & $ 108.6  $ & $  70.1  $ & $  45.3  $ & $  78.6  $ & $  53.8  $ & $  0.21  $ & $  0.29  $ & $  0.50  $ \\
$ \hmsm{08}{55}{40}{893} $ & $ \dmsm{58}{43}{34}{99} $ & $ 111.8  $ & $  96.7  $ & $  55.5  $ & $  99.3  $ & $  58.1  $ & $  0.31  $ & $  0.21  $ & $  0.34  $ \\
$ \hmsm{08}{55}{28}{304} $ & $ \dmsm{58}{44}{37}{99} $ & $ 109.8  $ & $  75.5  $ & $  46.6  $ & $  87.0  $ & $  58.2  $ & $  0.17  $ & $  0.15  $ & $  0.54  $ \\
$ \hmsm{08}{55}{40}{893} $ & $ \dmsm{58}{43}{27}{98} $ & $ 109.1  $ & $  89.9  $ & $  50.8  $ & $ 102.8  $ & $  63.6  $ & $  0.22  $ & $  0.11  $ & $  0.40  $ \\
$ \hmsm{08}{55}{27}{405} $ & $ \dmsm{58}{44}{23}{98} $ & $ 103.8  $ & $  65.3  $ & $  40.5  $ & $  71.0  $ & $  46.2  $ & $  0.21  $ & $  0.31  $ & $  0.45  $ \\
$ \hmsm{08}{55}{41}{792} $ & $ \dmsm{58}{43}{41}{99} $ & $ 108.3  $ & $  76.8  $ & $  43.8  $ & $  86.0  $ & $  53.1  $ & $  0.25  $ & $  0.25  $ & $  0.45  $ \\
$ \hmsm{08}{55}{27}{405} $ & $ \dmsm{58}{44}{30}{98} $ & $ 105.4  $ & $  66.8  $ & $  42.1  $ & $  74.9  $ & $  50.2  $ & $  0.19  $ & $  0.18  $ & $  0.48  $ \\
$ \hmsm{08}{55}{41}{792} $ & $ \dmsm{58}{43}{34}{99} $ & $  91.5  $ & $  68.2  $ & $  43.4  $ & $  69.0  $ & $  44.2  $ & $  0.12  $ & $  0.30  $ & $  0.41  $ \\
$ \hmsm{08}{55}{27}{405} $ & $ \dmsm{58}{44}{37}{99} $ & $ 107.8  $ & $  69.3  $ & $  42.5  $ & $  83.6  $ & $  56.8  $ & $  0.13  $ & $  0.08  $ & $  0.46  $ \\
$ \hmsm{08}{55}{41}{792} $ & $ \dmsm{58}{43}{27}{98} $ & $ 102.1  $ & $  78.8  $ & $  43.8  $ & $  94.5  $ & $  59.5  $ & $  0.17  $ & $  0.09  $ & $  0.39  $ \\
\hline
\multicolumn{2}{c}{mean} &$ 108.3  $ & $  80.3  $ & $  48.3  $ & $  87.8  $ & $  55.7  $ & $  0.22  $ & $  0.27  $ & $  0.40  $ \\
\multicolumn{2}{c}{standard deviation} &$   5.3  $ & $  12.6  $ & $   5.9  $ & $  11.8  $ & $   5.7  $ & $  0.06  $ & $  0.14  $ & $  0.10  $ \\
\hline
\end{tabular}
\end{center}
\end{table*}
%
%
\clearpage
\begin{landscape}
\section{Tilted-ring parameters}
\label{Sect_4.9}
\begin{table}[hbp]
\begin{center}
\caption{
Radially dependent best-fit parameters.
{\bf (1)}  Radius ($\,\arcsec$).
{\bf (2)}  Radius ($\rm kpc$, according to Table~\ref{Tab_1}).
{\bf (3)}  Error of $ r_{\rm t} $ ($\rm kpc$)
{\bf (4)}  Face-on surface brightness (${\rm Jy}\,{\rm km}\,{\rm s}^{-1}$).
{\bf (5)}  Error of $I_{\rm tot,f}$ (${\rm Jy}\,{\rm km}\,{\rm s}^{-1}$).
{\bf (6)}  Face-on {H\,{\small I}} column-density ($10^{19}\,{\rm atoms}\,{\rm cm}^{-2}$).
{\bf (7)}  Error of $N_{\rm {H\,{\small I}}}$ ($10^{19}\,{\rm atoms}\,{\rm cm}^{-2}$).
{\bf (8)}  Face-on surface density (${\rm M}_{\odot}\,{\rm pc}^{-2}$).
{\bf (9)}  Error of $\sigma$(${\rm M}_{\odot}\,{\rm pc}^{-2}$).
{\bf (10)} Rotation velocity (${\rm km}\,{\rm s}^{-1}$).
{\bf (11)} Error of $V_{\rm rot}$ (${\rm km}\,{\rm s}^{-1}$).
{\bf (12)} Inclination ($\,\deg$).
{\bf (13)} Error of $i$ ($\,\deg$).
{\bf (14)} Position angle ($\,\deg$).
{\bf (15)} Error of $pa$ ($\,\deg$).
{\bf (16)} Inclination ($\,\deg$).
{\bf (17)}-{\bf (23)} are Cartesian components of the spin normal
vector ${\bf n}$ of the ring and their errors.
{\bf (17)} Spin normal vector component towards W (natural units).
{\bf (18)} Error of $n_{\rm W}$ (natural units).
{\bf (19)} Spin normal vector component towards N (natural units).
{\bf (20)} Error of $n_{\rm N}$ (natural units).
{\bf (21)} Spin normal vector component towards observer (natural units).
{\bf (22)} Error of $n_{\rm LOS}$ (natural units).
}
\label{Tab_5}
\begin{tabular}{r r r r r r r r r r r r r r r r r r r r r r}
\hline
\hline
$ r_{\rm p} $ & $ r_{\rm t} $ & $\Delta\,r_{\rm t}$ & $ I_{\rm tot,f} $ & $\Delta\,I_{\rm tot,f}$ & $ N_{\rm {H\,{\small I}}} $ & $ \Delta\,N_{\rm {H\,{\small I}}} $ & $\sigma$ & $\Delta\,\sigma$ & $V_{\rm rot}$ & $\Delta\,V_{\rm rot}$ & $i$ & $\Delta\,i$ & $pa$ & $\Delta\,pa$ & $n_{\rm W}$ &  $\Delta\,n_{\rm W}$ & $n_{\rm N}$ & $\Delta\,n_{\rm N}$ & $n_{\rm LOS}$ & $\Delta\,n_{\rm LOS}$ \\
(1)           & (2)           & (3)                 & (4)               & (5)                     & (6)                         & (7)                                 & (8)      & (9)              & (10)          & (11)                  & (12)          & (13)                & (14)              & (15)                    & (16)           & (17)                   & (18)     & (19)            & (20)          & (21)\\
\hline
$     0 $ & $  0.00 $ & $  0.00 $ & $   0.0 $ & $   2.0 $ & $   0.0 $ & $   2.5 $ & $   0.0 $ & $   0.2 $ & $   0.0 $  & $   0.0 $ &  &  &  &  &  &  &  &  &  & \\
$    12 $ & $  0.88 $ & $  0.22 $ & $   0.0 $ & $   2.0 $ & $   0.0 $ & $   2.5 $ & $   0.0 $ & $   0.2 $ & $ 154.2 $  & $  79.0 $ & $  35.3 $ & $  18.7 $ & $ 245.0 $ & $  68.7 $ & $ -0.524 $ & $  0.379 $ & $  0.244 $ & $  0.639 $ & $  0.816 $ & $  0.189 $\\
$    24 $ & $  1.77 $ & $  0.44 $ & $   8.8 $ & $  20.8 $ & $  10.9 $ & $  25.9 $ & $   0.9 $ & $   2.1 $ & $ 134.9 $  & $  13.1 $ & $  65.4 $ & $  11.7 $ & $ 217.7 $ & $  10.5 $ & $ -0.556 $ & $  0.142 $ & $  0.720 $ & $  0.122 $ & $  0.416 $ & $  0.186 $\\
$    36 $ & $  2.65 $ & $  0.66 $ & $  72.4 $ & $   7.3 $ & $  90.4 $ & $   9.1 $ & $   7.2 $ & $   0.7 $ & $ 131.2 $  & $   5.6 $ & $  69.7 $ & $   4.0 $ & $ 204.7 $ & $   4.0 $ & $ -0.392 $ & $  0.060 $ & $  0.852 $ & $  0.035 $ & $  0.348 $ & $  0.065 $\\
$    48 $ & $  3.54 $ & $  0.87 $ & $  57.2 $ & $  12.1 $ & $  71.5 $ & $  15.1 $ & $   5.7 $ & $   1.2 $ & $ 144.6 $  & $  10.0 $ & $  54.4 $ & $   4.9 $ & $ 197.9 $ & $   5.9 $ & $ -0.249 $ & $  0.081 $ & $  0.774 $ & $  0.054 $ & $  0.582 $ & $  0.070 $\\
$    60 $ & $  4.42 $ & $  1.09 $ & $  33.5 $ & $  11.8 $ & $  41.9 $ & $  14.7 $ & $   3.4 $ & $   1.2 $ & $ 163.9 $  & $  30.5 $ & $  47.8 $ & $   7.6 $ & $ 160.0 $ & $   8.9 $ & $  0.253 $ & $  0.112 $ & $  0.696 $ & $  0.093 $ & $  0.672 $ & $  0.099 $\\
$    72 $ & $  5.31 $ & $  1.31 $ & $  39.0 $ & $  11.5 $ & $  48.7 $ & $  14.3 $ & $   3.9 $ & $   1.1 $ & $ 146.1 $  & $  18.2 $ & $  41.5 $ & $   8.3 $ & $ 161.1 $ & $  10.9 $ & $  0.215 $ & $  0.125 $ & $  0.627 $ & $  0.111 $ & $  0.749 $ & $  0.096 $\\
$    84 $ & $  6.19 $ & $  1.53 $ & $  15.9 $ & $  11.2 $ & $  19.8 $ & $  14.0 $ & $   1.6 $ & $   1.1 $ & $ 183.1 $  & $  41.9 $ & $  31.0 $ & $  11.3 $ & $ 147.4 $ & $   4.1 $ & $  0.277 $ & $  0.096 $ & $  0.434 $ & $  0.144 $ & $  0.857 $ & $  0.101 $\\
$    96 $ & $  7.08 $ & $  1.75 $ & $  22.0 $ & $   9.9 $ & $  27.4 $ & $  12.4 $ & $   2.2 $ & $   1.0 $ & $ 184.0 $  & $  20.7 $ & $  33.6 $ & $  13.6 $ & $ 120.0 $ & $  12.5 $ & $  0.479 $ & $  0.182 $ & $  0.276 $ & $  0.144 $ & $  0.833 $ & $  0.132 $\\
$   108 $ & $  7.96 $ & $  1.97 $ & $  14.0 $ & $   8.6 $ & $  17.5 $ & $  10.8 $ & $   1.4 $ & $   0.9 $ & $ 141.5 $  & $   6.4 $ & $  60.6 $ & $   4.0 $ & $ 132.9 $ & $   6.4 $ & $  0.638 $ & $  0.071 $ & $  0.593 $ & $  0.075 $ & $  0.490 $ & $  0.061 $\\
$   120 $ & $  8.85 $ & $  2.19 $ & $  21.9 $ & $   3.9 $ & $  27.4 $ & $   4.9 $ & $   2.2 $ & $   0.4 $ & $ 160.0 $  & $   8.2 $ & $  59.5 $ & $   4.0 $ & $ 126.1 $ & $   4.0 $ & $  0.696 $ & $  0.046 $ & $  0.508 $ & $  0.053 $ & $  0.507 $ & $  0.060 $\\
$   140 $ & $ 10.32 $ & $  2.55 $ & $  16.8 $ & $   2.0 $ & $  21.0 $ & $   2.5 $ & $   1.7 $ & $   0.2 $ & $ 152.2 $  & $  11.6 $ & $  67.3 $ & $   4.0 $ & $ 122.6 $ & $   4.0 $ & $  0.777 $ & $  0.041 $ & $  0.497 $ & $  0.056 $ & $  0.385 $ & $  0.064 $\\
$   160 $ & $ 11.80 $ & $  2.91 $ & $  17.5 $ & $   2.0 $ & $  21.8 $ & $   2.5 $ & $   1.7 $ & $   0.2 $ & $ 151.0 $  & $   8.2 $ & $  70.0 $ & $   4.0 $ & $ 125.5 $ & $   4.0 $ & $  0.765 $ & $  0.043 $ & $  0.545 $ & $  0.055 $ & $  0.342 $ & $  0.066 $\\
$   180 $ & $ 13.27 $ & $  3.28 $ & $  15.2 $ & $   2.0 $ & $  18.9 $ & $   2.5 $ & $   1.5 $ & $   0.2 $ & $ 156.4 $  & $   8.2 $ & $  69.1 $ & $   4.2 $ & $ 126.7 $ & $   4.0 $ & $  0.749 $ & $  0.044 $ & $  0.558 $ & $  0.055 $ & $  0.358 $ & $  0.068 $\\
$   200 $ & $ 14.75 $ & $  3.64 $ & $   5.0 $ & $   2.0 $ & $   6.2 $ & $   2.5 $ & $   0.5 $ & $   0.2 $ & $ 152.1 $  & $   8.2 $ & $  64.5 $ & $   7.0 $ & $ 125.8 $ & $   4.0 $ & $  0.732 $ & $  0.057 $ & $  0.527 $ & $  0.060 $ & $  0.431 $ & $  0.111 $\\
$   240 $ & $ 17.69 $ & $  4.37 $ & $   2.4 $ & $   2.0 $ & $   3.0 $ & $   2.5 $ & $   0.2 $ & $   0.2 $ & $ 147.5 $  & $   8.2 $ & $  63.6 $ & $   4.0 $ & $ 124.5 $ & $   4.0 $ & $  0.738 $ & $  0.044 $ & $  0.507 $ & $  0.054 $ & $  0.445 $ & $  0.063 $\\
$   280 $ & $ 20.64 $ & $  5.10 $ & $   2.0 $ & $   2.0 $ & $   2.5 $ & $   2.5 $ & $   0.2 $ & $   0.2 $ & $ 137.5 $  & $   8.2 $ & $  62.5 $ & $   4.0 $ & $ 121.3 $ & $   4.2 $ & $  0.758 $ & $  0.044 $ & $  0.460 $ & $  0.059 $ & $  0.462 $ & $  0.062 $\\
$   320 $ & $ 23.59 $ & $  5.83 $ & $   0.5 $ & $   2.0 $ & $   0.6 $ & $   2.5 $ & $   0.0 $ & $   0.2 $ & $ 143.9 $  & $  15.0 $ & $  64.8 $ & $   4.3 $ & $ 123.2 $ & $  13.7 $ & $  0.757 $ & $  0.122 $ & $  0.496 $ & $  0.182 $ & $  0.425 $ & $  0.068 $\\
$   360 $ & $ 26.54 $ & $  6.56 $ & $   0.8 $ & $   2.0 $ & $   1.0 $ & $   2.5 $ & $   0.1 $ & $   0.2 $ & $ 139.6 $  & $  15.0 $ & $  66.7 $ & $   4.0 $ & $ 121.9 $ & $   7.1 $ & $  0.780 $ & $  0.065 $ & $  0.485 $ & $  0.098 $ & $  0.396 $ & $  0.064 $\\
$   400 $ & $ 29.49 $ & $  7.29 $ & $   0.2 $ & $   2.0 $ & $   0.2 $ & $   2.5 $ & $   0.0 $ & $   0.2 $ & $ 146.7 $  & $  15.0 $ & $  66.9 $ & $   8.4 $ & $ 119.5 $ & $  10.0 $ & $  0.801 $ & $  0.093 $ & $  0.454 $ & $  0.143 $ & $  0.392 $ & $  0.134 $\\
$   420 $ & $ 30.96 $ & $  7.65 $ & $   0.9 $ & $   2.0 $ & $   1.2 $ & $   2.5 $ & $   0.1 $ & $   0.2 $ & $ 168.8 $  & $  15.0 $ & $  67.4 $ & $  10.6 $ & $ 119.0 $ & $  10.0 $ & $  0.807 $ & $  0.100 $ & $  0.448 $ & $  0.145 $ & $  0.384 $ & $  0.171 $\\
\hline
\end{tabular}
\end{center}
\end{table}
\end{landscape}
\end{document}